\documentclass{aa}
\usepackage{CJK}
\usepackage{enumitem}
\usepackage[varg]{txfonts}
\usepackage{lipsum}
\usepackage{amsmath}
\usepackage{physics}
\usepackage{tablefootnote}
\usepackage{supertabular}
\usepackage{pifont}
\usepackage{multirow}

\newcommand{\Ni}{$^{56}$Ni}

\newcommand{\Ms}{M$_{\odot}$}

\newcommand      \grays       {$\gamma$-rays}
\newcommand      \gray       {$\gamma$-ray}

\newcommand{\mic}{$\mu$m}

\def\gsim{\mathrel{\raise.5ex\hbox{$>$}\mkern-14mu
             \lower0.6ex\hbox{$\sim$}}}

\def\lsim{\mathrel{\raise.3ex\hbox{$<$}\mkern-14mu
             \lower0.6ex\hbox{$\sim$}}}

\newlength\mylen
\newlist{mycases}{enumerate}{1}
\setlist[mycases,1]{label=\textbf{Case~\arabic*.}, 
labelwidth=\dimexpr-\mylen-\labelsep\relax,leftmargin=0pt}         

\newlist{newcases}{enumerate}{1}
\setlist[newcases,1]{
  label={\textbf{Case~\arabic*}.},
  leftmargin=*,
  align=left,
  labelsep=1mm,
  itemindent=\dimexpr\labelsep+\labelwidth+10pt\relax
}

\begin{document}

\title{Formation, distribution, and IR emission of dust in the clumpy ejecta of Type II-P core-collapse supernovae, in isotropic and anisotropic scenarios }
\titlerunning{Dust in supernova ejecta}
\author{Arkaprabha Sarangi}
\institute{DARK, Niels Bohr Institute, University of Copenhagen, Jagtvej 128, 2200 Copenhagen, Denmark \\
\email{sarangi@nbi.ku.dk}}
\authorrunning{Sarangi, A.}

   \date{Received 1 July 2022 / Accepted 28 September 2022}

 
  \abstract{

Large discrepancies are found between observational estimates and theoretical predictions when exploring the characteristics of dust formed in the ejecta of a core-collapse supernovae. We revisit the scenario of dust production in typical supernova ejecta in the first 3000 days after explosion, with an improved understanding of the evolving physical conditions and the distribution of the clumps. The generic, nonuniform distribution of dust within the ejecta was determined and using that, the relevant opacities and fluxes were calculated. The dependence of the emerging fluxes on the viewing angle was estimated for an anisotropic, ellipsoidal geometry of the ejecta that imitate SN~1987A. We model the He core from the center to its outer edge as 450 stratified, clumpy, annular shells, uniquely identified by their distinct velocities and characterized by their variations in abundances, densities, and gas and dust temperatures. We find that the formation of dust starts between day 450 and day 550 post-explosion, and it continues until about day 2800, although the first 1600 days are the most productive. The total dust mass evolves from $\sim$ 10$^{-5}$ \Ms\ at day 500 to  10$^{-3}$ \Ms\ at day 800, finally saturating at about 0.06 \Ms. The masses of the O-rich dust (silicates, alumina) dominates the C-rich dust (amorphous carbon, silicon carbide) at all times; the formation of carbon dust is delayed beyond 2000 days post-explosion. We show that the opacities are largest between days 800 and 1600, and the characteristic spectral features of O-rich dust species are suppressed at those times. The fluxes emerging along the smallest axes of the ellipsoidal ejecta are found to be the most obscured, while a viewing angle between 16 to 21$^\circ$ with that axis appears to be in best agreement with the fluxes from SN~1987A at days 615 and 775. 

}

   \maketitle

\section{Introduction}
\label{sec_intro}

The rapidly evolving ejecta of core-collapse supernovae (SNe) are known sites for the synthesis of cosmic dust \citep{bou93, woo93, dwek_2006, sza13, matsuura2017, sarangi2018book}, and are often proposed to be the primary source of dust in high-redshift galaxies \citep{dwe07, zhukovska_2008, cherchneff2009, gal11, gall_2018}. 
The mass and properties of dust present in SN ejecta are quantified by analyzing the spectral energy distribution (SED) at various wavelengths, where the mid-infrared (IR) and far-IR part of the SED is often dominated by emission from dust formed in the ejecta. 
Observations of several SNe in the mid-IR wavelengths over the past couple of decades found the upper-limit of dust formed in the first 1000 days of the SN explosion to be constrained to within a few times 10$^{-3}$ \Ms\ \citep{sug06, erc07, kot09, wesson2015, priestley_2020, sza13, szalai_2019}. 

In addition to the emission in the IR, the presence of dust in the ejecta also gives rise to an asymmetry between the redshifted and blueshifted emission lines in the optical and near-IR wavelengths, since the radiation coming toward us from the far side of the ejecta is subjected to increased absorption and scattering by the dust lying along its path \citep{bevan2016}. 
The quantity and composition of dust that is required to account for the progressive blueshift of the [OI] doublet in SN~1987A \citep{luc89} was estimated by  \cite{bevan2016}. Combining this to the fitting of the mid-IR spectra using a radiative transfer model, it was suggested that the mass of dust formed in the ejecta of SN~1987A was no more than 3 $\times$ 10$^{-3}$ \Ms\ at 1000 days, with a predominantly carbon-rich composition \citep{wesson2015, wesson_2021, bevan_2018}. 

Interestingly, a much larger mass of dust, of about $\sim$ 0.5 \Ms, was found to be present in SN~1987A after 23 years of its explosion, derived from the fluxes in the far-IR and submillimeter wavelengths \citep{mat11, ind14, matsuura_2015}. This trend is also evident in the cold ejecta of other SN remnants (SNRs) such as Cas A \citep{barlow_2010, arendt_2014, delooze2017, maria_2021} and the Crab Nebula \citep{gom12, owen_2015, delooze_2019}. The apparent dichotomy between the estimated dust masses in the mid-IR and submillimeter wavelengths is perceived as evidence of a gradual increase in dust masses due to continuous dust formation over a couple of decades \citep{gal14, wesson2015, gall_2018}. 

\cite{dwe15} and \cite{dwek_2019} propose an alternative scenario where large masses of dust condense within a couple of years after the explosion, when the ejecta are still hot. However, the large masses are not reflected in the SED due to the large opacities of the ejecta in the mid-IR wavelengths. 
\cite{dwe15} find a reasonably good fit to the spectra of SN~1987A at days 615 and 775 \citep{woo93}, with a rather large dust mass of $\sim$ 0.45 \Ms, dominated by silicates, formed in dense clumps. 
The early synthesis of O-rich dust, such as silicates, complies well with the characteristic 9.7- and 18-\mic\ feature of silicate dust in the interstellar medium (ISM) \citep{breemen_2011}, which requires high temperatures, typically larger than 1000~K, during its formation \citep{dwek_2019}. 
In addition, the sharp decrease in intensity of the [Mg I] or [Si I] emission lines \citep{dan91}, concurrent with the increase in the mid-IR emission after day 530, also supports the formation of Mg-silicates in abundance in the gas \citep{dwe15}. 
 On the contrary, a rapid formation of dust leading to an early saturation is not congruent with the observed trend of a continuous increase in the blueshift of the emission lines \citep{bevan2016}, if a spherical shape is assumed for the ejecta. 

Dust forms in the SN ejecta through simultaneous phases of nucleation and condensation \citep{cherchneff2008, sarangi2018book}. Dust formation models, in general, support the scenario of efficient dust formation in the first 2-3 years after the explosion \citep{noz13, sar13, sar15, sluder2018}, since the gas densities in the ejecta are high at the early times, while the warm gas temperatures allow nucleation pathways to overcome activation energy barriers. The final dust masses predicted by the models depend on the assumed density of the gas and on the condensation scheme used. 

To account for the extreme nature of the environments typical of SN ejecta, recent models have relied on a chemical kinetic approach to address the chemistry of molecule and dust formation in SNe \citep{cherchneff2009, sar13, sar15, sarangi2018book, sluder2018}. 
The most recent models of both \cite{sar13, sar15} and \cite{sluder2018} adopt a similar network of reactions for the gas phase chemistry. 
For the growth of grains, \cite{sar15} assume growth via coagulation and rule out accretion on the surface of the small grains, while \cite{sluder2018} consider both the processes. \cite{sar15} model homogeneous as well as clumpy ejecta, where the geometry is assumed to consist of annular shells with unique abundances. In \cite{sluder2018}, the ejecta is considered as a three-phase system made up of \Ni-rich blobs, the swept-up mass by the blobs in the form of a shell, and the ambient gas. In strong contrast with the observational trends, for SN~1987A, the clumpy model of \cite{sar15}, and \cite{sluder2018}, both predict a formation of more than 0.01 \Ms\ of dust within a year of the explosion. 

When modeling the fluxes from the ejecta dust \citep{dwe15, wesson2015}, the distribution of dust within the He core is always assumed to be uniform or to scale with the densities of the clumps. This therefore ignores the fact that, from the center of the ejecta to the outer edge along any random line of sight, different clumps may have a different efficiency, composition, and timescale for dust production, affecting the overall opacities and fluxes. In addition, the evolution of dust composition, in other words, how the relative masses of each dust type can evolve differently, is also not accounted for. The impact of these factors are further compounded when the geometric shape of the ejecta is not a spherical, isotropic one; for instance, the elongated structure of the evolving ejecta of SN~1987A \citep{matsuura2017} is estimated to fit well with an ellipsoidal geometry \citep{kjaer_2010}. 

The advanced radiative transfer models that address the nebular phase of SNe enable us to better quantify the ejecta conditions, such as the distribution of clumps or the distribution of gas temperatures within the He core \citep{dessart_2018, dessart_2020}, with a much improved understanding of the molecular cooling, especially by CO molecules \citep{liljegren_2020}. 
Based on that, in this paper, we adopt a more realistic and higher-resolution model for the evolution of the physical conditions in SN ejecta. Applying the gas phase chemistry developed by \cite{sar13, sar15} to the upgraded model of the physical conditions, we derive the distribution and evolution of dust masses in clumpy SN ejecta in the spatial as well as temporal space. Using this generic distribution of dust, we calculate the opacities of the ejecta, and derive the resulting luminosities and fluxes. 
Having SN~1987A in mind, we also present a case where an ellipsoidal shape is assumed for the ejecta, to estimate the dust distribution, column densities, and fluxes, as a function of the viewing angle. This is the first study to account for the effect of anisotropy on the chemistry of dust formation. Moreover, this is also the first attempt to calculate the fluxes from the dust formed in the SN ejecta using a generic, nonuniform distribution of dust. 

The paper is arranged as follows. In Sect. \ref{sec_ejecta} we explain the evolution of physical conditions in the ejecta, in terms of gas densities (Sect. \ref{sec_gasden}), gas temperatures (Sect. \ref{sec_gasT}), projected dust temperatures (Sect. \ref{sec_dustT}), and the velocity-dependent stratification (Sect. \ref{sec_stratification}). 
Following that, Sect. \ref{sec_dustform} deals with the chemistry of dust formation, and the distribution and evolution of different dust species. 
In Sect. \ref{sec_opacities}, we discuss the opacities resulting from the dust formed in the ejecta and the expected SEDs from the distribution at various epochs. 
Section \ref{sec_elliptical} presents the case study of dust formation in anisotropic ejecta of ellipsoidal shape. 
In Sect. \ref{sec_compare}, we compare the dust distribution and masses between the isotropic and anisotropic cases, while Sect. \ref{sec_compare_spec} discusses the resulting SEDs as a function of viewing angle. 
After that, in Sect. \ref{sec_sn1987a} we present the predicted fluxes at days 615 and 775, in comparison to SN~1987A at the same epochs. Finally, we summarize and discuss our results in Sects. \ref{sec_summary} and \ref{sec_discussion}, respectively. 

\begin{figure*}
\vspace*{0.3cm}
\centering
\includegraphics[width=3.4in]{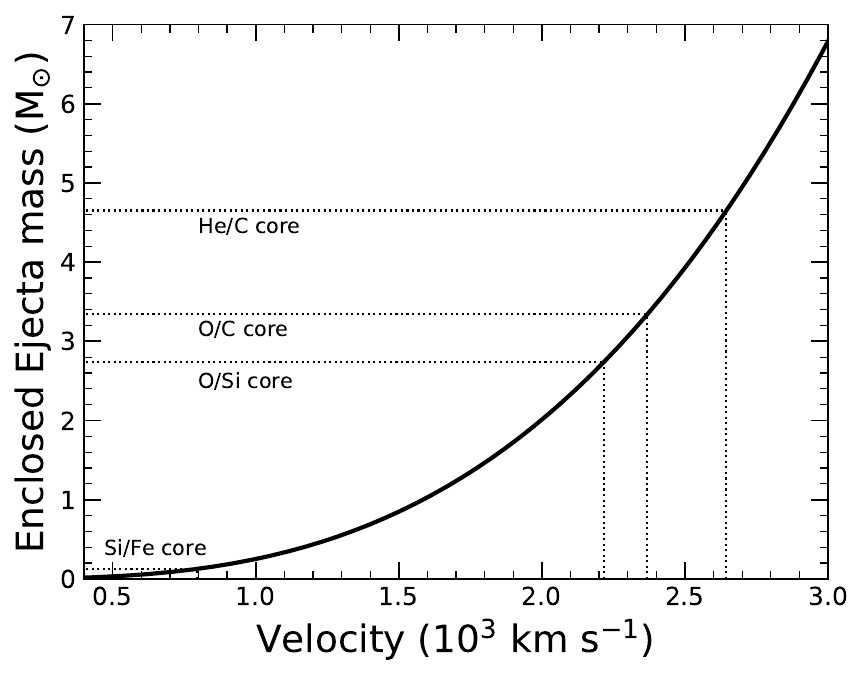}
\includegraphics[width=3.6in]{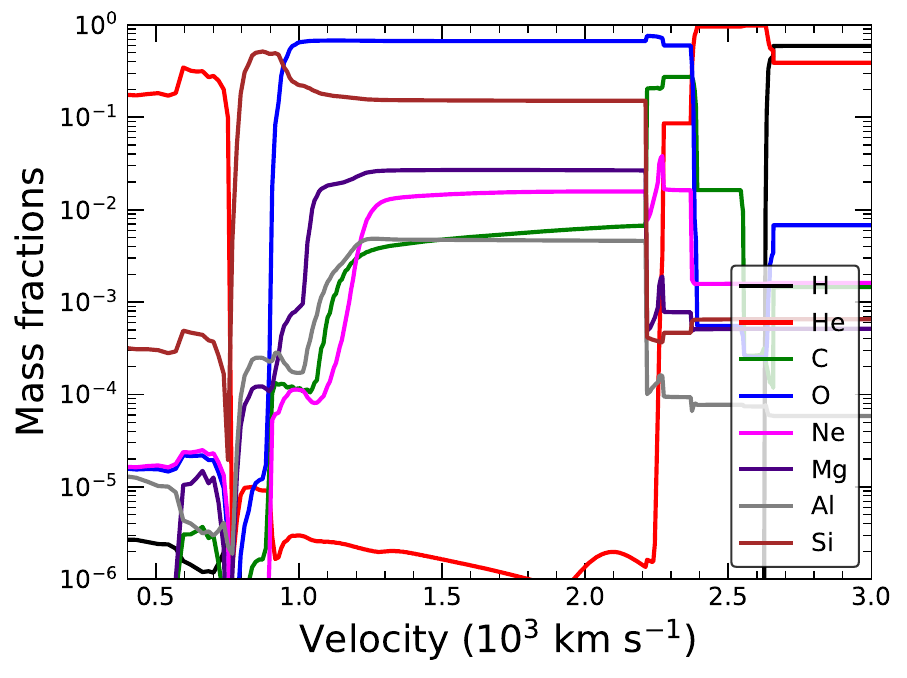}
\includegraphics[width=3.5in]{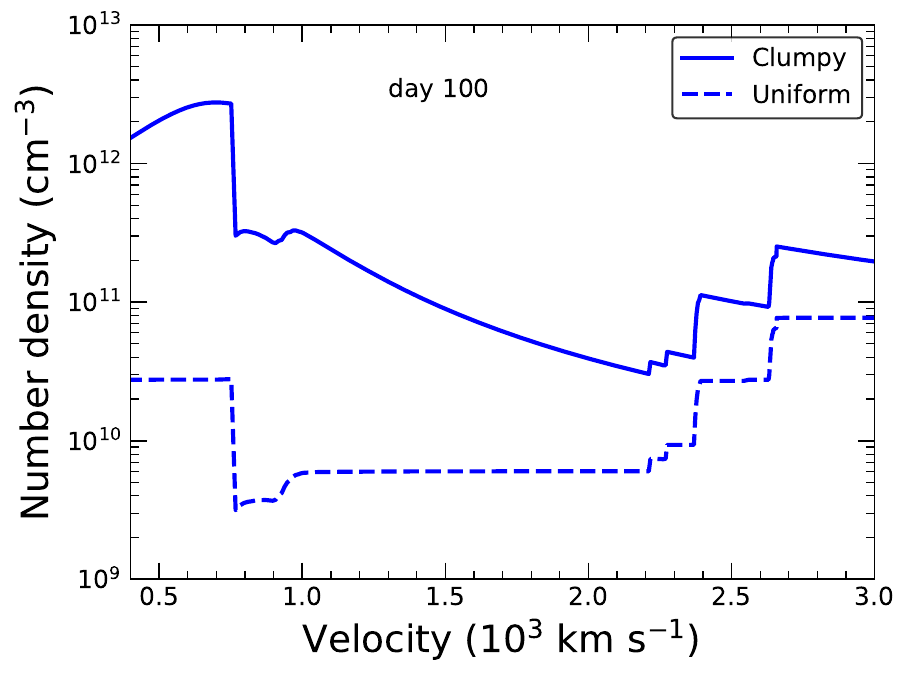}
\includegraphics[width=3.5in]{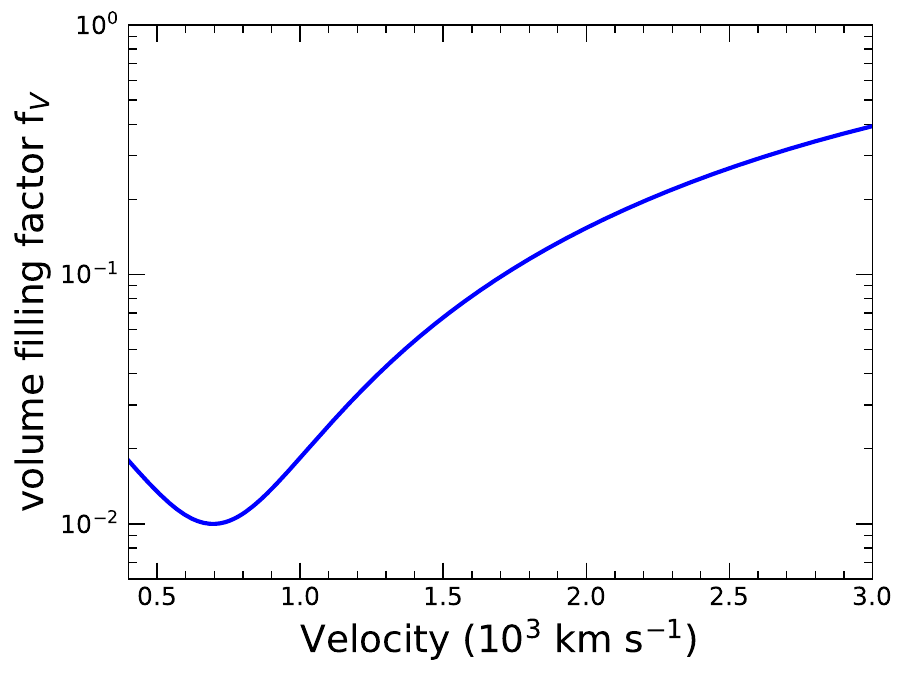}
\caption{\label{fig_mass_vel} Mass and density distribution of SN ejecta are presented in the velocity space. \textit{Top left}: Enclosed mass of the ejecta as a function of the ejecta velocity is shown. The He, O, and Si-cores are marked on the figures, with the respective mass-velocity boundaries (see Sect. \ref{sec_ejecta}). \textit{Top right}: Fractional abundances of the elements (H, He, C, O, Ne, Mg, and Si), taken from \cite{rau02}, are shown as a function of ejecta velocity, where the mass-velocity relation can be derived from the figure on the left, as shown in Sect. \ref{sec_ejecta}. \textit{Bottom left}: Distribution of gas (number) densities in the ejecta, as the function of velocity at 100 days post-explosion (see Sect. \ref{sec_gasden}). The densities for uniform ejecta are shown for reference, but we only model the scenario of clumpy ejecta. \textit{Bottom right}: Volume filling factor for the clumps are shown with reference to Eq.~\ref{eqn_fillingfact}. }
\end{figure*}

\section{Physical conditions of the ejecta}
\label{sec_ejecta}

In this section, we describe the physical conditions of SN ejecta with respect to the parameters adopted in this study. 
After the explosion, typical SN ejecta evolved through a free-expansion phase (homologous expansion), characterized by $v \cong r/t$. The ejecta are assumed to be made up of a core of uniform density, and an envelope where the density falls rapidly as a power-law \citep{tru99, dwa07}.  Following the model of \cite{tru99}, we define the velocity of the ejecta core, $v_c$, as 

\begin{equation}
\label{eqn_ejectacore}
v_c = \Big[\frac{10(n-5)}{3(n-3)} \frac{E}{M_{ej}}\Big]^{1/2},
\end{equation}
where $E$ is the explosion energy of the SN, $M_{ej}$ is the total mass of the ejecta, and $n$ is the power-law exponent that characterizes the density of the ejecta envelope. We model SN ejecta whose progenitor was a 20 \Ms\ main-sequence star, using the post-explosion yields from \cite{rau02}\footnote{\url{https://nucastro.org/nucleosynthesis/expl_comp.html}}. The mass of the ejecta, M$_{ej}$, in our model, is 12 \Ms\ (of which the He core is 4.65~\Ms), and the explosion energy is taken as 10$^{51}$ ergs, with reference to SN 1987A \citep{ens92}. The power law for the envelope, $n$, is taken as 12, typical for red supergiants \citep{moriya_2013}. Using these in Eq \ref{eqn_ejectacore}, $v_c$ is calculated to be 3300 km s$^{-1}$. The outer edge of the He core, which is the main site of dust formation in the ejecta, has a velocity of 2650 km s$^{-1}$. Therefore, the He core is entirely confined within the core of the ejecta and we only focus on that for the purpose of this study.

\subsection{Gas densities in the clumps}
\label{sec_gasden}

We assume that the ejecta are clumpy and the gas densities in any clump ($\rho_g$) are defined by a homogeneous density ($\rho_c$) and a volume filling factor ($f_V$). Following the prescription of \cite{tru99}, the density of the ejecta core is given by 

\begin{equation}
\label{eqn_rhocore}
\rho_h = \frac{3}{4 \pi} \frac{n-3}{n} \frac{M_{ej}}{v_{c}^3} t^{-3}, \ \ \ \   \\
\end{equation}
and therefore the density of a clump is simply 

\begin{equation}
\label{eqn_volfilling}
\rho_g = \rho_h f_V^{-1}.
\end{equation}

\cite{rau02} provide the abundances of all the elements (and their isotopes) in closely stratified layers, as a function of the enclosed mass of the ejecta below that layer (assuming spherical geometry). For a freely expanding ejecta, where the velocity of a layer is proportional to $r$, the enclosed mass can therefore be translated to the velocity of each layer using Eq. \ref{eqn_rhocore}, as 

\begin{equation}
\label{eqn_velocity}
v =  \Big[\frac{3}{4 \pi}\frac{m_{enc}}{t^3 \rho_h(t)}\Big]^{1/3} = \Big[\frac{n}{n-3} \frac{m_{enc}}{M_{ej}} \Big]^{1/3} v_c\ ,
\end{equation}
where $m_{enc}$ is the enclosed mass below a given layer. Using this, in Fig. \ref{fig_mass_vel}, the relation between the enclosed mass and the velocity is shown, and the abundances of elements, which are relevant for dust formation, are presented as a function of velocity. 

For SN ejecta, it is often estimated that the degree of clumpiness increases toward the inner parts of the He core, which is synonymous with smaller velocities \citep{wong15, dessart_2018}. To quantify that, we use the relation between the velocity and the volume filling factor given by \cite{dessart_2018} in our study, so that 

\begin{equation}
\label{eqn_fillingfact}
f_V =  1 + (f_{min} - 1) \ e^{-(v-v_{min})^2/v_c^2}\ .
\end{equation}

The minimum value for the filling factor, $f_{min}$, was taken as 0.01, with reference to studies on SN 1987A \citep{jerkstrand2011, erc07}, and $v_{min}$ corresponds to the velocity of the inner edge of the so-called Si core, which in this case is $\sim$ 750 km s$^{-1}$ (see Fig. \ref{fig_mass_vel}, top right). The volume filling factor is shown as a function of velocity in the bottom right panel of the same figure. 

The total number density of the gas in the clumps, calculated using the abundances of elements, the gas densities, and the filling factors, are shown in the bottom left panel of Fig \ref{fig_mass_vel}. The densities are consistent with those predicted by the 3D SN explosion models \citep{wong15, utrobin_2017}, when they are extrapolated to the nebular phase. 

Due to the much higher densities in the clumps, compared to the inter-clump medium, we assumed that the formation of dust occurs within the clumps only. Therefore, in the following text, where we deal with the dust formation chemistry and the corresponding dust opacities, the inter-clump region was ignored.

\begin{figure}
\vspace*{0.3cm}
\centering
\includegraphics[width=3.5in]{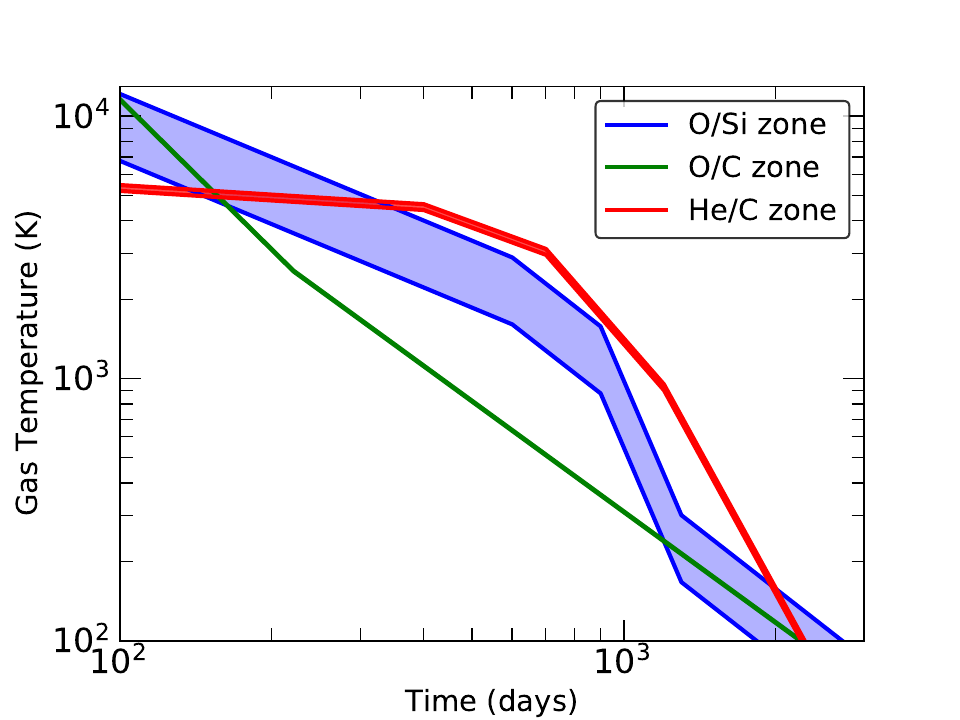}
\caption{\label{fig_gasT} Range of gas temperatures in different zones and their variation with post-explosion time is shown (see Sect. \ref{sec_gasT} for details). Even though our model does not depend on these coarse zone divisions of the ejecta, this is particularly helpful to understand how the gas temperatures evolve. }
\end{figure}

\subsection{Gas temperatures}
\label{sec_gasT}

The temperature of the gas in the ejecta evolves through collisional line cooling, initially by atoms and ions, and later by molecules \citep{kozma_1998, liu92, liu95, liljegren_2020}. Given the rapid evolution of the ejecta, along with the presence of nonthermal processes induced by the radioactivities, it is not straightforward to define an analytical expression for the evolution of gas temperatures. 

The cooling of a parcel of gas is often controlled by abundances of the elements in that parcel. Based on the most abundant elements, the He core of the ejecta can be overall subdivided into zones, namely the Si/Fe zone, the O/Si/Mg zone, the O/C zone, and the He/C zone (in order of increasing velocity). The extent of these zones within the He core are shown in Fig. \ref{fig_mass_vel} (top left panel). CO molecules form abundantly in the O/C zone \citep{sar13, liljegren_2020}, which is responsible for the rapid cooling of the gas in that region. As \cite{liu95} find, for SN~1987A the temperature of the O/C zones at day 800 is close to 700 K, while at the same time the O/Si zone is still hot with temperatures of $\sim$ 2200~K. 

According to \cite{kozma_1998}, who calculated the evolution of temperatures in each of these zones between day 200 and day 2000, the zones undergo a rapid cooling phase after a few hundred days of evolution, followed by another phase of leveling-off of the temperature. The inner layers experience the rapid cooling phase earlier, while for the outermost He zone, the rapid cooling phase almost never transpires. This is not completely unexpected, since the heavier elements (inner layers have heavier elements) are characterized by faster radiative cooling rates \citep{sutherland_1995, sarangi_2022_1_arxiv}. 

\cite{dessart_2020} calculate the distribution of temperature in the SN ejecta (type II) at a post-explosion epoch of 300 days, as a function of velocity. Their results are also in agreement with \cite{kozma_1998}, showing that the ejecta at a given epoch are cooler toward the inner layers and the temperature increases with velocity. 

For the purpose of this study, we obtained the temperatures as a function of time and abundances from \cite{kozma_1998}, and as a function of velocity from \cite{dessart_2020}; combining them, we constructed the matrix of gas temperatures in the ejecta as a function of velocity and time. For the CO forming regions, we consider the temperature to evolve as suggested by \cite{liljegren_2020} (consistent with \citealt{liu95}) once the CO molecules start to form in the ejecta.  

The range of gas temperatures that corresponds to each zone and their evolution from day 100 to day 3000 is shown in Fig. \ref{fig_gasT}. The O/Si zone shows the largest stretch in the range of temperature, as expected, since it is spread over a wide velocity range (see Fig. \ref{fig_mass_vel}, top left). The He/C zone, despite being larger than 1 \Ms\ in mass, is spread over a relatively small velocity range, given that it is at a larger radius. Therefore, the spread of gas temperature in this zone (at any given epoch) is also limited.

\begin{figure}
\vspace*{0.3cm}
\centering
\includegraphics[width=3.5in]{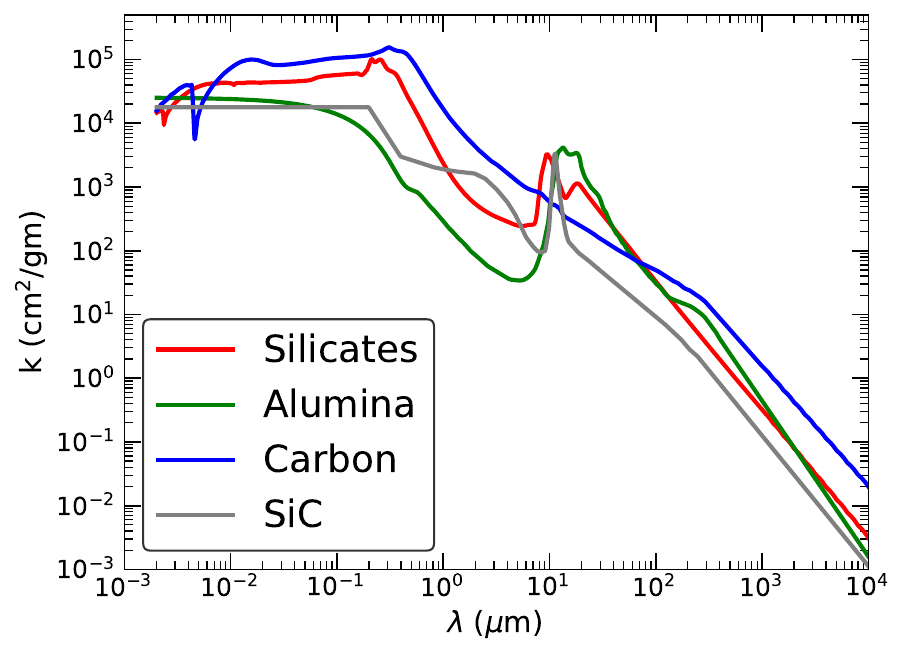}
\caption{\label{fig_kappa} Mass absorption coefficients for silicates \citep{wei01}, alumina \citep{koike_1995}, amorphous carbon \citep{zub04}, and silicon carbide \citep{pegourie_1988} are presented, for a grain size of 0.01 \mic. For alumina at wavelengths smaller than 0.3 \mic\  and for silicon carbide at wavelengths smaller than 0.2 \mic, the values are extrapolated due to lack of data. In addition, for these two dust types, we do not have data for various grain sizes, and therefore these are generalized values for all sizes. } 
\end{figure}

\subsection{Dust temperatures}
\label{sec_dustT}

Once the dust grains form in the ejecta through nucleation and coagulation \citep{sar15}, the temperatures of the grains are not necessarily the same as the kinetic temperature of the ambient gas, since the dust grains are efficient in cooling down via emission, lowering their internal energy. The temperatures of the grains determine their IR fluxes, as well as their growth (in size) via coagulation. Our approach to determine dust temperatures is similar to that of \cite{sluder2018}. However, in this study, we also calculate the evolution and distribution of dust temperatures in the velocity space for each clump in the ejecta, for four different dust species (silicates, amorphous carbon, alumina, and silicon carbide). 

A dust grain in the SN ejecta is heated radiatively by the UV photons produced as a result of the decay of radioactive isotopes. In addition, dust is also heated through collisions with the ambient gas, while it cools down via emission. 
When radiatively heated, the rate of heating is simply the incident flux on a dust grain multiplied to the absorption cross-section, integrated over all the wavelengths \citep{temim_2013}. Then again, flux is basically the product of the energy density (at a given radius, at a given time) and the velocity. If we take the number density of photons with energy $E$ at a given time and velocity to be $n_{\gamma} (E, v, t)$, then the radiative heating rate ($H_r$) of a dust grain of species $S$, of radius $a$, can be expressed as 

\begin{equation}
\label{eqn_radheating}
H_{S,r}(a, v, t) = \int^\infty_0 n_{\gamma}(E, v, t) E \ c \ \pi a^2 Q_S(E,a) \mathrm{d}E, 
\end{equation}
where $Q_S$ is called the absorption coefficient of the grain for incident energy $E$ and size $a$. The mass absorption coefficient $k(\lambda, a)$ (= 3Q/4a$\Omega$, $\Omega$ being the solid-state density of a grain) of silicates ($\Omega$ = 3.3 g cm$^{-3}$), amorphous carbon ($\Omega$ = 2.2 g cm$^{-3}$), alumina ($\Omega$ = 3.95 g cm$^{-3}$), and silicon carbide ($\Omega$ = 3.21 g cm$^{-3}$) are shown in Fig. \ref{fig_kappa} for $a$ = 0.01 \mic\ \citep{wei01, koike_1995, zub04, pegourie_1988}. 

As assumed by \cite{sluder2018}, $n_{\gamma} (E)$ can be expressed as a Gaussian distribution with a mean energy $E_0$ and deviation $\sigma_0$. Then the energy density (at a given $v$ and $t$) can be expressed as 

\begin{equation}
\label{eqn_endensity}
\begin{gathered}
u_{UV} (v, t) = D \int^\infty_0 e^{-(E-E_0)^2/2\sigma_0^2} E \mathrm{d}E \\
= D \sigma_0^2 e^{-E_0^2/2 \sigma_0^2} + D (2\pi)^{1/2} \sigma_0 E_0,
\end{gathered}
\end{equation}
where $D$ is a normalization coefficient, which is a function of $v$ and $t$. Taking $E_0$ = 4.43 eV and $\sigma_0$ = 1 eV, with reference to \cite{jerkstrand2011} and \cite{sluder2018}, the first term of the above expression is much smaller than the second term, and thus can be ignored. 

The UV luminosity $L_{UV}$, responsible for the heating of the dust grains, is a fraction of the total radioactive energy deposited in the ejecta ($L_X$) at time $t$, which is converted to UV photons. The coefficient for the fractional conversion, $\alpha$, is taken as 0.35 \citep{sluder2018, kozma_1992}, so that $L_{UV} (t) = \alpha L_X (t)$. For a thin shell of width $\Delta r$ at radius $r$ (= $vt$) at time $t$, we can write the energy density as 

\begin{equation}
\label{eqn_lumdensity}
u_{UV} (v, t) = \frac{\alpha L_X (t)}{4 \pi r^2 \Delta r} \ \frac{\Delta r}{c} = \frac{\alpha L_X (t)}{4 \pi r^2 c}.
\end{equation} 

Combining Eqs. \ref{eqn_endensity} and \ref{eqn_lumdensity}, we have 

\begin{equation}
\label{eqn_Aconst}
D(v,t) = \frac{\alpha L_X (t)}{2^{5/2} \ \pi^{3/2} \ c \ v^2 \ t^2 \ E_0 \ \sigma_0 }.
\end{equation} 

\begin{table}
\centering
\caption{All the parameters required for Eq. \ref{eqn_lumradio} are listed, taken from \cite{seitenzahl_2014}}
\label{table_radioactivity} 
\begin{tabular}{cccccc}
\hline \hline
Species & m & $\tau_{X}$ & $\kappa^{\gamma}_i$ & $q^l$ & $q^{\gamma} $ \\
& \Ms & day & cm$^2$g$^{-1}$ & keV & keV \\
\hline
$^{56}$Co & 6.9(-2) & 111.4 & 0.033 & 119.4 & 3606 \\
$^{57}$Co & 2.4(-3) & 392.0 & 0.0792 & 17.8 & 121.6 \\
$^{60}$Co & 4.0(-5) & 2777.8 & 0.04 & 96.4 & 2504 \\
$^{44}$Ti & 3.8(-5) & 31037 &0.04 &  596.0 & 2275 \\
\hline

\end{tabular}
\end{table}

The energy deposited in the ejecta through radioactivities was calculated by \cite{cherchneff2009} following the formalism of \cite{woosley1989}, which was later applied to SN~1987A by \cite{sar13} and \cite{sluder2018}. In an identical approach, \cite{seitenzahl_2014} calculated the energy deposited in the ejecta and the masses of all the radioactive species, in accordance with the light curve of SN~1987A until the late epoch of day 4334 post-explosion. For any radioactive species $i$ with total mass $m_i$ produced at the time of explosion, we can express the rate of energy deposition in the ejecta through its radioactive decay \citep{woosley1989, cherchneff2009, seitenzahl_2014, sluder2018} as 


\begin{equation}
\label{eqn_lumradio}
\begin{gathered}
L_{Xi}(t) = \frac{m_i N_A}{A_i} \frac{e^{-t/\tau_{Xi}}}{\tau_{Xi}} (q^l_{i} + f(t) q_{i}^{\gamma}) \\
f(t) = 1 - e^{\tau_{\gamma}(t)}, \ \ \ \tau_{\gamma}(t) = \frac{3}{4\pi} \frac{\kappa^{\gamma}_i M_{core}}{v_c^2 t^2}, \ \ L_X = \sum_i L_{Xi} 
\end{gathered},
\end{equation}
where $A_i$ is the mass number, $\tau_{Xi}$ is called the e-folding decay time, and $\kappa^{\gamma}_i$ is the average \gray\ mass-absorption coefficient of the radioactive isotope $i$. 
$N_A$ is the Avogadro's number, $\tau_{\gamma} (t)$ is the \gray\ optical depth at time $t$, and $q^l_{i}$ and $q^{\gamma}_{i}$ are the energies of photons per decay, produced by the charged leptons and \grays,\ respectively. The mass of the ejecta core, $M_{core}$, derived from Eq. \ref{eqn_velocity} (using $v$ = $v_c$, $n$ = 12 and $M_{ej}$ = 12 \Ms) is about 9~\Ms. 

We chose the best-fit scenario of \cite{seitenzahl_2014} for the masses of the radioactive species $^{56}$Co, $^{57}$Co, $^{60}$Co, and $^{44}$Ti, listed in Table \ref{table_radioactivity}, along with all the parameters from Eq. \ref{eqn_lumradio}. With that, we have now defined all the quantities required to derive the heating rate of a dust grain, given by Eq. \ref{eqn_radheating}, as a function of the velocity of the ejecta where it is located, and the post-explosion time.

Besides radiative heating, dust grains are also subjected to heating by the ambient gas, collisionally. The collisional heating rate of a dust grain was calculated by \cite{dwe87, hol79}, and applied to SN environments by \cite{sarangi2018}. The rate of heating due to collision \citep{burke_1983, sluder2018}, $H_c$, can be expressed as

\begin{equation}
\label{eqn_colrate}
\begin{gathered}
H_{c} (a, v, t) = n_g (v, t)  \times \pi a^2 \times \Big[\frac{8 k_B T_g(v, t)}{\pi \langle\mu (v)\rangle}\Big]^{1/2}   \\
\times \ f_{ed} \times (2k_BT_g(v, t) - 2k_BT_d(v, t)) \\
f_{ed} = 0.1 + 0.35 \ e^{-[(T_d + T_g)/500]^{1/2}}
\end{gathered},
\end{equation} 

where $k_B$ is the Boltzmann constant, $\mu (v)$ is the mean particle mass of the gas in the ejecta, for the velocity bin $v$, and $T_g$ and $T_d$ are the gas and dust temperatures, respectively.

\begin{figure}
\vspace*{0.3cm}
\centering
\includegraphics[width=3.5in]{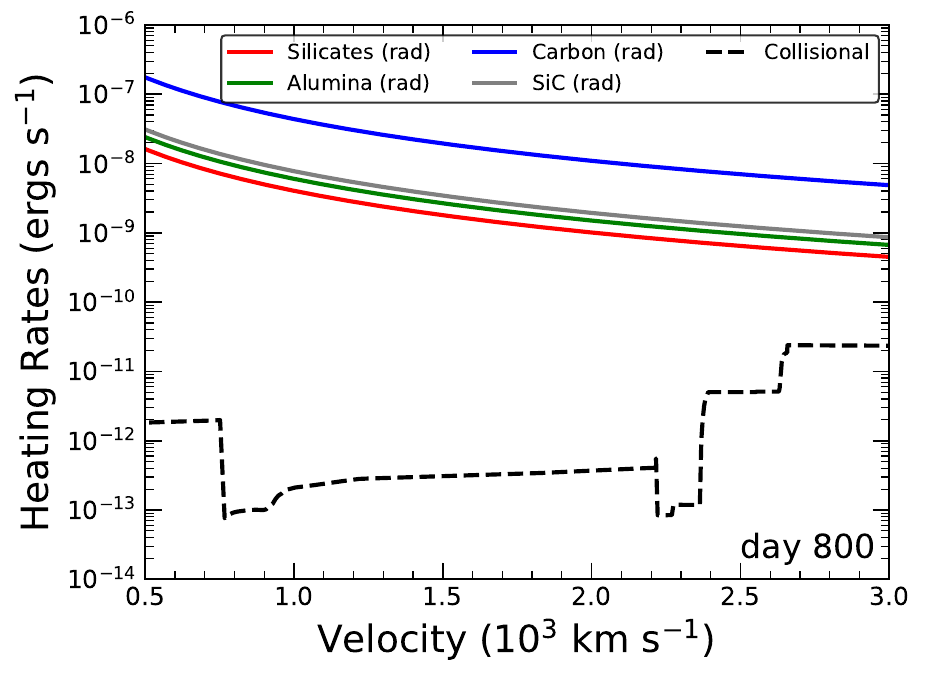}
\caption{\label{fig_heatingrates} Radiative heating rate (Eq. \ref{eqn_radheating}) for the four different dust species are shown compared to the (maximum) heating rate of the dust grains via collision with the ambient gas (Eq. \ref{eqn_colrate}), for day 800 and a grain size of 0.01~\mic. As it shows, the dust temperatures are controlled by the radiative heating only; that is also the case for other epochs and grain sizes as well. }
\end{figure}

\begin{figure*}
\vspace*{0.3cm}
\centering
\includegraphics[width=3.5in]{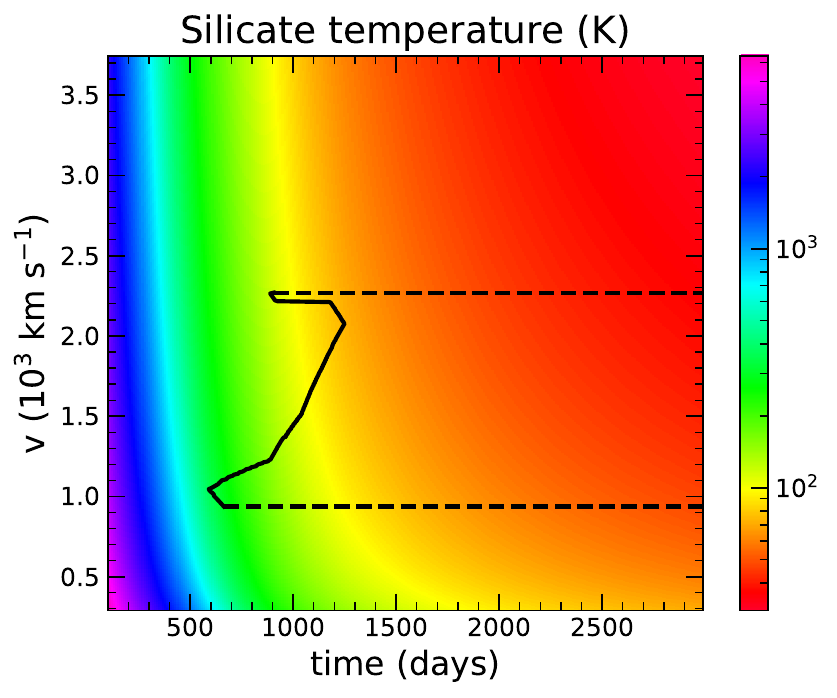}
\includegraphics[width=3.5in]{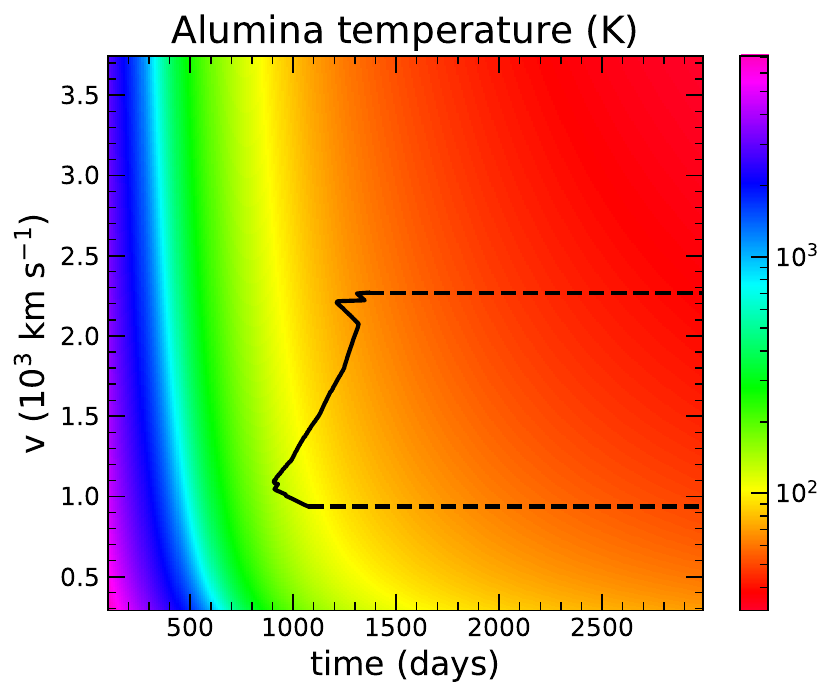}
\includegraphics[width=3.5in]{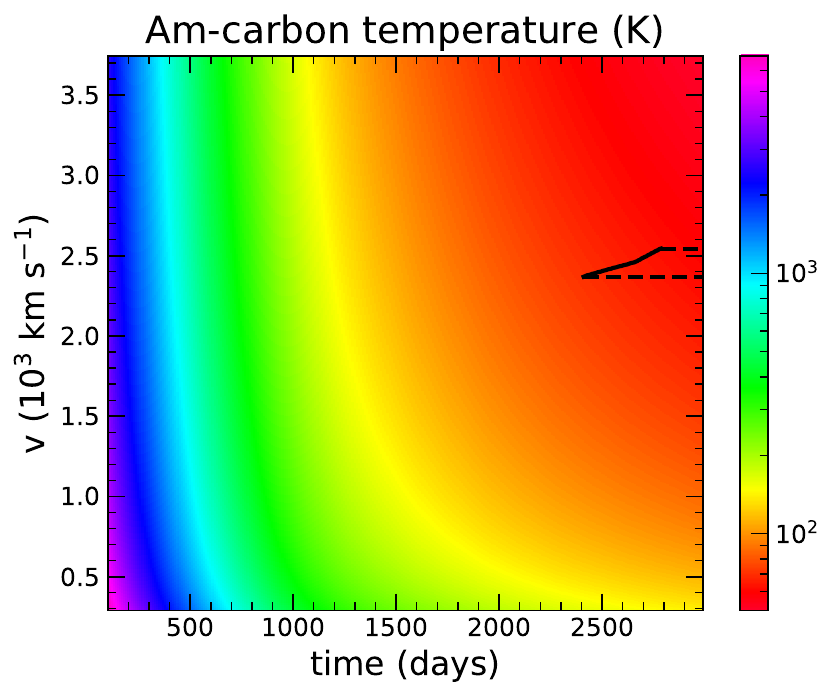}
\includegraphics[width=3.5in]{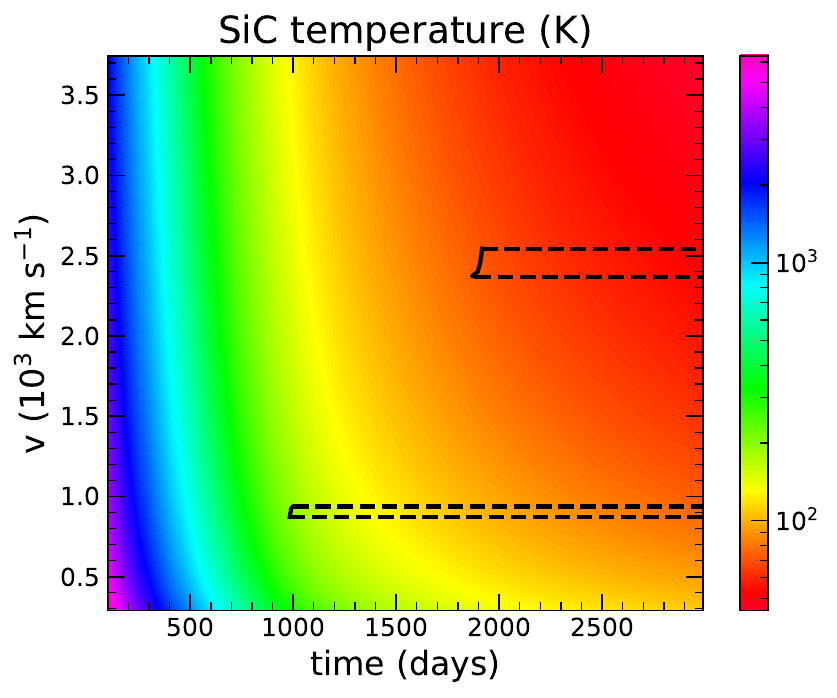}
\caption{\label{fig_dustT} Resulting dust temperatures for silicates (\textit{top left}), alumina (\textit{top right}), carbon (\textit{bottom left}), and silicon carbide (\textit{bottom right}) are shown for different days and different velocity bins. The solid black lines refer to loci in terms of time and velocity, when the dust formation starts in a particular bin. The enclosed area inside the contours (dashed black lines) represents the space where that dust species is present (see Sect. \ref{sec_dustform} for details). }
\end{figure*}

Comparing the collisional and radiative heating rates from Eqs. \ref{eqn_radheating} and \ref{eqn_colrate}, we find that the rate of radiative heating is far more dominant at all times and at all velocities, and therefore we can safely ignore the impact of collisional heating. In Fig. \ref{fig_heatingrates}, we show the comparison of the radiative heating rate with the maximum rate of heating due to collision, for all four dust species, at day 800. 

The rate of radiative cooling \citep{temim_2013} of a dust grain of species $S$ (with mass absorption coefficient $k_S$), at temperature $T_S$, which is analogous to the energy lost by the grain through emission, is given by

\begin{equation}
\label{eqn_coolingrate}
C_S(a, T_d) = 4 m_S(a) \int \pi B_{\lambda}(\lambda, T_{S}) k_S(\lambda, a) \mathrm{d}\lambda\ , 
\end{equation} 
where $m_S$ is the mass of the dust grain and $B_{\lambda}$ is the Planck function per unit wavelength. 

Comparing the radiative heating from Eq. \ref{eqn_radheating} to this cooling rate, the temperature of the dust grains are determined for each velocity bin and time. Figure \ref{fig_dustT} shows the obtained distribution of dust temperatures for silicates, alumina, amorphous carbon and silicon carbide for a grain size of 0.01~\mic. 
To be clear, this map of dust temperatures in the complete $v, t$ space only reflects the temperatures assuming that dust species were hypothetically present there; in other words, this map does not provide any information on the time and location of dust formation from the chemistry (see Sect. \ref{sec_dustform} for that). 


When accounting for the dust temperatures, we did not include the effect of secondary heating by the IR radiation produced by the dust itself. Given that the coefficient of absorption of IR radiation is much smaller than that of UV (seen in Fig. \ref{fig_kappa}), we do not presume this to be significant; we shall use a more sophisticated radiative transfer model in the future to verify this.

It is important to note that the rapidly evolving SN ejecta in their nebular phase are estimated to not be in chemical or collisional equilibrium \citep{don85, cherchneff2009, sarangi2018book}. However, they are still likely to be in radiative equilibrium given the high velocity of the photons; since the dust temperatures are essentially controlled by radiation only, the temperatures derived above are justified.

\begin{figure*}
\vspace*{0.3cm}
\centering
\includegraphics[width=3.5in]{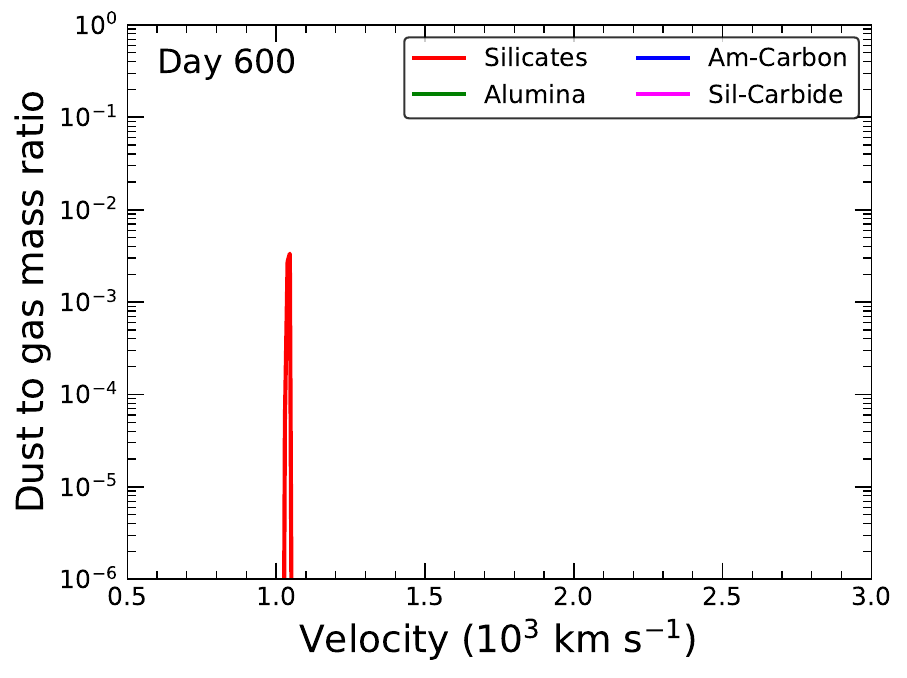}
\includegraphics[width=3.5in]{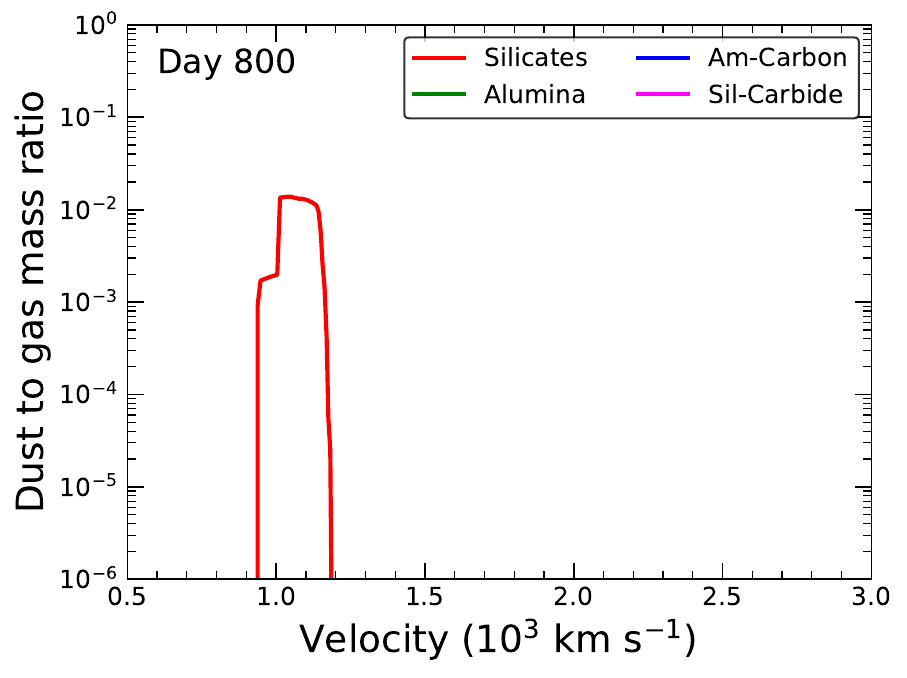}
\includegraphics[width=3.5in]{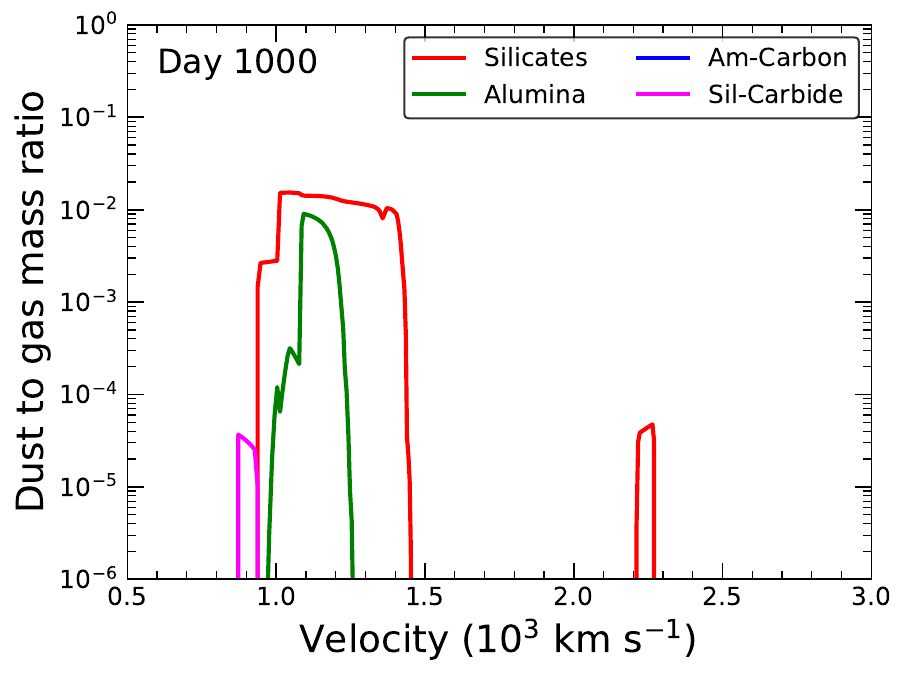}
\includegraphics[width=3.5in]{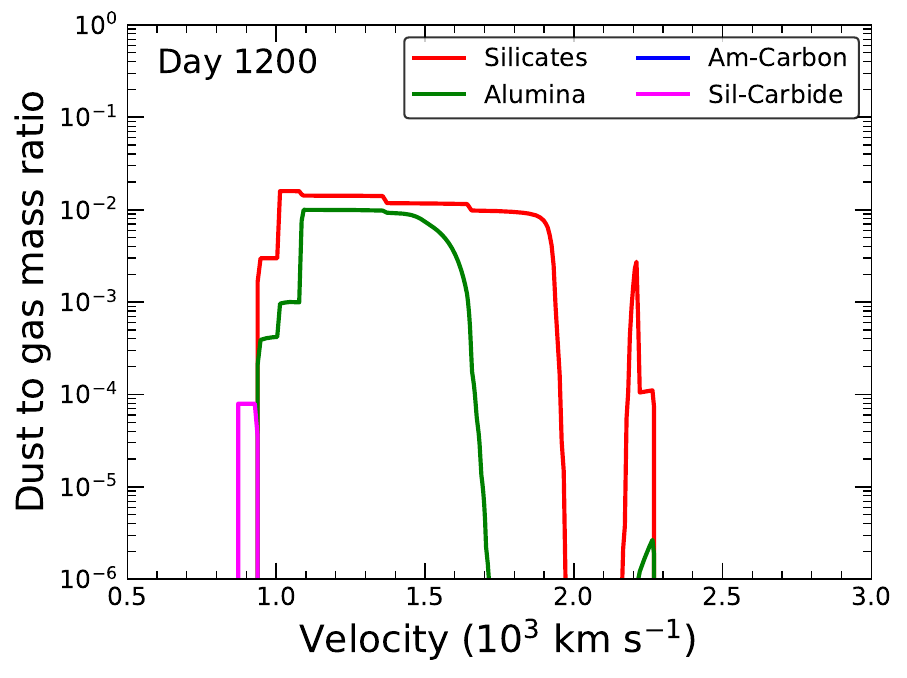}
\includegraphics[width=3.5in]{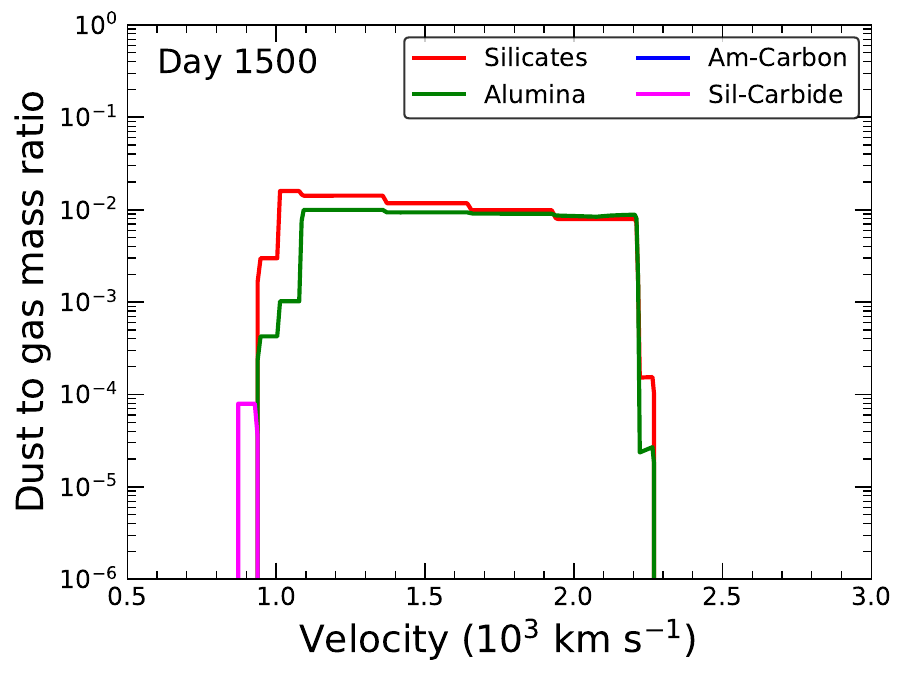}
\includegraphics[width=3.5in]{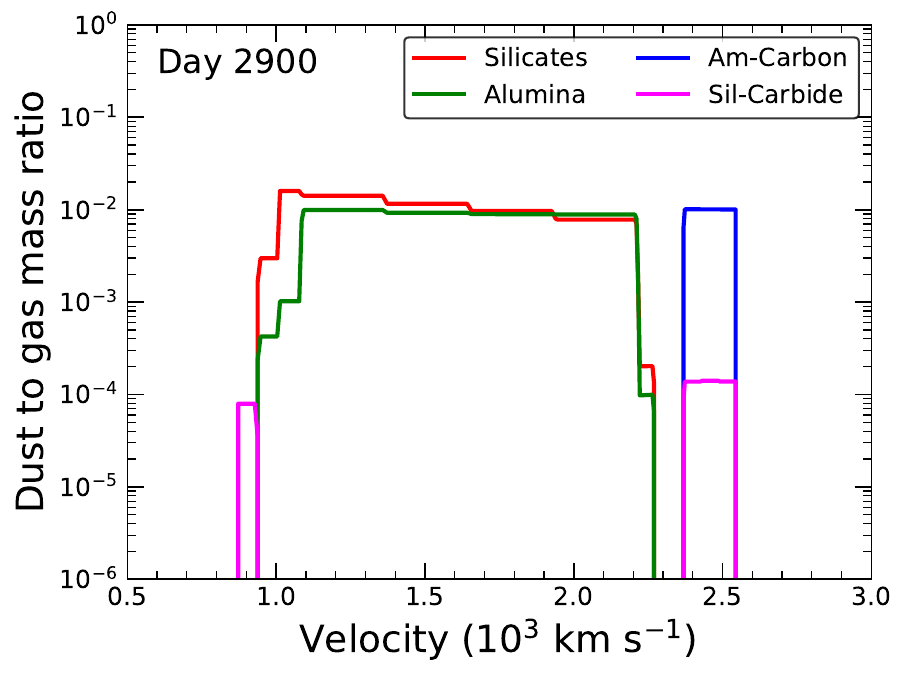}
\caption{\label{fig_dust_vel} Efficiency of dust formation, which is the same as the dust-to-gas ratio in the SN ejecta, is shown for days 600, 800, 1000, 1200, 1500, and 2900, as a function of velocity. There is no significant change in the ratios between day 1500 and 2500, and therefore we chose to present the case of day 2900 after day 1500.  }
\end{figure*}

\subsection{Stratification}
\label{sec_stratification}

In Sects \ref{sec_gasden}, \ref{sec_gasT}, and \ref{sec_dustT}, we have explained the physical conditions of the ejecta and their evolution. It is clear that in different parts of the ejecta, characterized by different velocities, there will be variations in the densities of the clumps, as well as in the gas and dust temperatures, all of which are key physical conditions that are critical to the formation and growth of dust grains. 

In our model, we stratify the ejecta into 670 thin shells, following the same stratification as that of the post-explosion nucleosynthesis data of a 20 \Ms\ main-sequence star of \cite{rau02}\footnote{\url{https://nucastro.org/nucleosynthesis/expl_comp.html}}. Out of those, the He core is confined within the first (inner) 450 shells and we uniquely follow the dust formation chemistry in those shells. In other words, it can be visualized as a set of 450 different groups of clumps, distinguished by their abundances, densities, and gas and dust temperatures. For reference, we identify the clumpy shells by their velocities, since that is unique to each shell throughout the time of their evolution. In the following section, we discuss the chemistry of dust formation and the net chemical yields in light of this model. 

\section{Dust formation in the ejecta}
\label{sec_dustform}

Dust formation in the SN ejecta is controlled by simultaneous phases of nucleation and condensation \citep{cherchneff2009, cherchneff2010, sar13, sar15}. The chemistry of the hot and dense SN ejecta was modeled by \cite{sar13}, where all the possible chemical reactions and pathways relevant to the atoms, molecules, and molecular clusters present in the gas phase were accounted for. In \cite{sar15}, the growth of dust grains through stochastic coagulation and coalescence was modeled. 
We adopted the same formalism in this study, and no changes were made to the chemical reaction network and pathways of dust growth, which were developed by \cite{sar13, sar15}; for a complete list of all chemical reactions, we refer the reader to \cite{cherchneff2009} and \cite{sar15}. 
This means that, by applying the chemical network in this study, we derive the chemical yields in the 450 different groups of clumps, which are identified by their unique abundances of species and unique evolution of gas and dust conditions. 

The gas phase nucleation processes lead to the formation of molecules and molecular clusters \citep{sar13}. As the He cores of the ejecta are predominantly rich in O, a lot of O-rich diatomic molecules are synthesized in the nebular phase of the SN. Specifically, molecules such as CO, SiO, AlO, O$_2$, and SO are efficiently produced. Following their formation after about 200 days from the time of explosion, CO molecules retain their large abundances in the ejecta and gradually increase in mass. The total mass of CO molecules formed in the ejecta is 0.37 \Ms. On the contrary, molecules such as SiO or AlO are increasingly depleted into dust grains with time. Using the formalism developed in \cite{sar15}, in this study we follow the profiles of the four most abundant dust species in the ejecta, namely, magnesium silicates (of chemical form [Mg$_2$SiO$_4$]$_n$), alumina (of chemical form [Al$_2$O$_3$]$_n$), amorphous carbon (of chemical form [C$_{28}$]$_n$) and silicon carbide (of chemical form [SiC]$_n$; n is an integer generally larger than four). More details on the types of dust species relevant for SN ejecta, and their individual properties, can be found in \cite{sar15}. 
For any clump of a given abundance, the required parameters to study the chemistry of dust formation are: the evolution of gas density, the gas temperature, the rates of all possible gas phase reactions, dust temperatures, the solid state density of the dust grains, and the coefficients for the Van der Waals interaction. 

The smallest grains of dust, or rather the precursors, are synthesized through gas phase chemical reactions, where the temperatures are expected to be analogous to the gas temperatures. They coagulate and grow larger than 10 \AA, and are labeled as dust grains. Coagulation is not well understood in high temperatures, and is generally perceived to be inefficient when the kinetic energies of collisions are large \citep{hou_2022}; we have only assumed coagulation when the gas temperatures are lower than 2000~K. 

In Fig. \ref{fig_dustT}, where the dust temperatures are shown, the contours (shown by solid and dashed black lines) refer to the region in velocity space where a particular dust type is present at a given epoch. The left boundary of that contour (shown as a solid black line) is therefore the time when dust formation commences in that velocity bin. This is further understood in Fig.~\ref{fig_dust_vel}, where we show the efficiency of dust formation at various epochs for each of the dust type, in the velocity space. 

As a general comment, the inner parts of the ejecta (smaller velocities) tend to be more clumpy as well as cooler in temperature, compared to the outer shells, at any give time. Both these factors enhance the dust formation efficiency and ensure an earlier onset of dust formation. As seen in Fig.~\ref{fig_dust_vel}, the dust first appears in the velocity bin of about 1000 km s$^{-1}$, while by day 2900, it is spread out over the entire velocity space of the He core (He-core boundary $\sim$ 2500 km s$^{-1}$). The majority of the CO molecules form in a narrow band between 2200 and 2350 km s$^{-1}$. Interestingly, due to faster cooling in those bands, induced by CO molecules, the formation of dust in those regions precedes the formation of dust in the neighboring velocity bins; this can be seen in the panel of day 1000 in Fig.~\ref{fig_dust_vel}. 
However, a large fraction of the O atoms are locked in CO molecules in that part of the ejecta; hence, the efficiency of O-rich dust production is largely compromised.  
There is a very narrow (almost) dust-free shell between 2300 and 2400 km s$^{-1}$, where silicates are formed only in a very small proportion. 

The overall efficiency is slightly compromised in the bins where dust formation commences later, which, in general, can be associated to larger velocities; this is also visible in Fig.~\ref{fig_dust_vel}. However, it is important to realize that in the velocity space, the smaller velocities are inner shells, so the total mass of those shells are smaller than the outer ones. Therefore, the inner shells start dust formation earlier with larger efficiencies, but those shells are low in total mass, and therefore end up forming a small mass of dust. When dust formation progresses in the outer shells, the total mass of the shells being larger compromises for the reduced efficiency.

\begin{figure*}
\vspace*{0.3cm}
\centering
\includegraphics[width=6.5in]{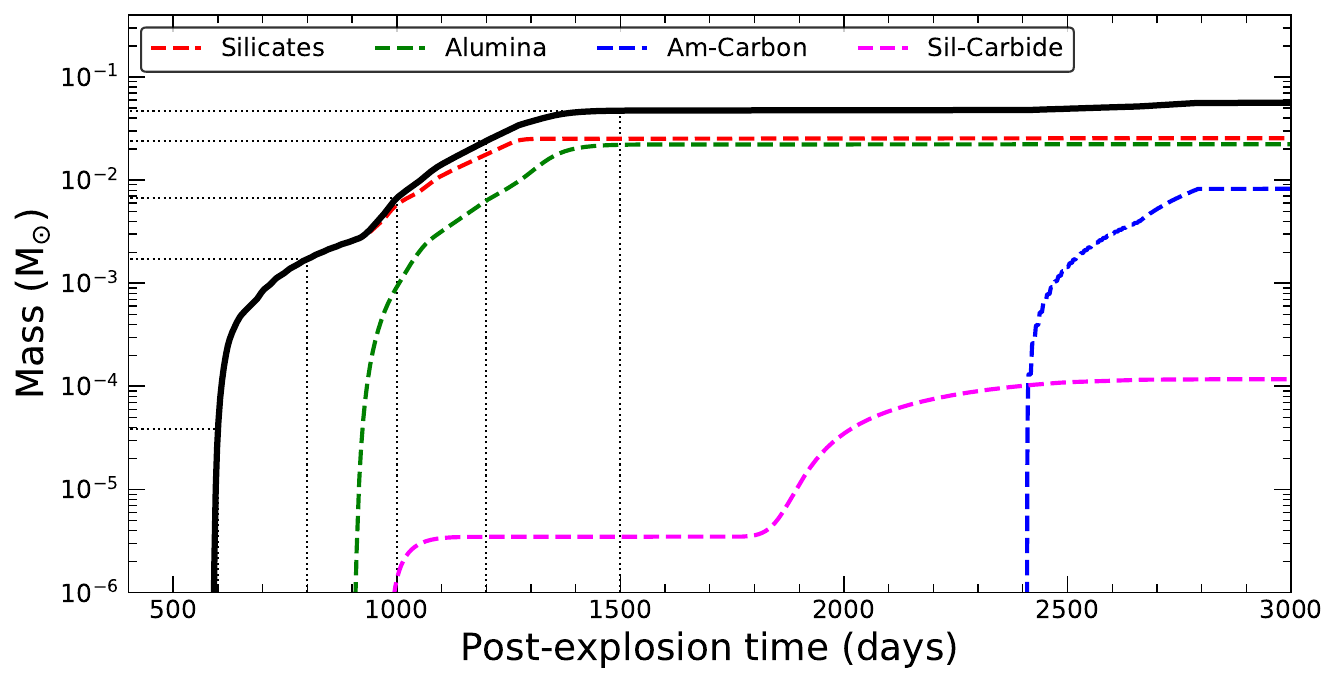}
\caption{\label{fig_dustmass} Total mass of grains of silicates, alumina, amorphous carbon, and silicon carbide, as a function of post-explosion time. The sum of all dust species is shown by the solid black line. The results correspond to clumpy, isotropic SN ejecta, whose progenitor was a  20 \Ms\ star (at main sequence).  }
\end{figure*}

\begin{figure}
\vspace*{0.3cm}
\centering
\includegraphics[width=3.5in]{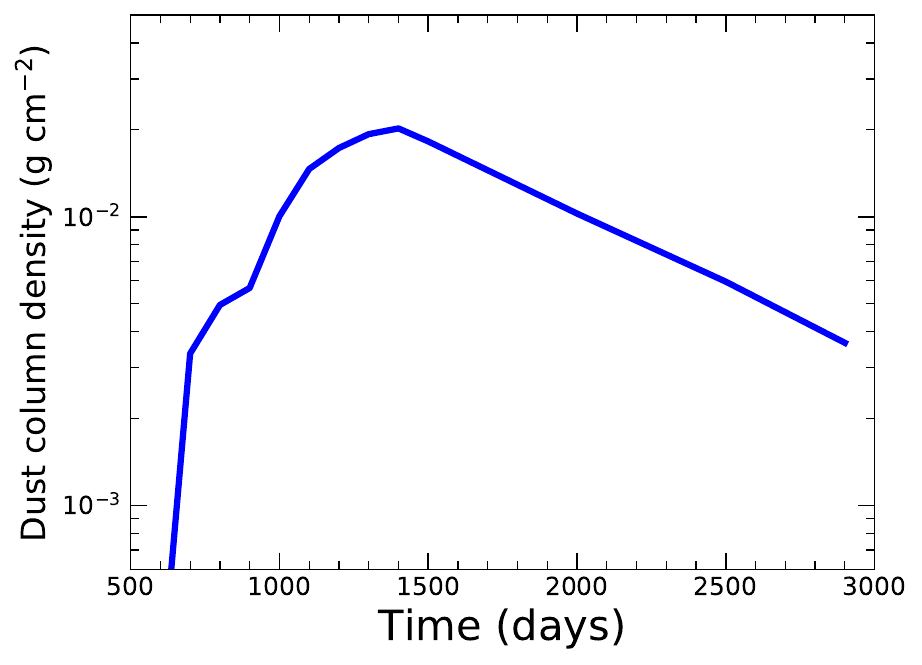}
\caption{\label{fig_dustcol} Column density of dust as a function of time, post-explosion. Despite the nonuniform distribution of dust, this can be correlated to the total opacities (from the center to the outer edge of the ejecta), using the relative abundances of each dust type at a given epoch and their corresponding optical constants (see Sect. \ref{sec_opacities}). }
\end{figure}

In Fig. \ref{fig_dustmass} we present the complete evolution of each dust type, summed over the all the clumps in the ejecta. We find that the formation of dust commences in the ejecta around 580 days post-explosion. There could be a very small mass of sulfides, or metallic clusters, formed a little earlier than that, but their mass is not expected to exceed 10$^{-6}$~\Ms; we did not follow those in this paper. The earliest grains to form in abundance are the magnesium silicates, followed by alumina, silicon carbide, and amorphous carbon at very late times. The mass of dust grows from 4$\times$10$^{-5}$ \Ms\ at 600 days to 1.8$\times$10$^{-3}$ \Ms\ at day 800, and to 6$\times$10$^{-3}$ \Ms\ at day 1000. The dust continues to form until $\sim$ day 2800, but after day 1600, the rate of dust formation is considerably reduced. The total dust mass in this scenario is 0.06 \Ms. Figure \ref{fig_dustcol} shows how the total column density (from the center to the outer edge of the ejecta) of the dust evolve with time, which combines the effect of dust formation with the expansion of the ejecta. 
Now, we shall briefly discuss the scenario of each dust species individually.

\subsection{Silicates}
\label{sec_silicates}

Silicates appear to form in the ejecta about 1.5 years after the explosion, which is the earliest among the four major dust species. Silicates form in the ejecta over a period that starts from about day 580 and continues until about day 1300. The formation zone spans over a large range in the velocity space (approximately between 900 and 2300 km s$^{-1}$; see Fig. \ref{fig_dust_vel}). As seen in Fig. \ref{fig_dust_vel}, the formation starts in a bin close to $v$ $\sim$ 1000 km s$^{-1}$; the formation zone generally follows a gradual inside to outside trend, other than for a few exceptions. For instance, the formation of CO molecules in large proportion and the cooling of the gas associated with it, often induce an earlier synthesis of silicates in that region of the ejecta. The mass of silicates grows from 4$\times$10$^{-5}$ \Ms\ at 600 days to 0.025 \Ms\ at $\sim$ 1400 days; thereafter its mass remains mostly unchanged (see Fig. \ref{fig_dustmass}). 

\subsection{Alumina}
\label{sec_alumina}

Alumina is the second most abundant dust species to form in the SN ejecta. The final masses are comparable to silicates, but they form much later in time. The synthesis starts at about day 860, and it continues until $\sim$~day 1500. Their formation regions in the velocity space overlap with those of silicates, since they are both formed where O atoms are the most abundant in the gas (see Fig. \ref{fig_dust_vel}). They end up at a final mass of about 0.022 \Ms (see Fig. \ref{fig_dustmass}).

\subsection{amorphous carbon}
\label{sec_am_carbon}

The synthesis of amorphous carbon dust provides an interesting scenario. It is found to form in the outer parts of the He core, where the velocities are between 2350 and 2550 km s$^{-1}$ (Fig. \ref{fig_dust_vel}). Those parts of the ejecta are He rich in nature, as seen in Fig \ref{fig_mass_vel}. The helium atoms, when ionized by the fast Compton electrons, produce He$^+$, which are extremely detrimental to any stable molecules or clusters present in the gas \citep{cherchneff2009, cherchneff2010, sar13}. Eventually, with the decrease in the number of the Compton electrons, ions of an inert gas such as helium tend to quickly recombine back to their neutral state. The timescale for this depends on the density of the gas in the clumps. In this scenario, we find the abundances of He$^+$ to be no longer significant after $\sim$ 900 to 1000 days. 

The formation of stable diatomic carbon, C$_2$, is a crucial bottleneck in the pathways to form carbon chains, which eventually lead to the formation of  carbon rings and fullerenes \citep{sar15}. In the first $\sim$ 1000 days, the formation of stable C$_2$ molecules is hindered by the presence of He$^+$. After that, stable C$_2$ molecules form via the radiative association of two C atoms: 

\begin{equation}
\label{reac_carbon}
\begin{gathered}
C + C \rightarrow C_2 + h\nu \\
O + C_2 \rightarrow C + CO 
\end{gathered}
\end{equation}
In the outer parts of the ejecta, the carbon atoms are more abundant in number compared to oxygen; however, there is a relatively large abundance of $^{18}$O compared to $^{16}$O in those regions \citep{rau02}. The reaction between neutral O and C$_2$ (Reaction \ref{reac_carbon}) is found to be a favored chemical path that results in the further breakdown of C$_2$ into atomic C. In this way, the C atoms end up in a loop with C$_2$ and CO. Depending on the density, this cycle continues for a rather long time, which does not allow for the formation of larger carbon clusters. Then finally, after a couple of years, all the O atoms end up depleted in CO molecules. Thereafter, rather quickly (within a few days), C chains and clusters are synthesized, leading to the condensation of amorphous carbon dust. Moreover, this series of processes are found to be extremely sensitive to gas densities. Higher gas densities results in an earlier recombination of He ions, and thereby an earlier onset of CO formation that leads to a much faster formation timescale. This also means that a small change in the abundance of oxygen, especially of $^{18}$O, can be significant in altering the timescale of the appearance of carbon dust in the ejecta, and its total mass. There is, of course, a certain degree of uncertainty associated with the ratio of  $^{18}$O to $^{16}$O predicted by the SN nucleosynthesis models; on the other hand, observationally it could be possible to identify the O isotope in the CO. This can then be used as a tool to predict the mass and epochs of carbon dust formation in the ejecta. 

The formation of carbon dust starts around 2400 days after the explosion and continues until day 2800 (Fig \ref{fig_dustmass}). In this time a mass of about 8 $\times$ 10$^{-3}$ \Ms\ is formed, which is strictly controlled by the C-to-O ratio. It found to be always less abundant than O-rich dust species in the gas. 

\subsection{Silicon carbide}
\label{sec_sil_car}

Silicon carbide is the least abundant dust species, which reaches a maximum of about 10$^{-4}$ \Ms\ (Fig. \ref{fig_dustmass}). SiC dust forms in two phases and in two distinct layers. In the inner layer of the ejecta, where the velocity range is between $\sim$ 850-950 km s$^{-1}$, silicon carbide forms around day 900, in a small mass of about 4 $\times$ 10$^{-6}$ \Ms\ (Fig. \ref{fig_dust_vel}). Thereafter, post 1900 days, more SiC dust is synthesized in the outer regions of the He core, in the same parts of the ejecta where C dust forms. The production of SiC at times later than 2000 days are potentially in agreement with the isotopic signature of SiC grains (of SN origin) found in meteorites \citep{liu_2018,ott_2019}.

\begin{figure}
\vspace*{0.3cm}
\centering
\includegraphics[width=3.5in]{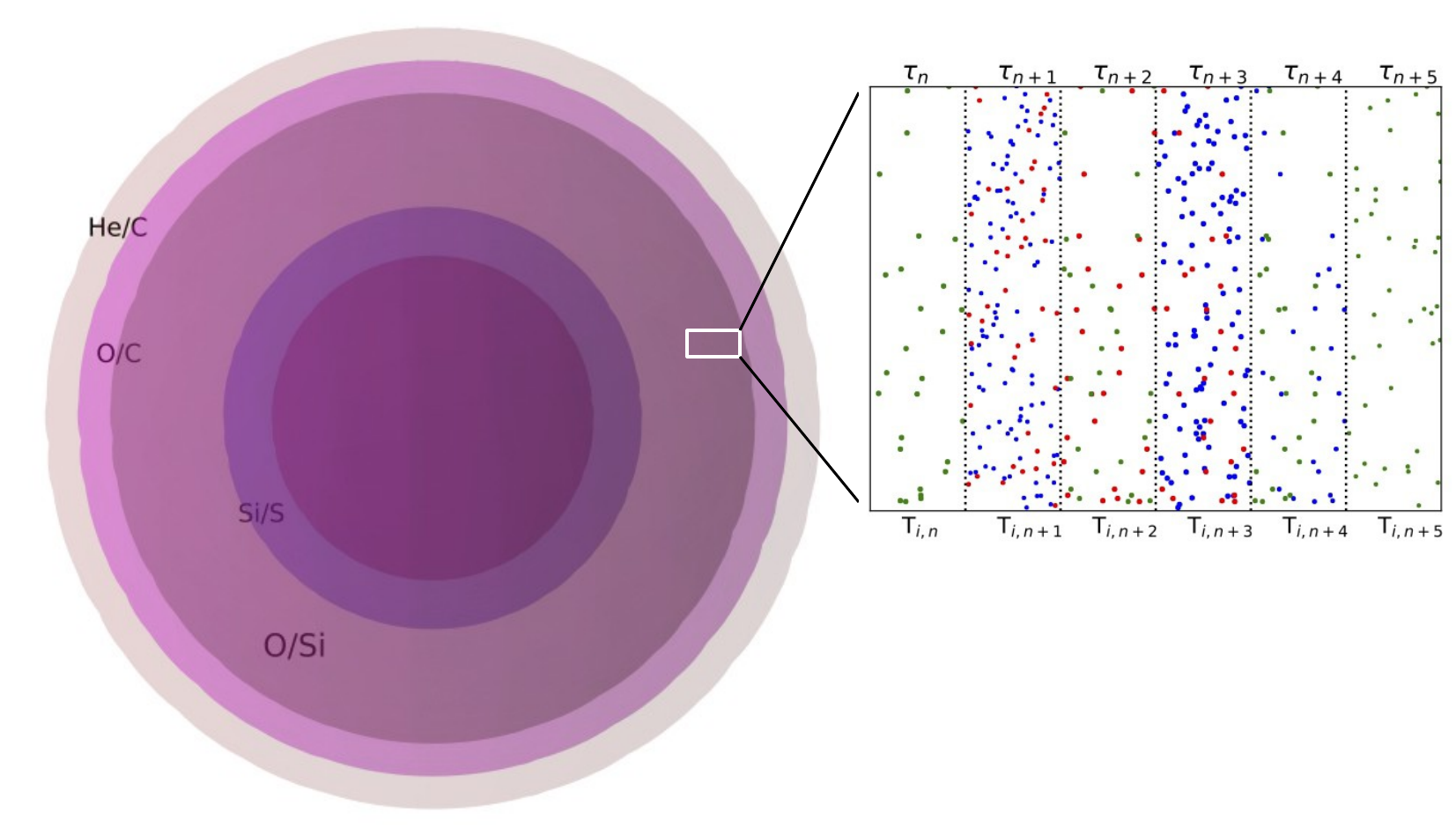}
\caption{\label{fig_cartoon_sp} Schematics of the geometry is shown, where the ejecta is characterized by different zones according to the abundances of elements, while each zone is stratified into several layers, defined by unique gas densities in the clumps, as well as unique dust and gas temperatures. As a result, the efficiency of dust production varies widely among the layers, and so do the opacities. The zoomed-in view of the small random rectangular cross-section of the gas indicates the unique dust densities, dust temperatures, and opacities in its different layers (despite being entirely in the O core). }
\end{figure}

\begin{figure*}
\vspace*{0.3cm}
\centering
\includegraphics[width=3.5in]{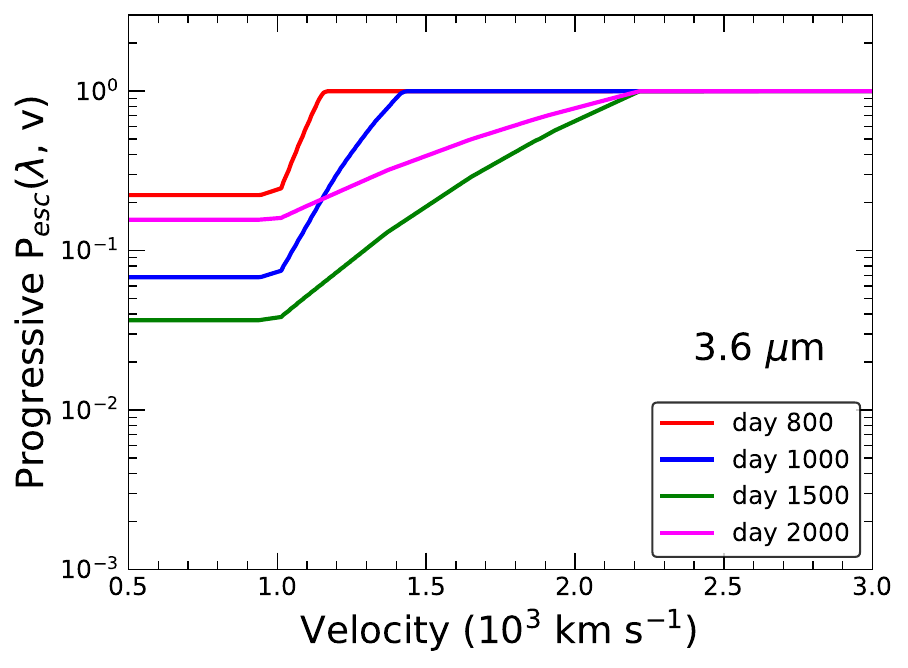}
\includegraphics[width=3.5in]{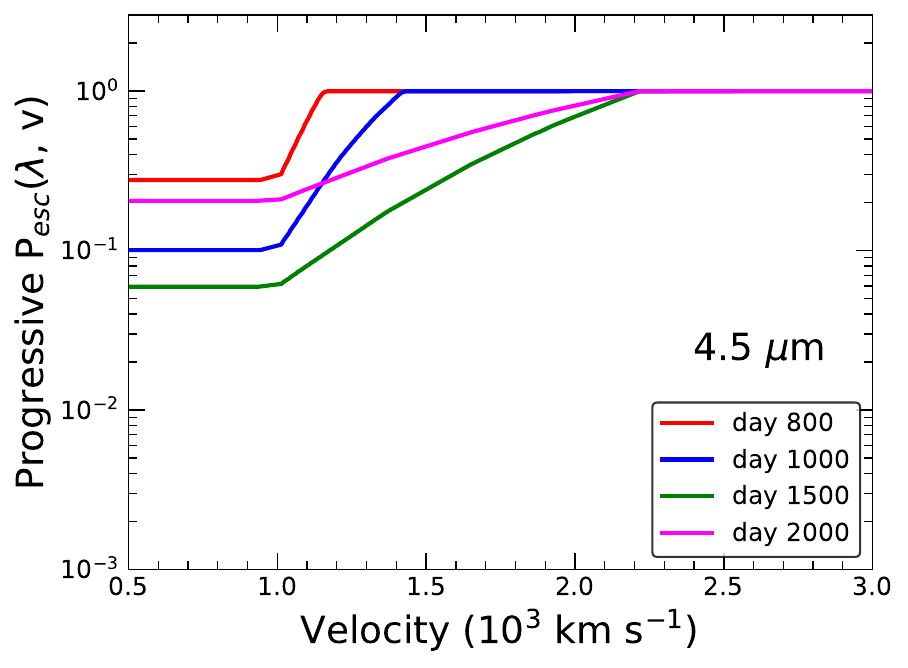}
\includegraphics[width=3.5in]{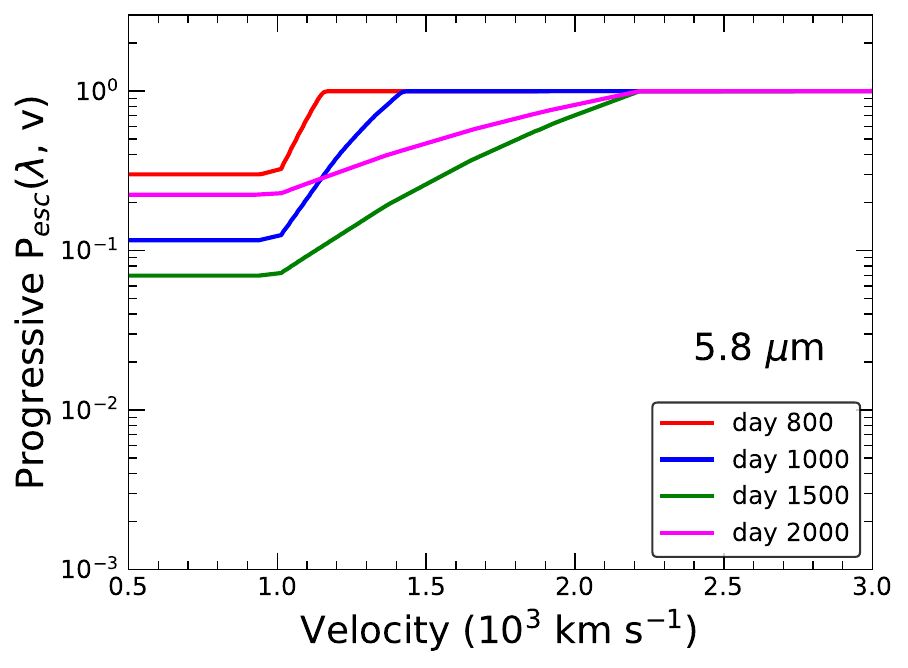}
\includegraphics[width=3.5in]{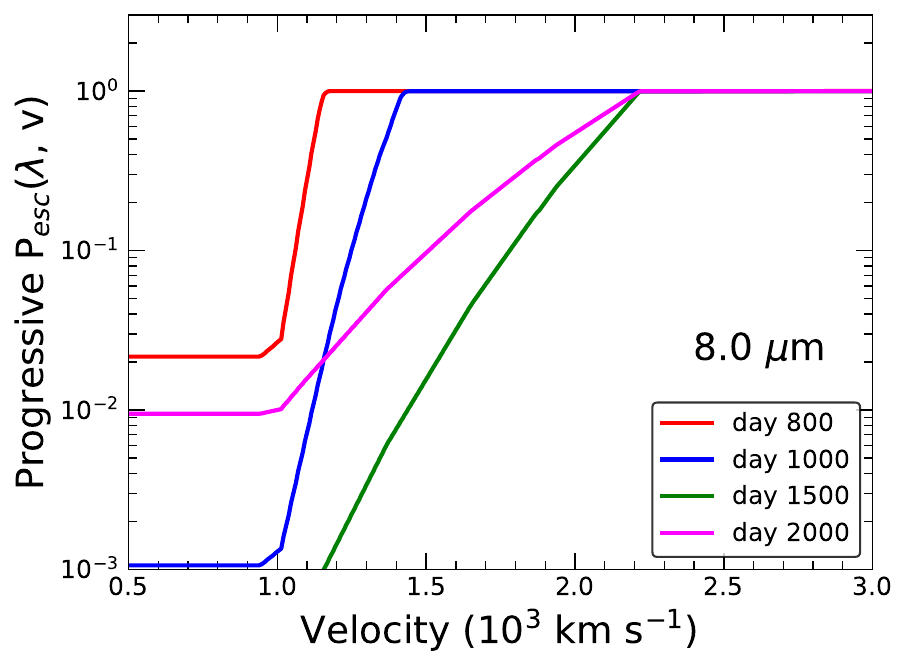}
\caption{\label{fig_Pesc} Wavelength-dependent progressive escape probabilities (Eq. \ref{eqn_pesc_full}), which are a function of the local velocity, are presented for days 800, 1000, 1500, and 2000, for four different wavelengths (3.6, 4.5, 5.8, and 8.0 \mic) that correspond to the \textit{Spitzer Space Telescope} IRAC bands. }
\end{figure*}

\begin{figure*}
\vspace*{0.3cm}
\centering
\includegraphics[width=2.33in]{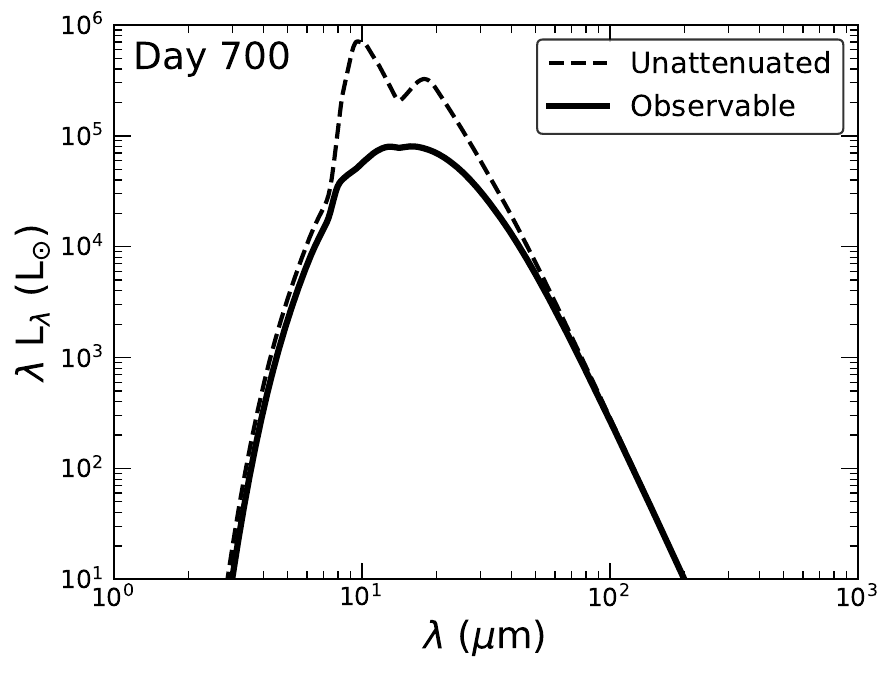}
\includegraphics[width=2.33in]{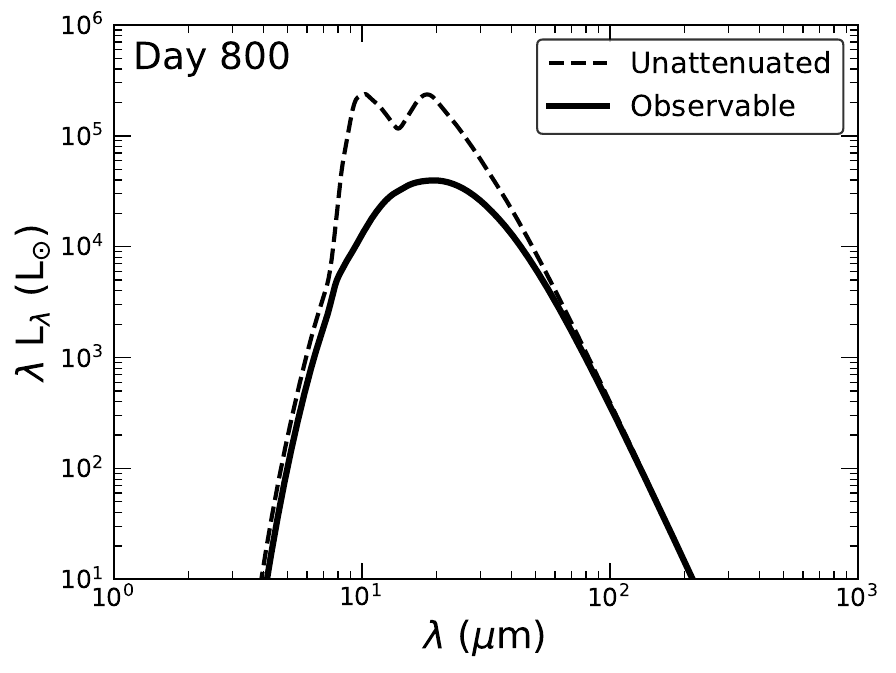}
\includegraphics[width=2.33in]{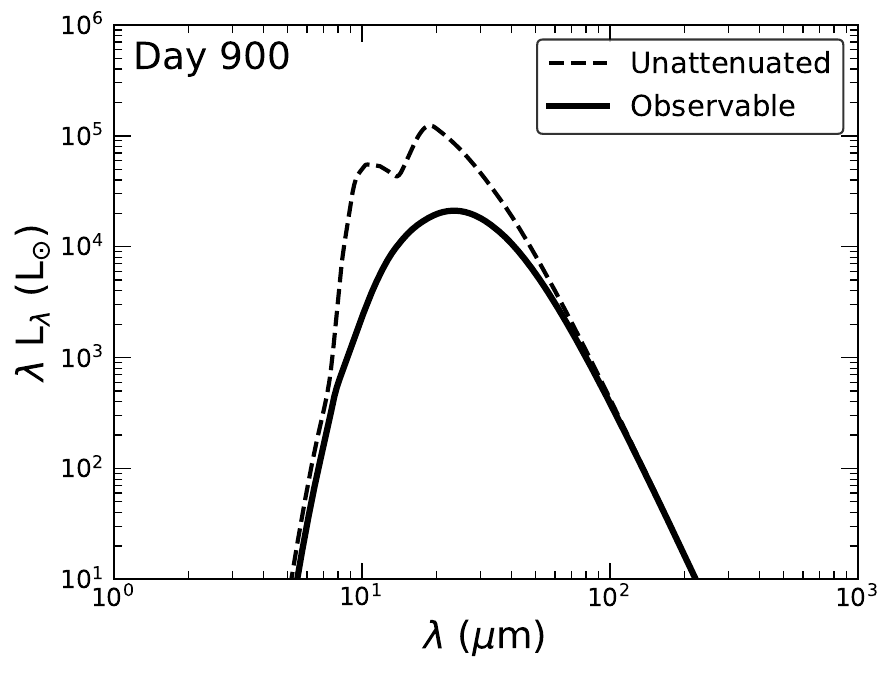}
\includegraphics[width=2.33in]{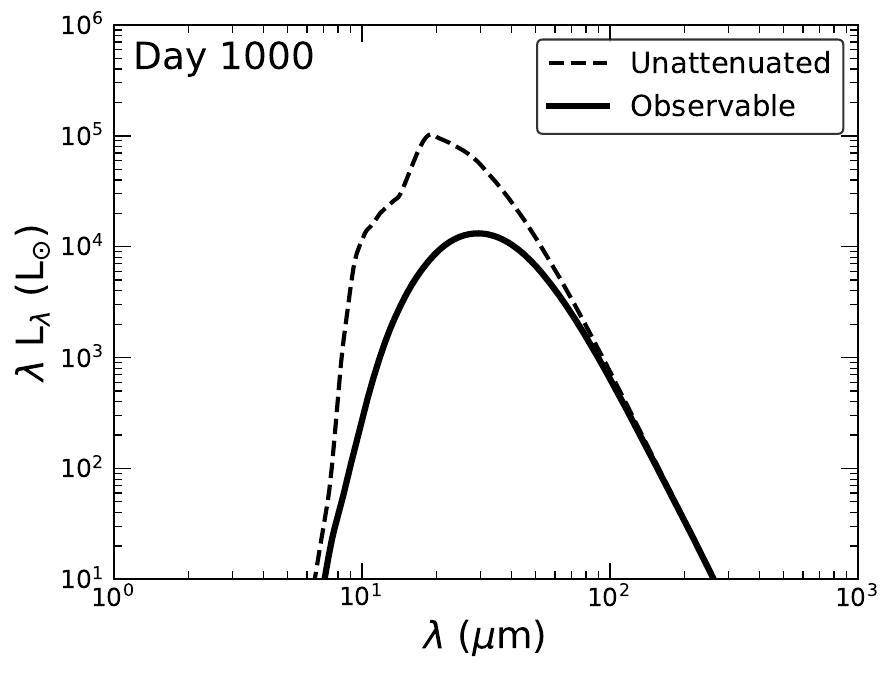}
\includegraphics[width=2.33in]{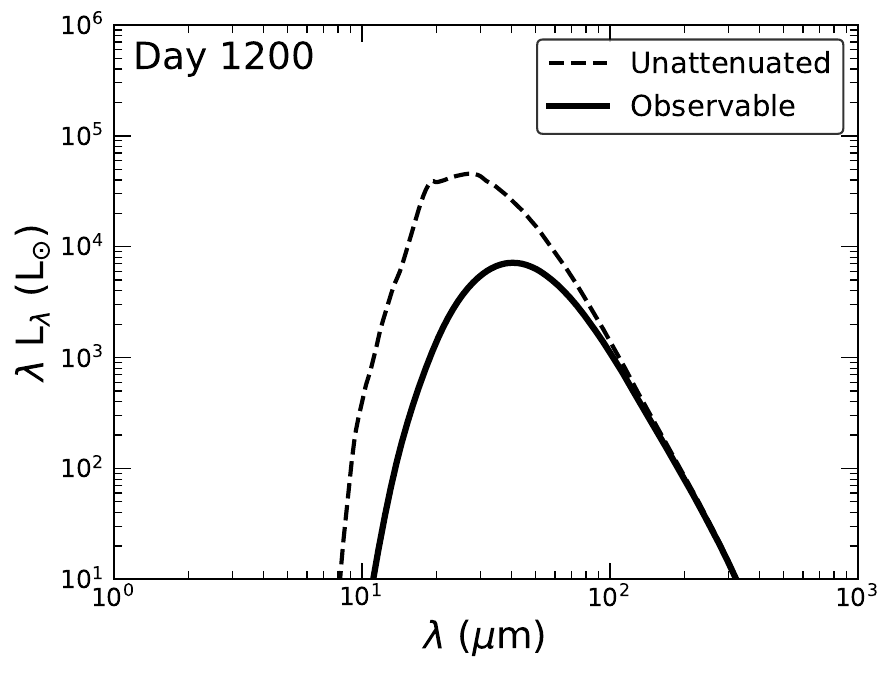}
\includegraphics[width=2.33in]{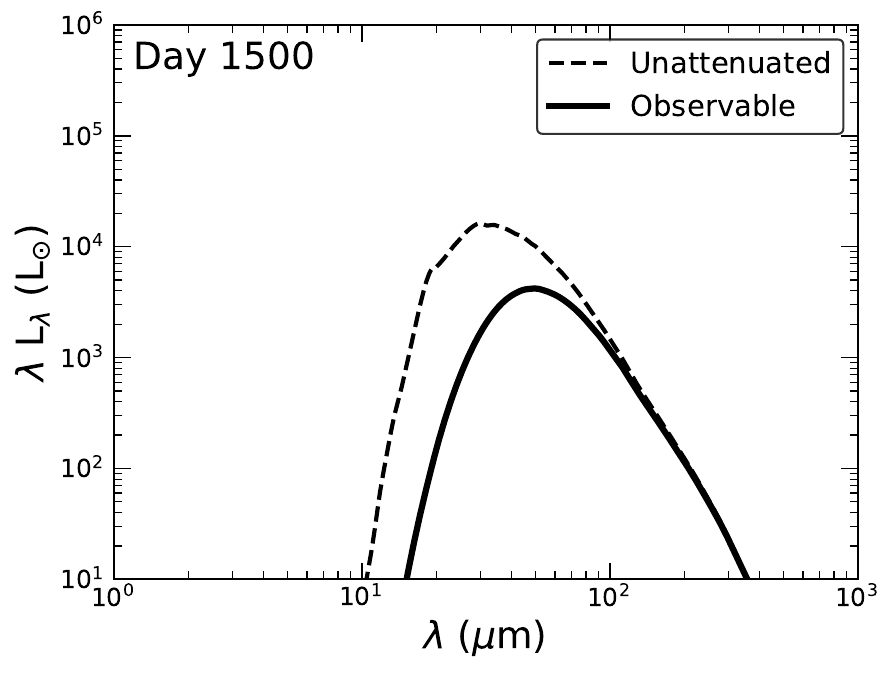}
\includegraphics[width=2.33in]{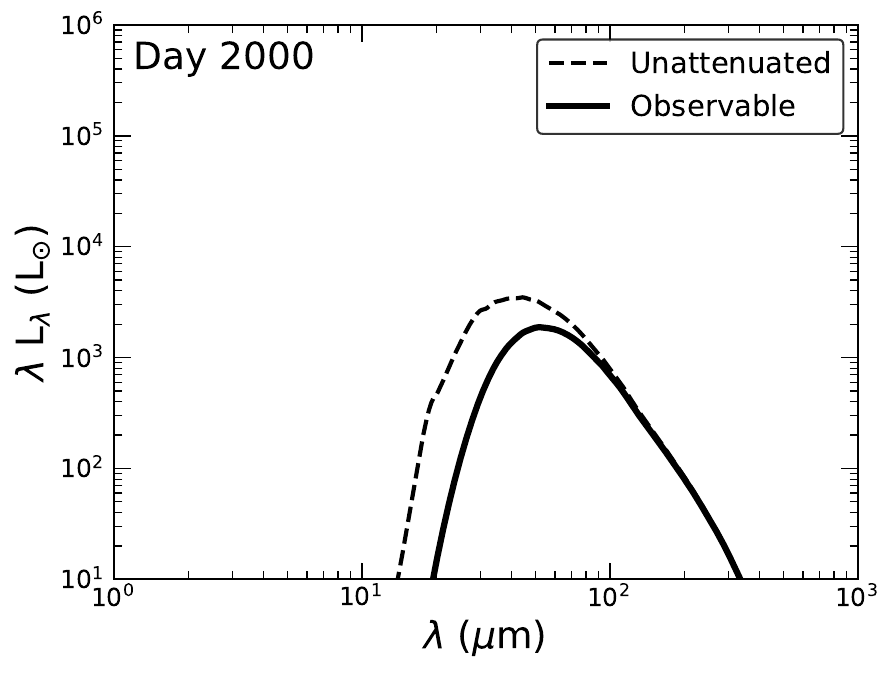}
\includegraphics[width=2.33in]{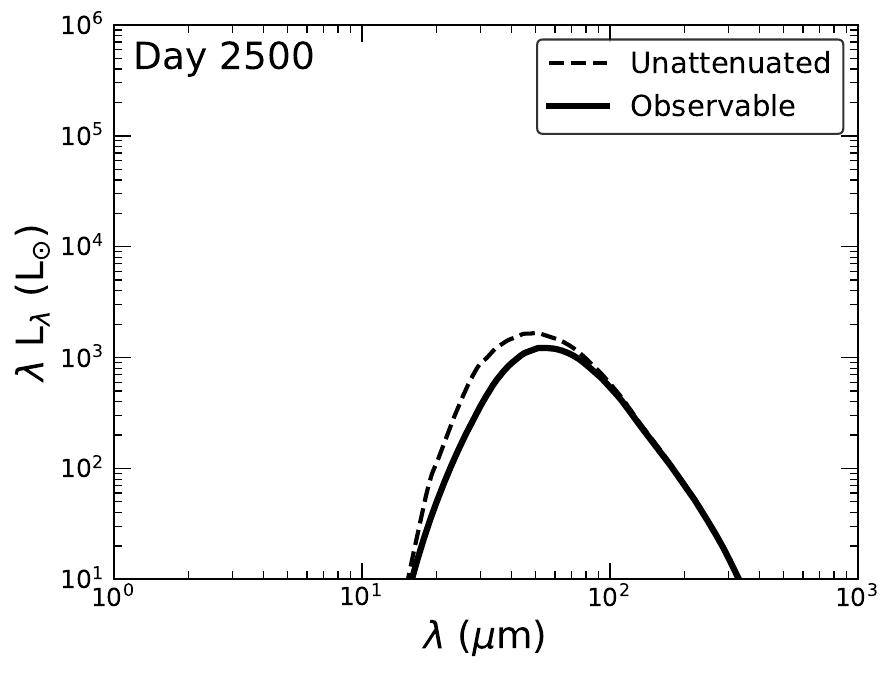}
\includegraphics[width=2.33in]{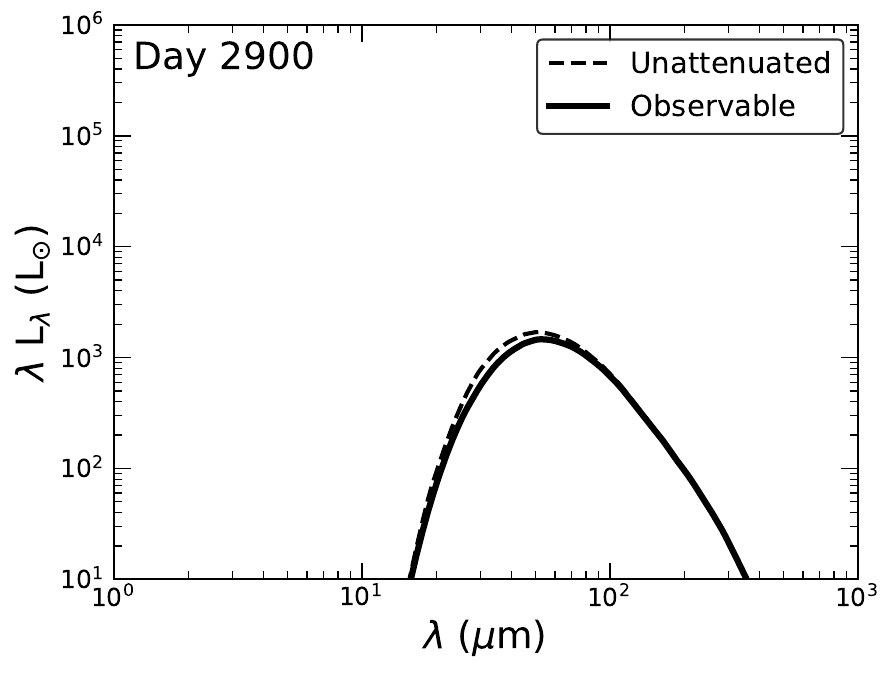}
\caption{\label{fig_Lum} Radiated energy per unit time (in solar luminosities) emitted by dust in the SN ejecta as a function of wavelengths, for nine different epochs (see Eq. \ref{eqn_lum}). The unattenuated luminosity (dashed line) corresponds to the total energy per unit time (from the dust), while the observable luminosity (solid line) is the luminosity that escapes the system, after taking the progressive escape probabilities into account, and will be the one that can be observed (see Sect. \ref{sec_opacities} and Eqs. \ref{eqn_pesc_full} and \ref{eqn_lum} for details). }
\end{figure*}

\section{Progressive opacities and infrared spectra}
\label{sec_opacities}

In this section, we discuss the IR radiation and the fluxes generated by the distribution of dust in the SN ejecta. 
The rate of emission per unit wavelength produced by a single dust grain is shown in Eq. \ref{eqn_coolingrate}, where it is integrated over all wavelengths to find the cooling rate. The emerging radiation from a distribution of dust is, however, not always the total radiation of all dust grains combined, since it also depends on the opacities of the medium as a function of wavelength. Assuming a uniform distribution of dust within a spherical volume, the probability of escape of the IR photons, $P_{esc}(\lambda)$, was estimated by \cite{dwe15} with respect to SN~1987A. 

In this study, by modeling the formation of dust in the thinly spaced, clumpy shells in the ejecta, we estimated the generic distribution of each dust species individually, along with their respective distribution of temperatures. This is illustrated in Fig. \ref{fig_cartoon_sp}, where we zoom into a small cross-section of the O-rich zone of the ejecta, to show thinly spaced shells characterized by different optical depths (owing to their different dust densities and compositions) and different dust temperatures. 

For a thin shell of homogeneous density and thickness $\Delta r$, the optical depth $\tau_h(\lambda)$ and the corresponding escape probability $P_{esc}^h$ \citep{inoue_2020} of photon at wavelength $\lambda$, can be expressed as

\begin{equation}
\label{eqn_shelltau}
\begin{gathered}
\tau_{h}(\lambda, v) = \sum_S \sum_a \int \rho_{d,S}(v, a) k_S(\lambda, a) \mathrm{d}r \\
 = \sum_S \sum_a \frac{M_{d,S}(v, a) }{4 \pi r^2 \Delta r} k_S(\lambda, a) \Delta r = \frac{\sum_{S,a} M_d k}{4 \pi r^2} \\
 P_{esc}^h = e^{-\tau_h},
\end{gathered}
\end{equation}

where $\rho_{d,S}$ and $M_{d,S}$  are the density and total mass of the dust, respectively, of a given species $S$ and radius $a$, in that shell. We have summed over all grain sizes $a$ and all dust species $S$. $P_{esc}^h$ defined in this way, of course, is a function of the velocity bin ($v$) of the ejecta; in other words, each shell will have a unique $P_{esc}^h$.  

For a clumpy medium, a suitable approach for deriving the opacities and escape probabilities is to consider a clump as one large dust grain or mega-grain, defined by its individual absorption and emission properties \citep{varosi_1999, inoue_2020}. In that scheme, following \cite{inoue_2020}, the optical depth of a clump can be written in terms of the corresponding homogeneous optical depth: 

\begin{equation}
\label{eqn_taueff}
\tau_{cl}(\lambda) = \tau_{h}(\lambda) \ \frac{r_{cl}}{r} f_V^{-1}; \ \ r_{cl} \simeq \ \ \frac{r f_V}{10}; \ \  \tau_{cl} \simeq 0.1 \tau_h,
\end{equation}

where we have used the limiting relations for a radius of a clump, $r_{cl}$, from \cite{dessart_2018}. Assuming most of the dust is located inside the clumps, we can derive the escape probability of the individual shells, $P_{esc}^{sh}$ \citep{inoue_2020}, in the following form:  

\begin{equation}
\label{eqn_pesc}
\begin{gathered}
\tau_{eff}(\lambda) = \tau_{h}(\lambda) \ P_{esc}^h (\tau_{cl}) = \tau_{h}(\lambda) e^{- 0.1 \tau_h(\lambda)} \\
P_{esc}^{sh}(\lambda, v) = P_{esc}^{h}(\tau_{eff}) = e^{-\tau_{eff}(\lambda, v)}
\end{gathered}.
\end{equation}

Finally, using Eq. \ref{eqn_pesc}, we calculate the probability of escape of an IR photon from the entire ejecta, to reach the observer. If the photon originates from a clump located in a given shell identified by velocity $v$, then $P_{esc}$ can be expressed as the cumulative product of the probability of escape of all the shells lying along its path, starting from that shell to the outer edge of the He core. Therefore, we can write 


\begin{equation}
\label{eqn_pesc_full}
P_{esc}(\lambda, v) = \prod_v^{v_c} P_{esc}^{sh}(\lambda, v) = \exp[- \sum_v^{v_c} \tau_{eff}(\lambda, v)].
\end{equation}

We have termed $P_{esc}(\lambda, v)$ as the progressive escape probability, since it is expressed in the velocity space (along with $\lambda$ and $t$). 
Considering the generic distribution of dust is nonuniform, this formulation is the way to account for the dust density distribution at any give time, when estimating the chances of a photon originating from a given clump in a shell, to reach the observer. 

We find that the sizes (radii) of all dust species range between 0.001 to 0.1 \mic. In the infrared, the wavelengths are much larger compared to the average grain radii ($\lambda_{IR} \gg a$). Within this Rayleigh limit, the luminosities of the dust grains are not dependent on the grain sizes. The same can be said for the optical depths in the IR regime of the individual clumps. For simplicity, a single grain of radius 0.01 \mic\ was assumed. 

In Fig. \ref{fig_Pesc} we show $P_{esc}$ for days 800, 1000, 1500, and 2000 for the \textit{Spitzer Space Telescope} Infrared Array Camera (IRAC) bands ($\lambda$ = 3.6, 4.5, 5.8, 8.0 \mic). As the figures show, at the earlier epochs, the opacities are large ($P_{esc}$ small) only in the inner parts of the ejecta. With time, the opacities are found to increase more and more in the outer velocity bands, as the formation zones of silicates and alumina shift outward. Overall, the escape probabilities are shown to decrease consistently until about day 1500, after which there is a decline in the opacity, and therefore an increase in $P_{esc}$. This results from a combined effect of the slowing down of the dust production and the expansion of the ejecta. Figure \ref{fig_dustcol} shows the dust column densities, which also present similar trends and can easily be translated to the total opacities from the center of the ejecta to the outer edge of the He core. The 8-\mic\ band clearly shows larger opacities than the other three wavelengths, owing to the larger values of the absorption coefficients of O-rich dust grains close to that wavelength (see Fig. \ref{fig_kappa}).

The IR flux from SN ejecta is the IR luminosity (amount of energy that escapes the ejecta per unit time), divided by $4\pi L^2$, with $L$ being the distance of the SN from the Earth. The observable IR luminosity is expressed as 

\begin{equation}
\label{eqn_lum}
\begin{split}
L_{\lambda}^{obs}(t) = \sum_v \sum_{S, a} & \ 4 \ M_{d,S}(v,t,a) k_S(\lambda, a) \\
& \times \pi B_{\lambda}(\lambda, T_S(v, t, a)) P_{esc}(\lambda, v, t) ,
\end{split}
\end{equation}

where we have summed over all dust species $S$, and all grain sizes and temperatures, over the complete velocity space. The total energy emitted by dust per unit time, which can also be called the unattenuated luminosity, is given by the same expression, without the $P_{esc}$ term. 

In Fig. \ref{fig_Lum}, the unattenuated as well as the observable luminosities produced by the dust grains in the SN ejecta are shown for days 700, 800, 900, 1000, 1200, 1500, 2000, 2500, and 2900. As expected, the ratio of the two luminosities changes over time, and the contrast is largest between 1200 and 1500 days. At longer wavelengths, typically larger than $\sim$ 100 \mic, the unattenuated and the observable luminosities almost always concur, since the dust opacities are much lower in the far-IR wavelengths compared to the mid-IR bands. 

Observationally, the detection or non-detection of the strong silicate feature at 9.7 \mic\ \citep{simpson_1991} often motivates the estimation of the possible dust composition \citep{erc07, wesson2015, wesson_2021}. Our results suggest that the composition of dust in the ejecta is heavily dominated by O-rich dust at all times. However, we want to stress that, when there is a local maximum in the absorption coefficient at any wavelength band (for example around 9.7 \mic\ for silicates), while the IR luminosity in that band increases, the optical depths at that wavelength also increase simultaneously. Therefore, this leads to a strong attenuation at those wavelengths. The combination of the two (increased luminosity, increased attenuation), often leads to a complete suppression of the strong features in the spectra. This was also pointed out by \cite{dwe15}. Figure \ref{fig_Lum} confirms this as well, where we can clearly see the strong features originating from the O-rich dust components in the unattenuated luminosities until day 1000, while the observable luminosity never shows any noticeable feature.  After day 1000, dust formation is accelerated in the outer parts of the O/Si zone, and the mass-weighted dust temperatures are already too cool to radiate strongly in the 10-20 \mic\ range. 

As we see in Fig.~\ref{fig_Lum}, post day 2500, the ejecta becomes increasingly thin to the self-emitted IR radiation. amorphous carbon dust appears in the outermost layers of the He core at about the same time. However, they do not form an outer dusty shell that is thick enough to block the radiation coming from the inner O-rich dust grains. 

\begin{figure}
\vspace*{0.3cm}
\centering
\includegraphics[width=3.5in]{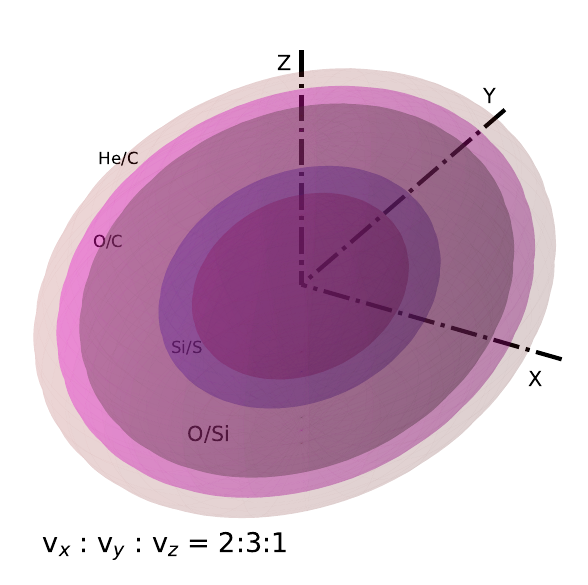}
\caption{\label{fig_ellipsoid} Geometry of anisotropic ejecta and the corresponding zones, in the form of an ellipsoid. To emulate SN~1987A, we chose the velocity ratios to be 2:3:1 (with reference to \citealt{kjaer_2010}), where the velocity distribution along the mean axes (taken as the x axis in the figure) remain the same as the isotropic model. }
\end{figure}

\begin{figure}
\vspace*{0.3cm}
\centering
\includegraphics[width=3.5in]{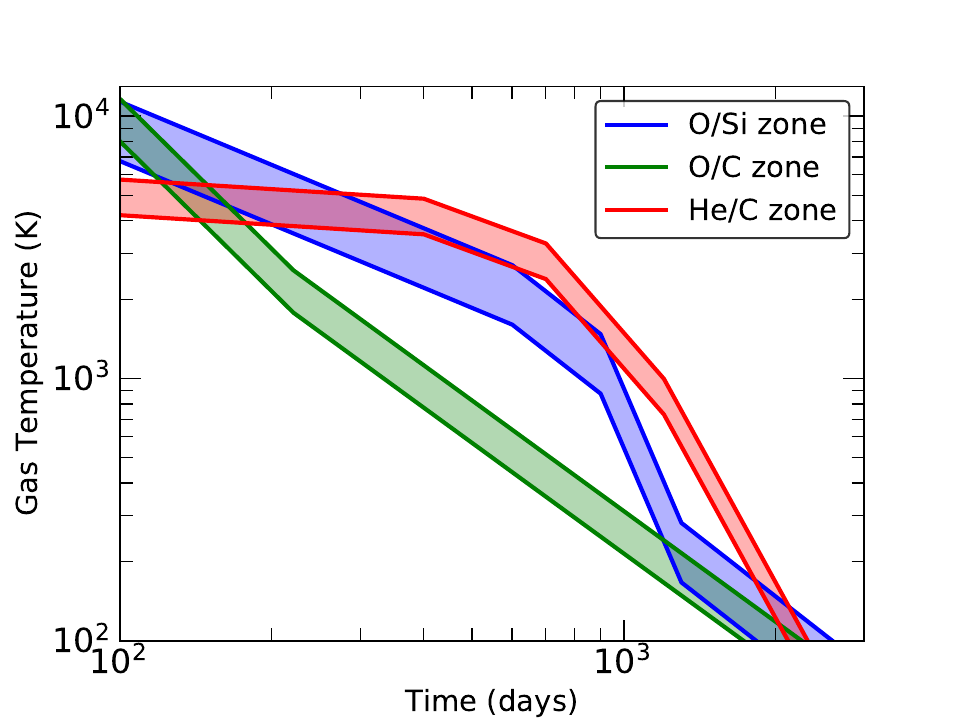}
\caption{\label{fig_ellipsoid_gasT} Distribution and evolution of temperature with respect to the various zones are shown for the ellipsoidal ejecta. Comparing to the same distribution and evolution (Fig. \ref{fig_gasT}) for the spherical model, it is evident that the temperatures are spread to wider ranges in this case, as expected}
\end{figure}

\begin{figure*}
\vspace*{0.3cm}
\centering
\includegraphics[width=6.5in]{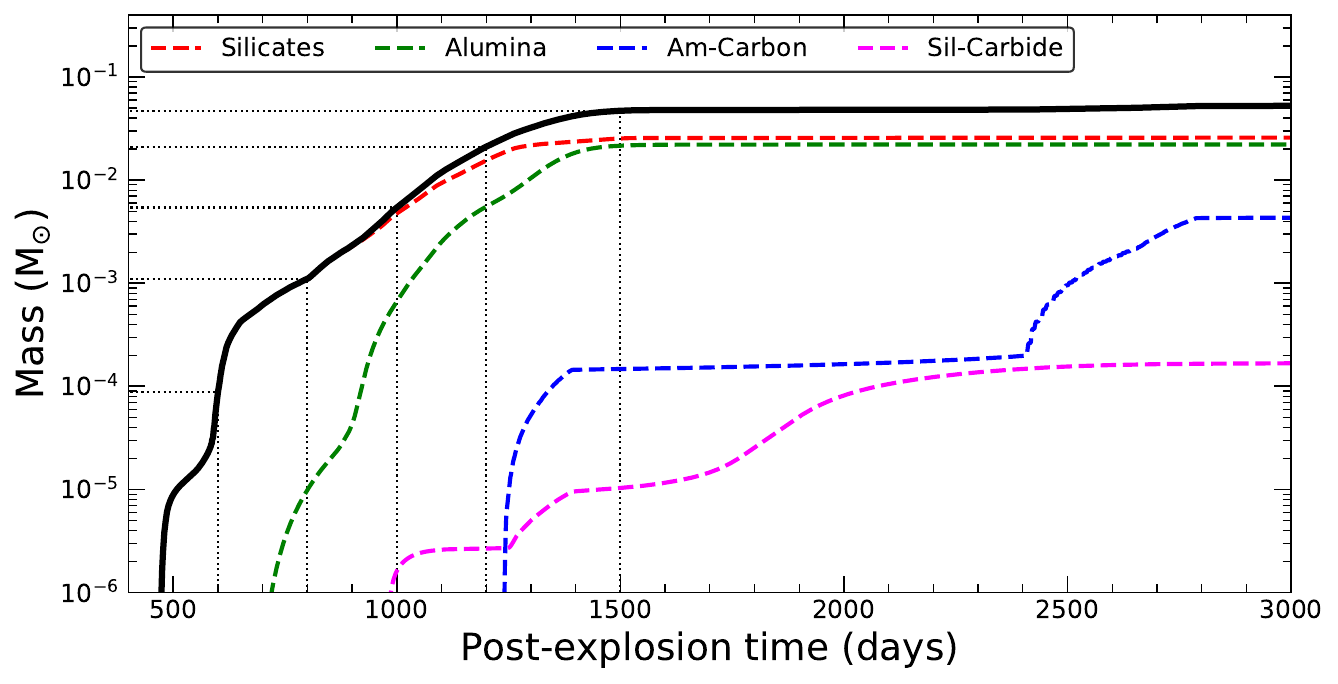}
\includegraphics[width=6.5in]{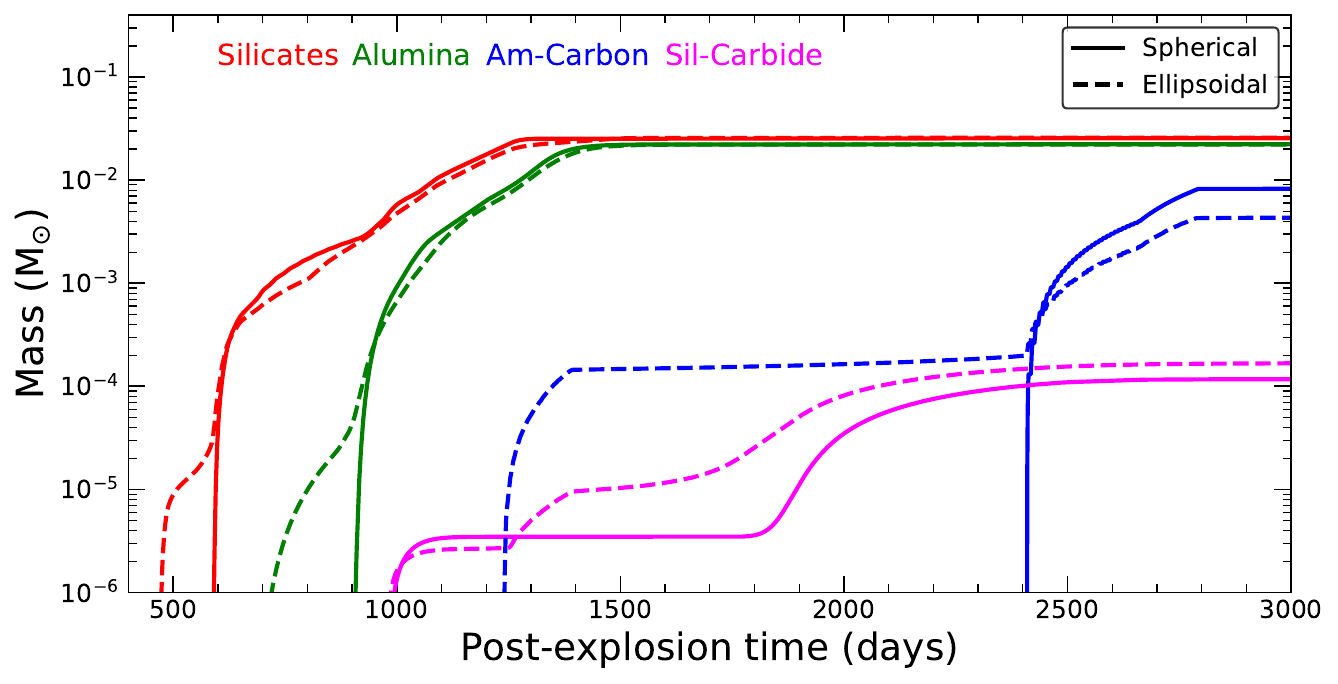}
\caption{\label{fig_dustmass_el} Mass of dust for the ellipsoidal case is presented. \textit{Top}: Evolution of total dust masses, in the case of anisotropic ejecta, are shown for each dust species, and the sum is shown by the solid-line. \textit{Bottom}: Mass of dust produced in the spherical and ellipsoidal cases are compared for each dust species. }
\end{figure*}

\begin{figure*}
\vspace*{0.3cm}
\centering
\includegraphics[width=3.5in]{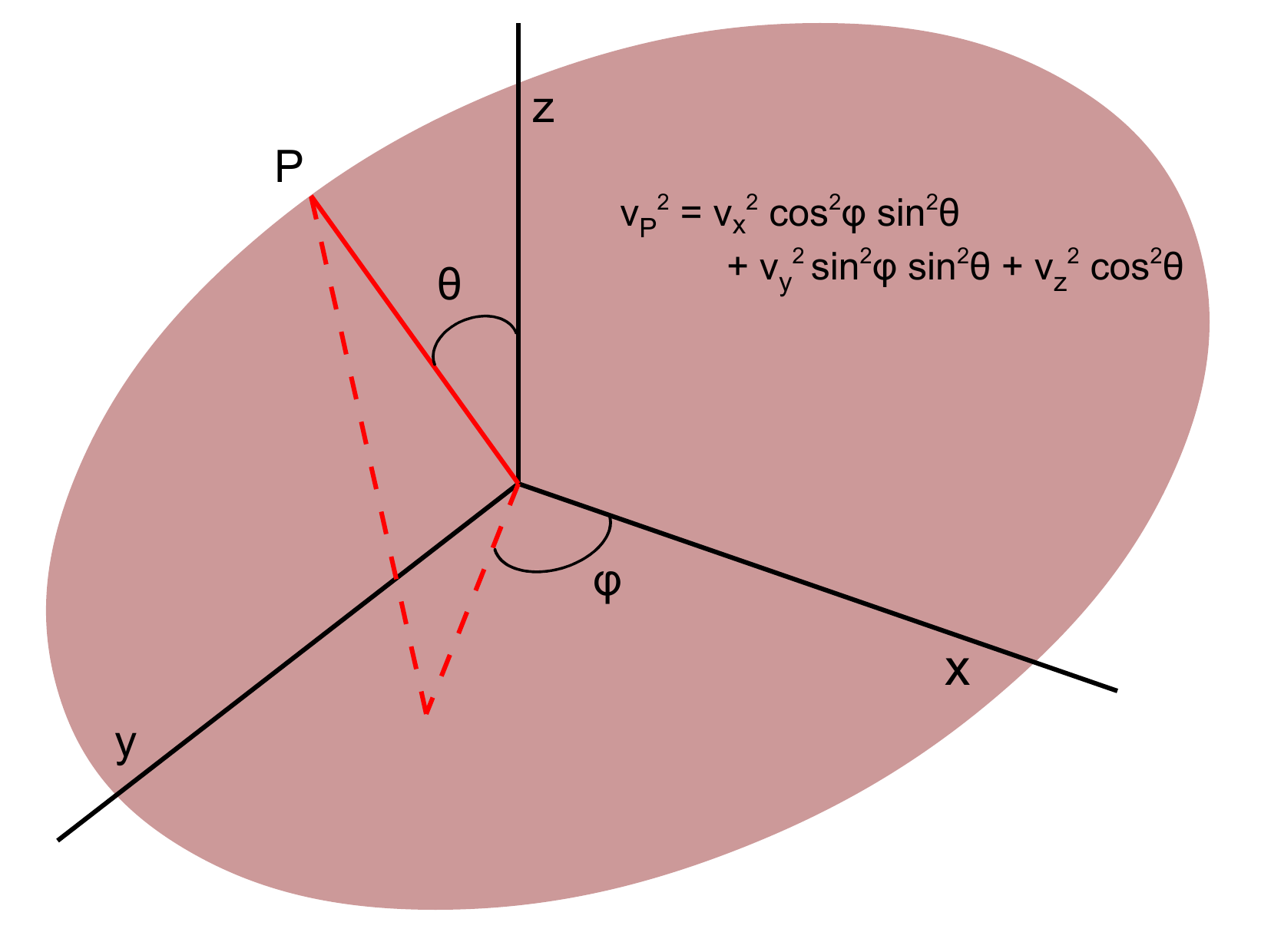}
\includegraphics[width=3.5in]{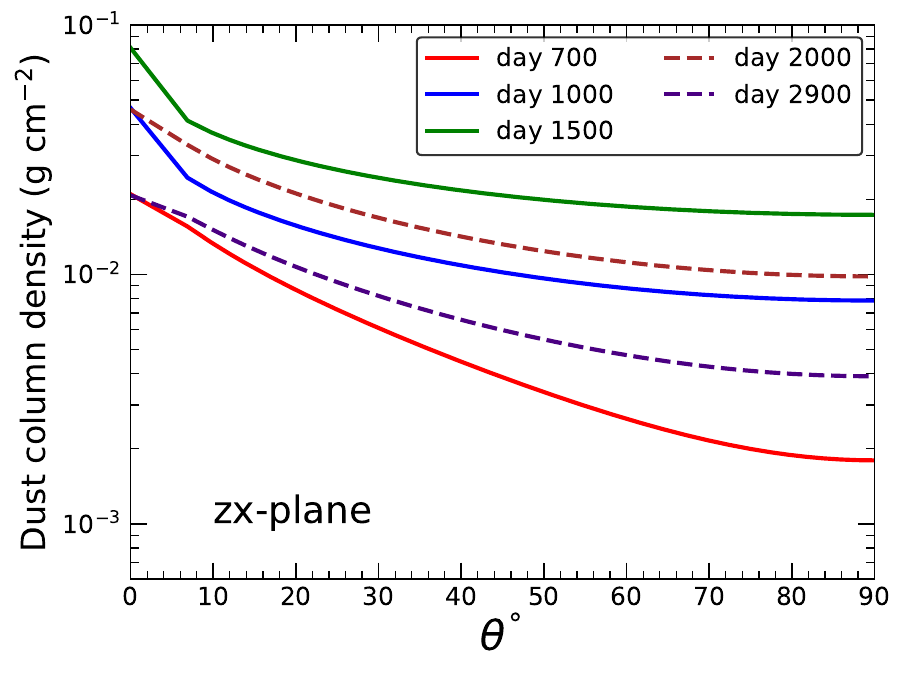}
\includegraphics[width=3.5in]{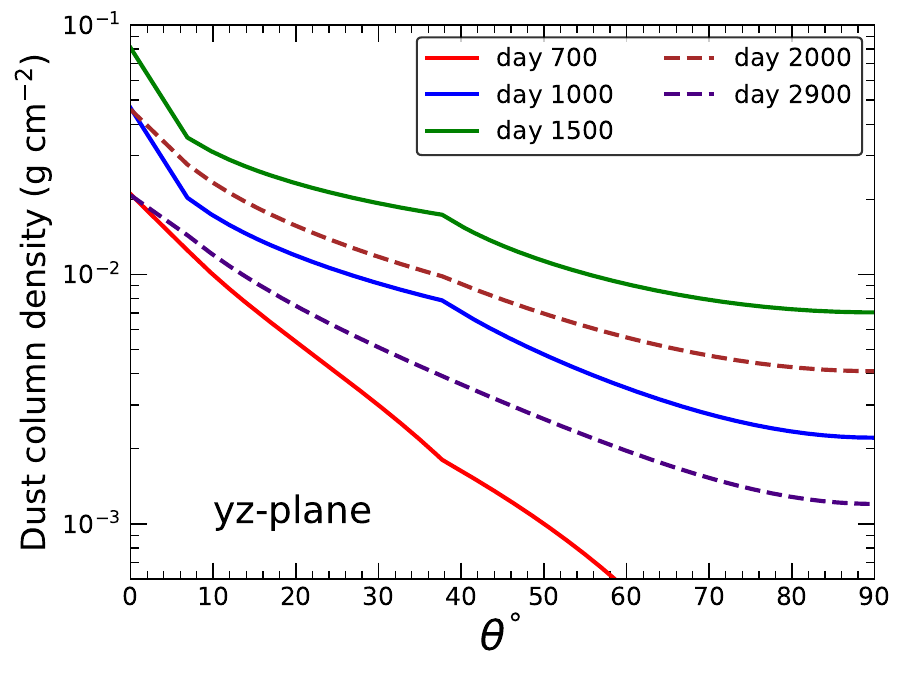}
\includegraphics[width=3.5in]{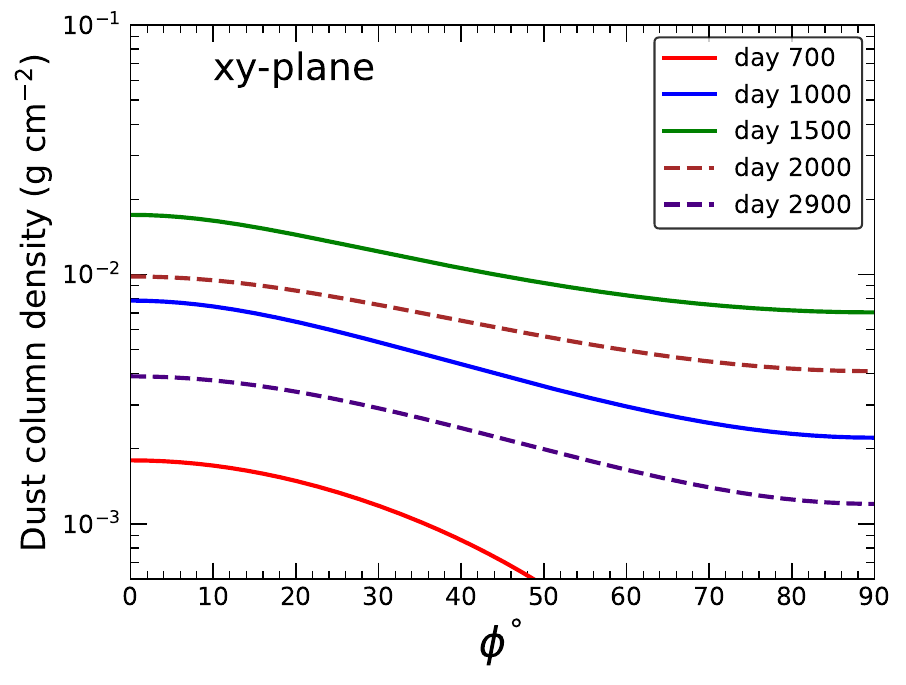}
\caption{\label{fig_dustcolumn} Dust column densities are shown for the ellipsoidal model for five different epochs. In the top left panel, the chosen polar coordinates are shown in terms of $\theta$ and $\phi$ for an ellipsoidal shell. The velocity of any clump at point $P$ inside the shell is calculated. In the top right panel and the bottom panel, we show the evolution of dust column densities (from the center to the outer edge of the ejecta) along various directions as a function of the viewing angle, in the three axial planes. See Sect. \ref{sec_compare} for details.  }
\end{figure*}

\section{Elliptical geometry}
\label{sec_elliptical}

In this section, we discuss the formation and distribution of dust in the ejecta, and the resulting IR emission, when the geometry of the ejecta is assumed to be anisotropic. 

By analyzing the spectral information obtained from the 3D spectroscopy of SN~1987A, \cite{kjaer_2010} found that the 2D projection of the ejecta is elliptical in shape with an axial ratio of approximately 1:1.8. The observations can be fit well using an ellipsoidal geometry where the axes are in ratios of 1:2:3. 

We adopted the same geometry for the ejecta, where the ellipsoid is defined by the ratio of velocities along the $x$, $y$, and $z$ axes, which are taken to be 2:3:1, as shown in Fig.~\ref{fig_ellipsoid}. The velocities along the $x$ direction are kept identical to the velocities of the spherical ejecta that were derived in Sect. \ref{sec_ejecta}. With respect to that, the velocities along the $y$ axis are then given by $v_y = 1.5v_x$, and along $z$ axis by $v_z = 0.5 v_x$. Assuming a conservation of column densities, the gas density therefore scales along the $y$ and $z$ directions as $n_y = (2/3) n_x$ and $n_z = 2n_x$, respectively, while $n_x$ (along the $x$ direction) remains the same as the gas densities in the spherical case.  

In Sect. \ref{sec_stratification}, we explained the stratification of the ejecta in the velocity space, where the He core is divided into 450 thinly spaced spherical, clumpy shells, identified by their velocities. Adopting the same approach, each shell in this case is converted into a thin ellipsoidal volume, where the velocities scale in the chosen ratio of 2:3:1, along the three axes. To be specific, the clumps at velocity $v$ of the spherical model are now distributed in a range of velocities from $0.5 v$ to $1.5 v$; all these clumps have identical abundances, but the gas densities, and the gas and dust temperatures are now distributed over a range as a function of their radial velocities. The spread of gas temperatures in each zone of the ellipsoidal ejecta are shown in Fig. \ref{fig_ellipsoid_gasT}, in comparison to Fig. \ref{fig_gasT} of the isotropic model. The O/C zone and the He/C zone, which are toward the outer part of the ejecta, clearly shows a larger range in gas temperatures compared to Fig. \ref{fig_gasT}. In the isotropic model, the O/C layer was a thin shell, so the range of temperatures was also narrow. However, that thin shell in an ellipsoidal geometry now covers a range of velocities along different axes, which explains the larger spread in the temperature distribution. The same is true also for the He/C zone as well. 

Within each ellipsoidal shell, the individual clumps can be identified by their polar coordinates, as shown in the top left panel of Fig. \ref{fig_dustcolumn}. We associate the clumps with their respective velocities within each shell, as shown for a hypothetical clump at point P in the figure. A bin size of 80 was chosen for each shell, which means the ejecta were stratified into 450 $\times$ 80  different groups of clumps (450 ellipsoidal shells, each divided into 80 bins). Fortunately, for each of these shells, all the clumps within a single shell have the same mass fraction of elements, and only vary in physical conditions. Hence, the computation is relatively less strenuous, compared to an actual 3D hydrodynamic model, where any random parcel of gas has unique abundances. In the following sections, we discuss the chemical yields, estimated assuming this anisotropic geometry.

\begin{figure*}
\vspace*{0.3cm}
\centering
\includegraphics[width=3.5in]{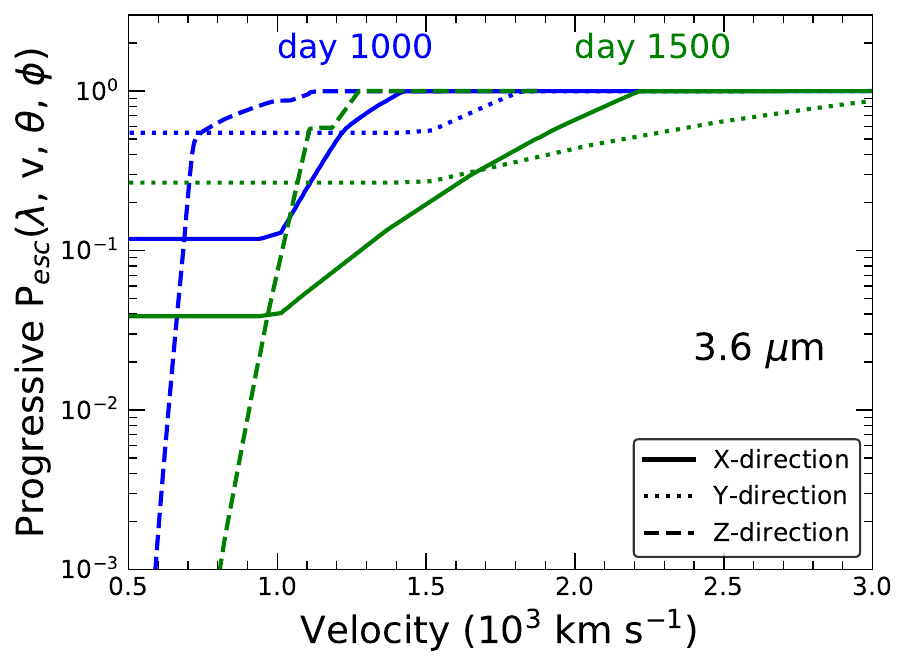}
\includegraphics[width=3.5in]{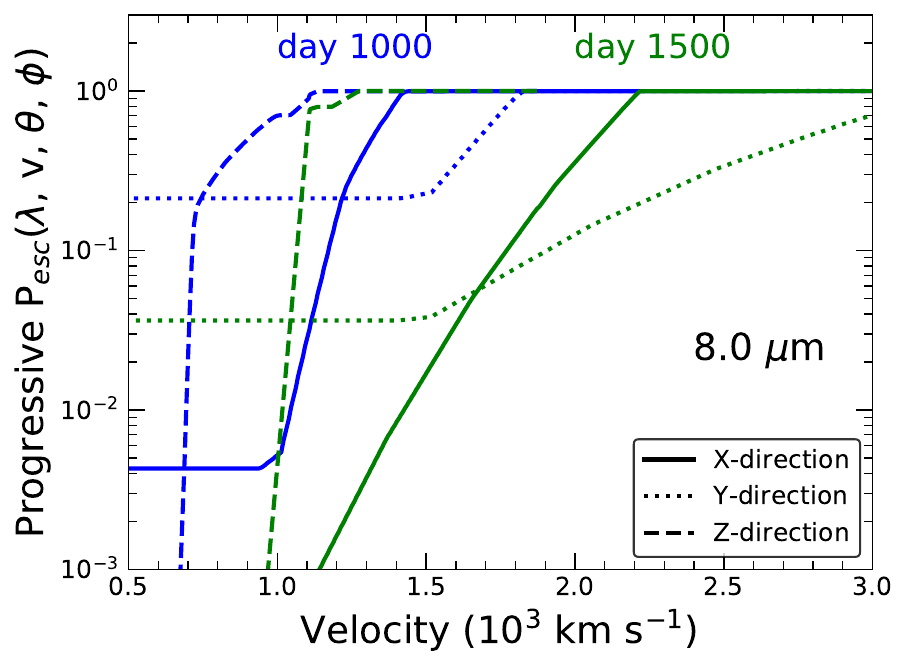}
\includegraphics[width=3.5in]{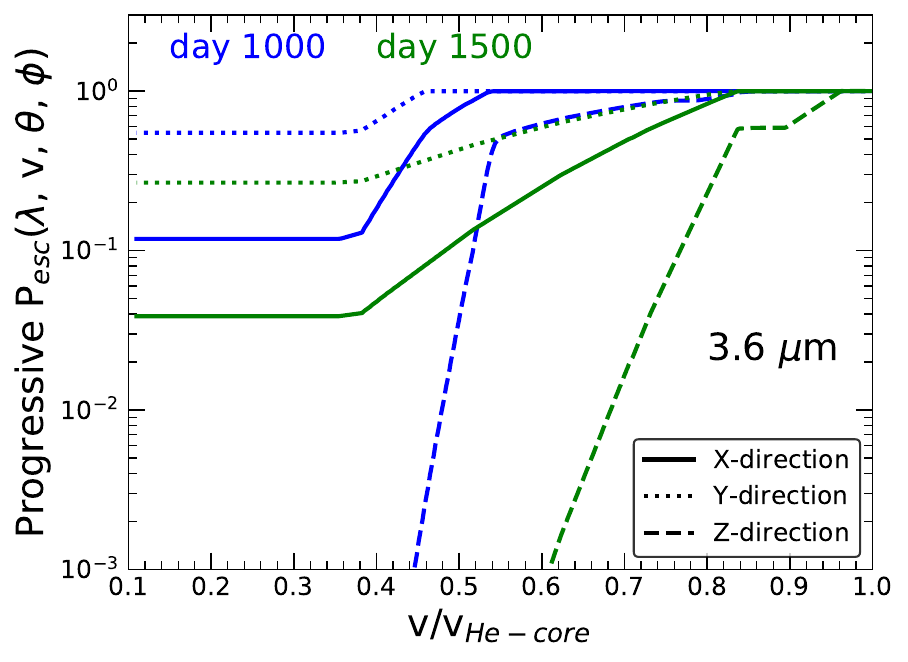}
\includegraphics[width=3.5in]{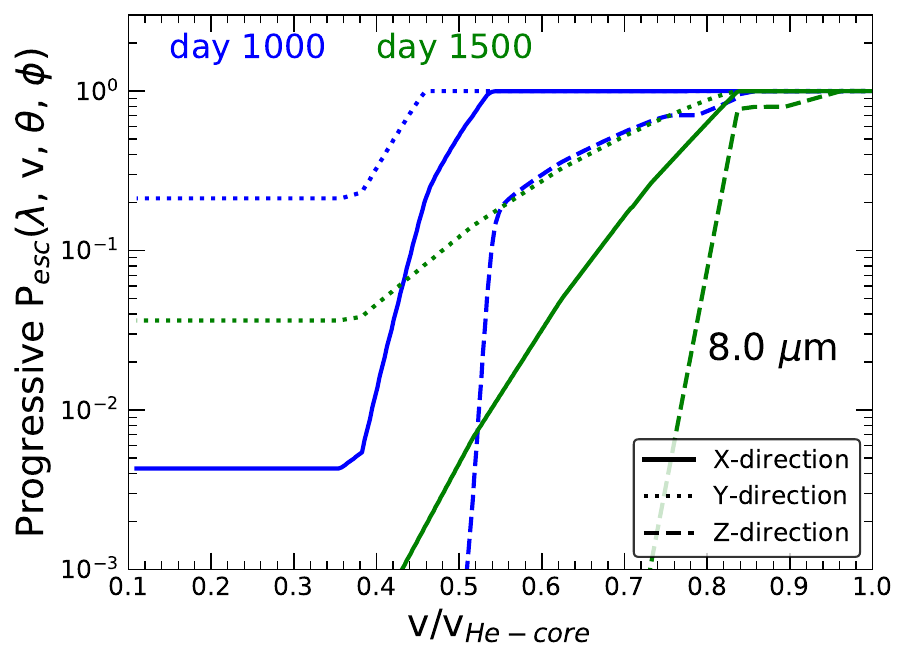}
\caption{\label{fig_Pesc_e} Progressive escape probabilities, corresponding to the anisotropic case, are shown for two epochs, days 1000 and 1500, for the 3.6- and 8.0-\mic\ \textit{Spitzer} IRAC bands. The other two IRAC bands, 4.5 and 5.8 \mic,\ also show a similar behavior. The escape probabilities are presented in the top panel as a function of ejecta velocities, while in the bottom panel, we show the probabilities as functions of the normalized He-core velocities (normalized with respect to the velocity of the outer edge of the He core in that direction). The escape probability in this scenario is dependent on the viewing angle; here we present the three axial directions. See Section \ref{sec_compare_spec} for more details.} 
\end{figure*}

\subsection{Comparing dust masses}
\label{sec_compare}

Here we describe the chemistry and dust masses formed in the ejecta. The same formalism, as described in Sect. \ref{sec_dustform} is used, where we trace the distribution and evolution of dust grains of silicates, amorphous carbon, alumina, and silicon carbide.

Along the smallest axis (which is the $z$ axis in our model) of the ellipsoidal ejecta, the clumps are denser and the gas is found to cool faster. Both these factors considerably enhance the rate and efficiency of dust production along that axis, which has the smallest velocities. Conversely, along the large axis, the gas is found to be hotter (compared to the other two directions), as well as less dense, leading to a delayed dust formation. 

The top panel of Fig. \ref{fig_dustmass_el} shows the evolution of dust masses in this scenario and its comparison with the isotropic model. The onset of dust formation is about 440 days, leading to the formation of a small amount of dust, typically less than 10$^{-5}$ \Ms. By 600 days the total mass of dust increases to 9 $\times$ 10$^{-5}$ \Ms, and then to $\sim$10$^{-3}$ \Ms\ by the end of 800 days post-explosion. After 1000 days, the total mass of dust in the ejecta is 5.5 $\times$ 10$^{-3}$ \Ms, and finally it ends up forming about 0.059 \Ms\ of dust.

The dust composition is dominated by silicates in the first $\sim$ 1400 days, after which alumina and silicates are found in almost equal proportion. The final mass of both these species are between 0.02 and 0.03 \Ms. Carbon dust appears as soon as 1200 days after the explosion, and continues to form until 2800 days. For silicon carbide, the formation starts after about 900 days and continues slowly over a long time, finally ending up at a mass close to 2 $\times$ 10$^{-4}$ \Ms. 

In comparison to the isotropic case, the evolution of dust masses differs in the following way: (a) the formation of dust starts at about 450 days post-explosion in the ellipsoidal ejecta, compared to 580 days in the spherical scenario; (b) the trend of an earlier synthesis of dust is valid for both the O-rich dust species in the ellipsoidal case; (c) the timescale for the onset of carbon dust production changes significantly from 2400 days in the isotropic model to 1200 days in the anisotropic scenario; (d) the final mass of carbon dust produced in the anisotropic ejecta is 5 $\times$ 10$^{-3}$ \Ms, compared to 8 $\times$ 10$^{-3}$ \Ms\ in case of the isotropic model, and so even though carbon dust forms earlier in the ellipsoidal scenario, the final mass is smaller; (e) the mass of silicon carbide in the ellipsoidal ejecta is larger than the mass in the spherical case by a factor of 1.5.

The bottom panel of Fig. \ref{fig_dustmass_el} compares the evolution of each dust species for the two cases. We find that the final dust masses in the two cases are almost identical. The epochs for the onset of formation and the initial rate of production vary to some extent for all the dust species. This is mainly induced by the geometry, where within the same ellipsoidal shell, some of the clumps see much earlier dust production than others. When this is aggregated over all the shells, the earliest and the latest epochs for dust formation in the ejecta get spread out over a larger span of time compared to an isotropic ejecta model. In Sect. \ref{sec_am_carbon}, we stressed the fact that the formation of amorphous carbon is highly sensitive to the physical conditions of the ejecta. In the same context, here we see that in the denser clumps, which are along the slowest axis ($z$ axis), carbon dust forms as soon as 1200 days post-explosion, while in the clumps along the fastest axis ($y$ axis), the formation of C dust does not take place even at the end of day 3000, when we terminate our simulations. This results in an overall smaller mass of C dust at 3000 days in the ellipsoidal scenario. 

Even though the masses of dust are not significantly different in the two cases, the distribution of dust inside the ejecta shows strong anisotropy. In Fig. \ref{fig_dustcolumn} the evolution of dust column densities (from the center to the edge of the He core) are presented for days 700, 1000, 1500, 2000, and 2900 along all the directions in each of the three axial planes. The figures clearly indicate that along the $xz$ or the $yz$ plane, the column densities increase rapidly with a decrease in $\theta$ ($\theta$ = 0 is the $z$ direction). The velocity $v_z$ being the slowest velocity, the dust is more concentrated in that direction, as previously explained. The contrast is more visible in the earlier epochs, because the onset of dust formation also varies widely along the planes. In the $yz$ plane, the variation in column density with $\theta$ is most prominent, since the velocities along the $y$ and $z$ axes are the largest and smallest, respectively. The $xy$ plane has relatively lower columns of dust and also later epochs of formation, compared to the other two.

\begin{figure*}
\vspace*{0.3cm}
\centering
\includegraphics[width=2.33in]{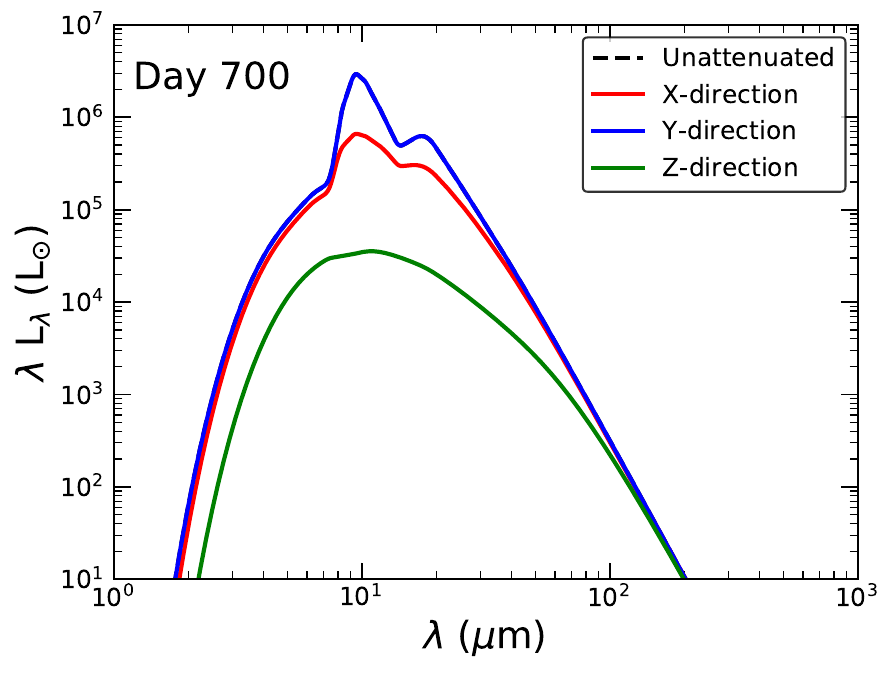}
\includegraphics[width=2.33in]{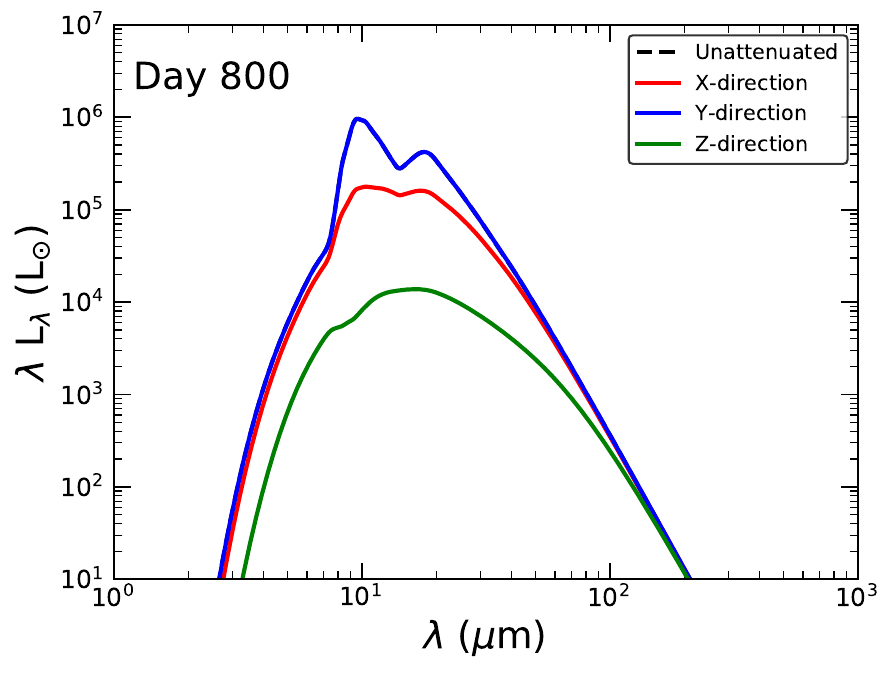}
\includegraphics[width=2.33in]{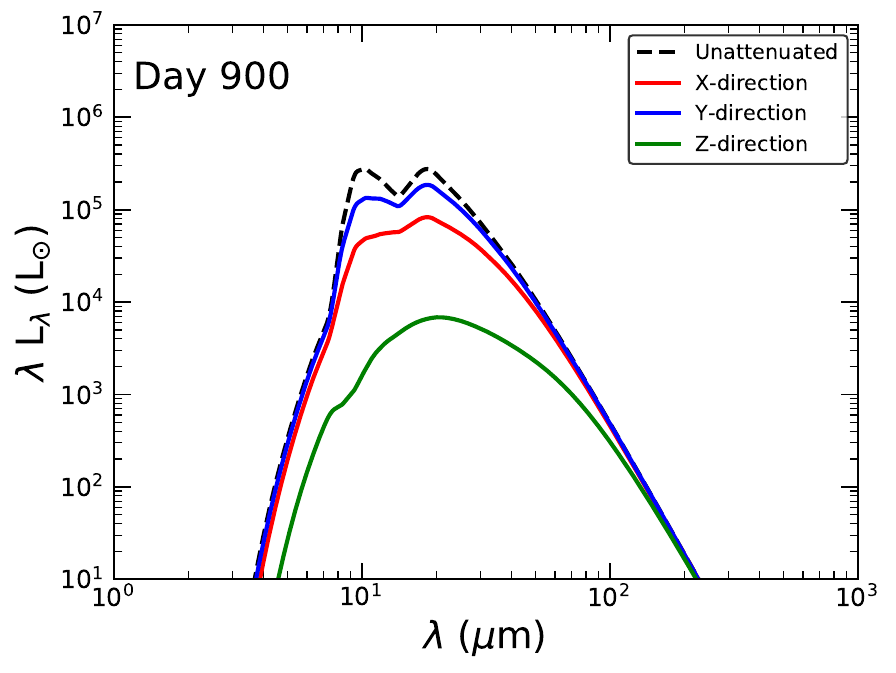}
\includegraphics[width=2.33in]{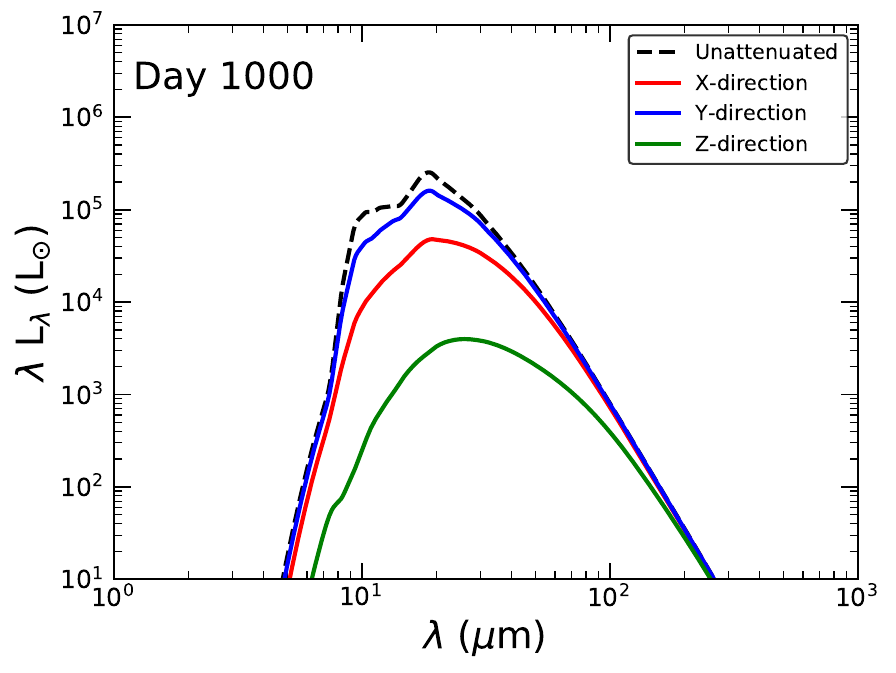}
\includegraphics[width=2.33in]{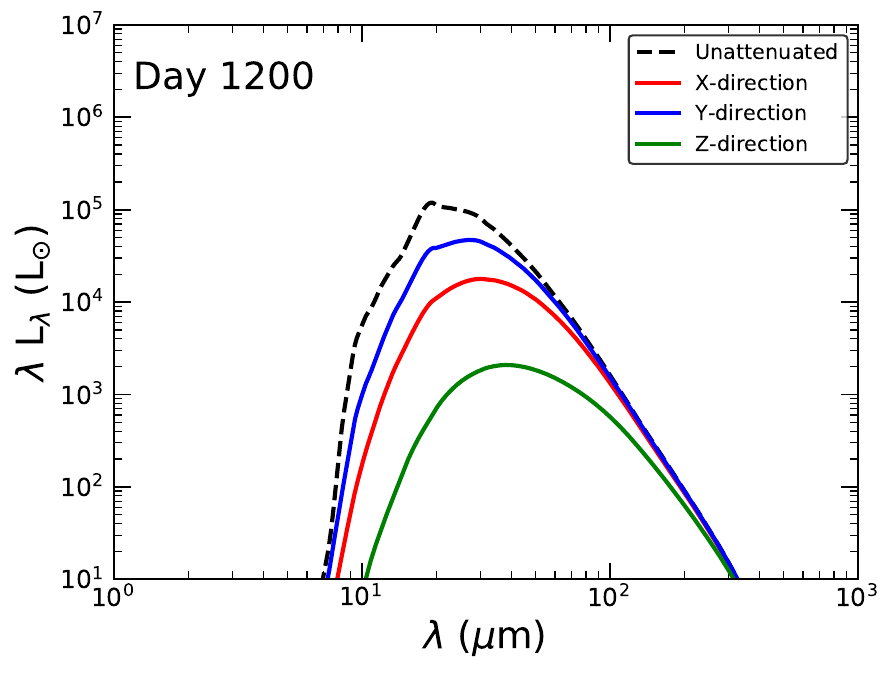}
\includegraphics[width=2.33in]{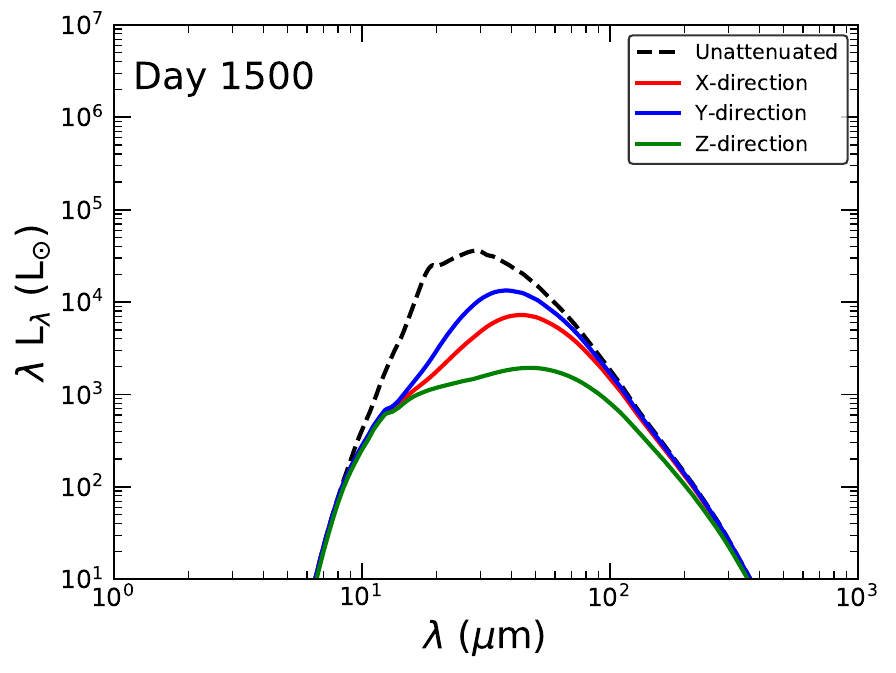}
\includegraphics[width=2.33in]{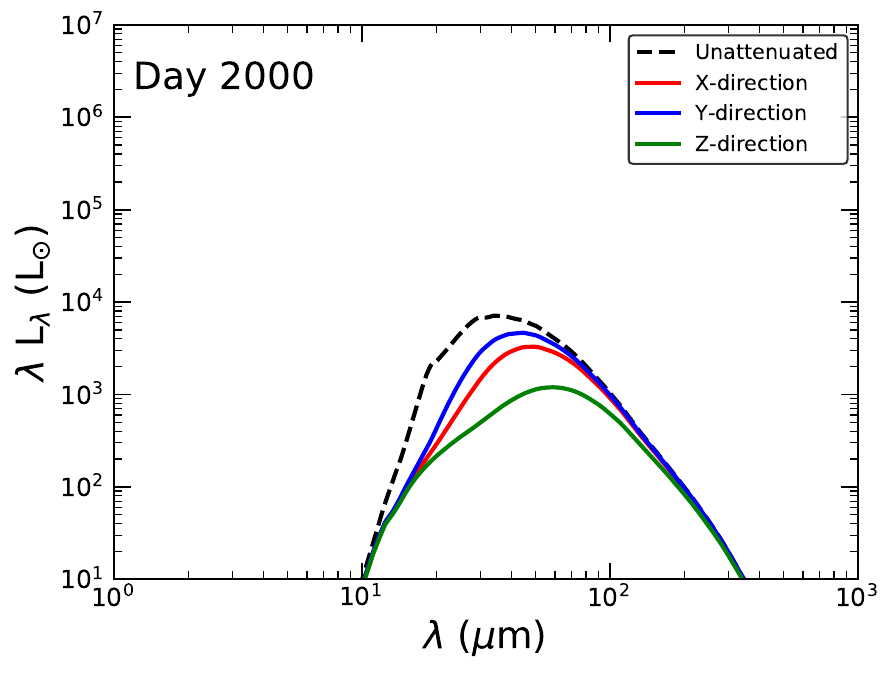}
\includegraphics[width=2.33in]{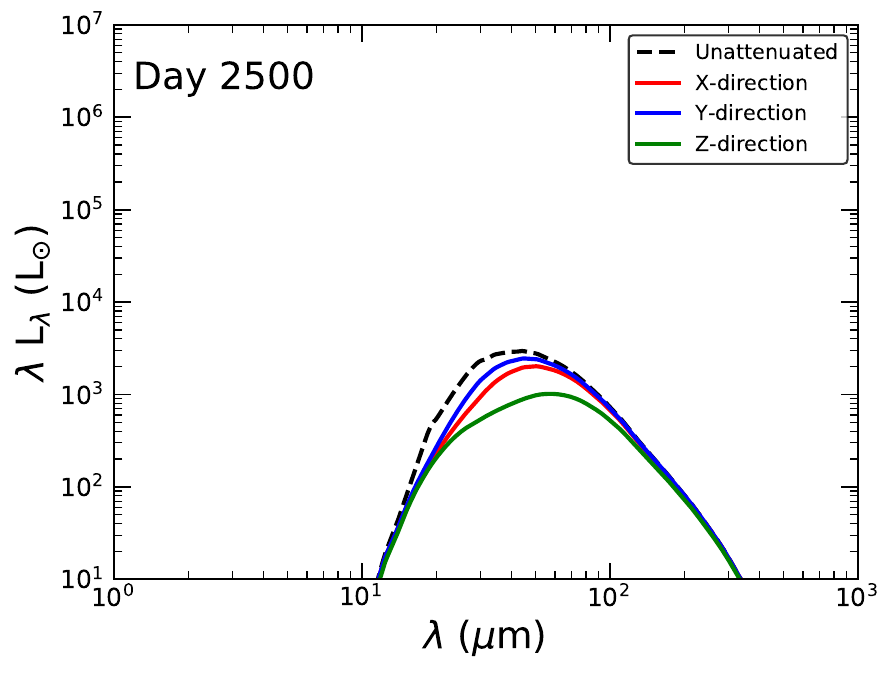}
\includegraphics[width=2.33in]{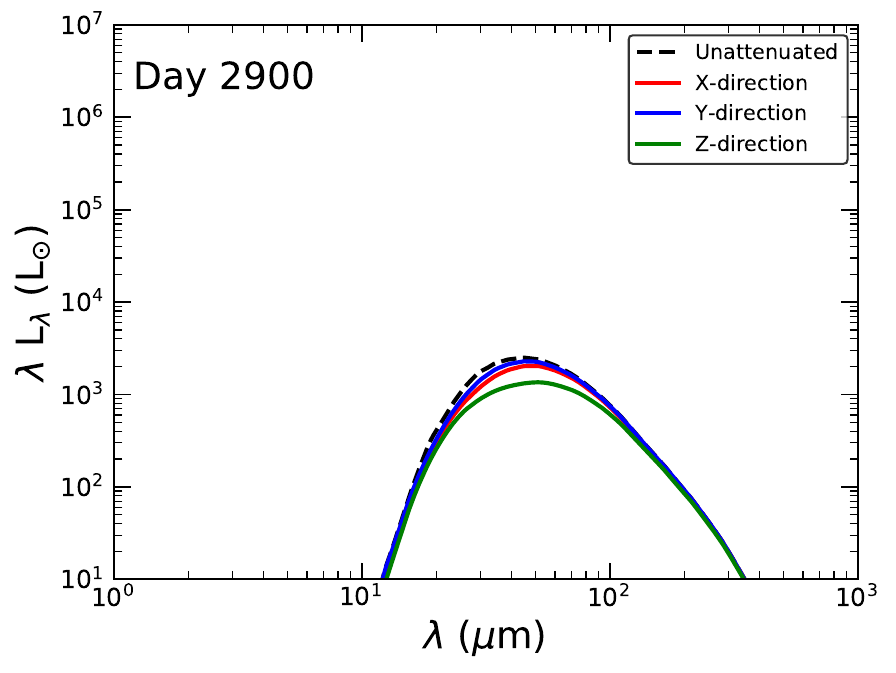}
\caption{\label{fig_lum_e} Unattenuated luminosities, as well the observable luminosities, along the three axial planes are shown for nine different epochs. For reference, in our chosen geometry, the $z$ axis is the smallest axis and the $y$ axis is the largest axis of the ellipsoid. See Section \ref{sec_compare_spec} for the description.  }
\end{figure*}

\begin{figure*}
\vspace*{0.3cm}
\centering
\includegraphics[width=3.5in]{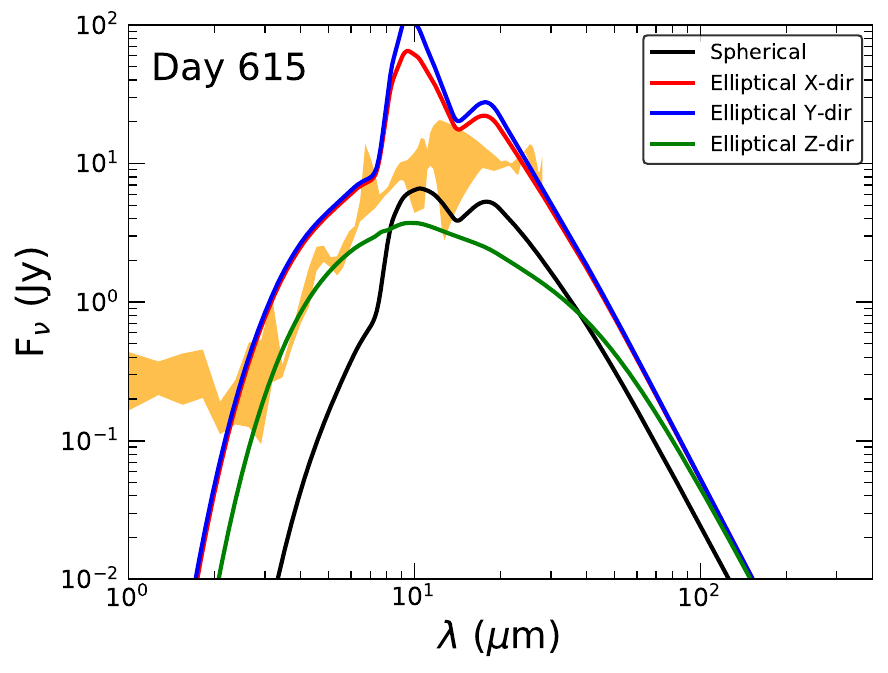}
\includegraphics[width=3.5in]{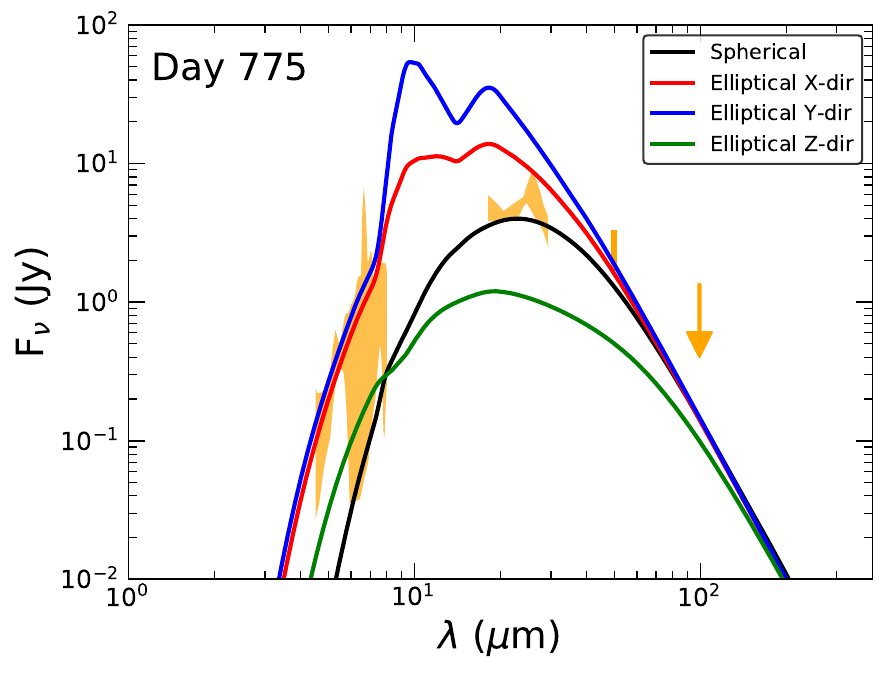}
\includegraphics[width=3.5in]{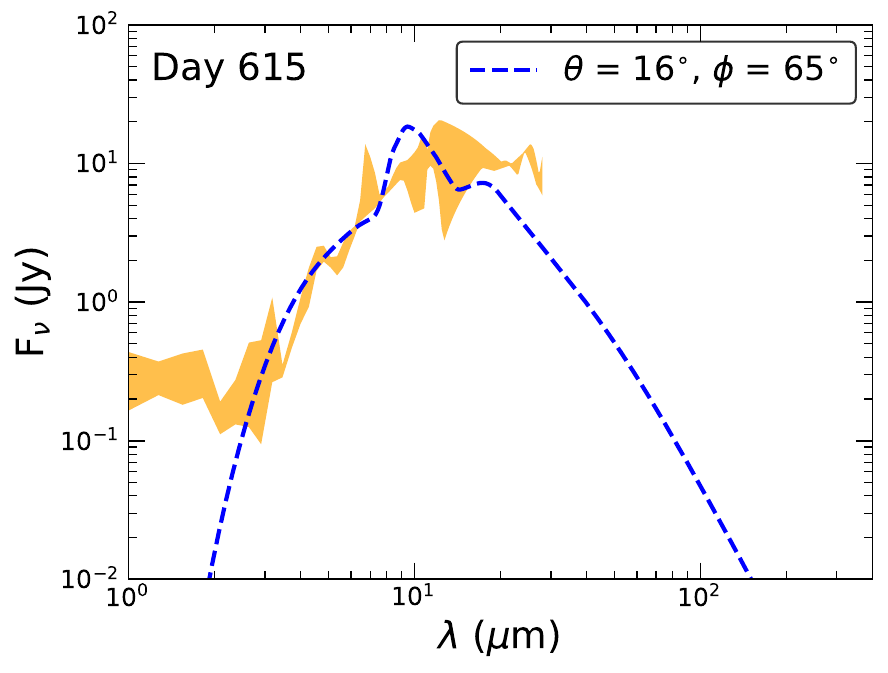}
\includegraphics[width=3.5in]{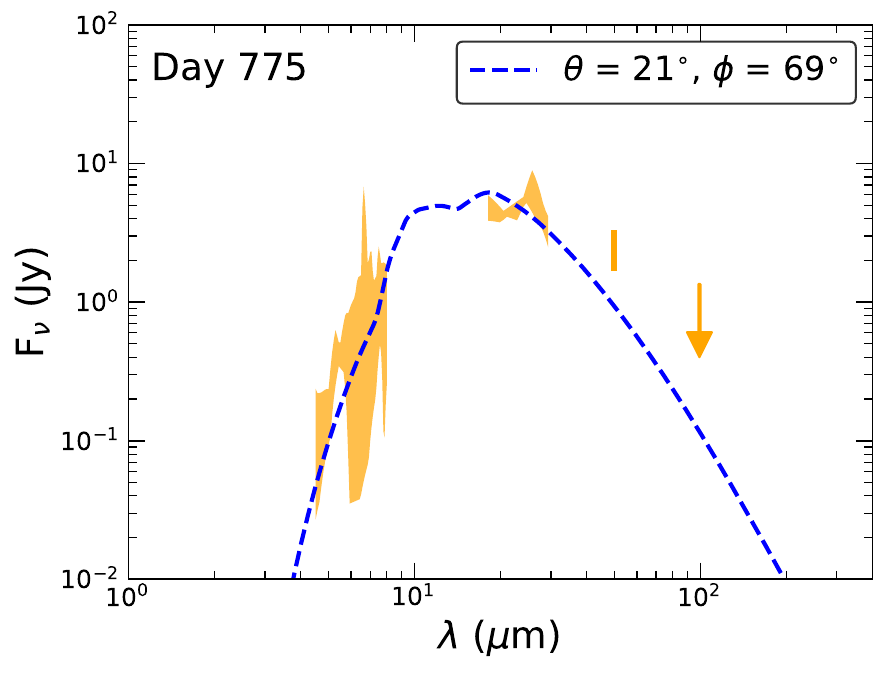}
\caption{\label{fig_87A} Fluxes obtained from our calculations (assuming the SN to be at the distance of SN~1987A) for days 615 and 775, are co-plotted with the near- and mid-IR data of SN~1987A (shaded region in yellow) from \cite{wesson2015}. In the top panel, the fluxes for the isotropic model, as well as the three axial directions of the anisotropic model, are shown for the two epochs. In the bottom panel, we show the viewing angle with respect to the coordinates shown in Fig. \ref{fig_dustcolumn}, where we find the fluxes emerging from the anisotropic ejecta match well with the observations. For the details, see Sect. \ref{sec_sn1987a}.}
\end{figure*}

\subsection{Anisotropic IR spectra}
\label{sec_compare_spec}

Based on the resulting distribution of dust, we calculated the opacities and progressive escape probabilities of the medium along various directions. Using Eq. \ref{eqn_pesc_full}, the $P_{esc} (\lambda, v_x,\theta, \phi)$ can be estimated, which tells how many of the IR photons of wavelength $\lambda$, originating in a thin ellipsoidal shell (defined by velocity along the $x$ axis, $v_x$), escape the ejecta to reach the observer who is in the ($\theta$, $\phi$) direction. 
When calculating the optical depth $\tau_h (\lambda, v_x, \theta, \phi)$ using Eq. \ref{eqn_shelltau}, the total mass of dust cannot be used in the expression, since the distribution of dust is anisotropic. Instead, the dust densities in a given direction should be integrated over all the shells along its path, as it is also shown in Eq. \ref{eqn_shelltau}. 

In Fig.\ref{fig_Pesc_e}, the progressive escape probability along the three axes of the ellipsoidal ejecta are shown for days 1000 and 1500, for the \textit{Spitzer Space Telescope} IRAC bands $\lambda$ = 3.6, 8.0 \mic. 
To summarize the trend, initially dust forms in the clumps centered around the $z$ axis. However, the emission from those dust grains cannot easily escape in the $z$ direction since the opacities become large very quickly. On the contrary, the same radiation can easily pass through the He core in the $x$ or $y$ direction, since the clumps along the path in those directions are not very dusty in the earlier epochs. With time and dust formation, the opacities along other directions also start to increase, although this trend remains true for later epochs as well, since the dust is always more concentrated toward the smaller-velocity axis. Moreover, along the faster axes, even if dust forms with equal efficiency, as the densities are lower and the clumps are more spread out, the opacities cannot get much larger. All these factors collectively result in the variation in $P_{esc}$ with velocities along different directions, which is seen in Fig. \ref{fig_Pesc_e}. For the spherical ejecta, the outer-edge of the He core is at 2650 km s$^{-1}$, which is the same for the $x$ direction in this scenario, while the velocities are half of that and 1.5 times  that along the $z$ and $y$ directions, respectively. In Fig. \ref{fig_Pesc_e} we also show the escape probabilities as a function of normalized He-core velocities, which actually reflect the efficiency of dust production along the three axes of the ellipsoidal ejecta, normalized to the same scale. 

The observable luminosity, given by Eq.\ref{eqn_lum}, is again a function of the direction ($\theta, \phi$), in correlation with the directional escape probabilities. However, the unattenuated luminosity has no direction dependence since it is just the total amount of energy radiated by all the dust grains together, per unit time. In Fig. \ref{fig_lum_e}, we show the observable luminosities along the three axes, compared to the unattenuated luminosities, for nine epochs. 
In qualitative terms, the variation in luminosity with wavelength, which is explained in Sect. \ref{sec_opacities}, is also applicable to the anisotropic ejecta. 

Looking at Fig. \ref{fig_lum_e}, it is evident that the observable luminosity in the $z$ direction is always much lower than in the other two axes, given the large opacities in that direction. Moreover, there are no particular spectral features visible along that axis. For days 700 and 800, the observable luminosity along the $y$ direction coincides with the unattenuated luminosity, which implies that the ejecta is optically thin along the $y$ axis, since the clumps aligned along that axis are yet to form dust. The 9.7- and 18-\mic\ spectral features of silicate dust \citep{simpson_1991} are therefore visible in the spectra along the $xy$ plane until about day 900, when the distribution of dust is very skewed toward the smallest velocity axis only. 

With time, dust forms in all the clumps, with varied efficiency. The expanding ejecta also become optically thinner. Typically, after 2400 days, we find the observable luminosity in all the directions to be within comparable range. After 2800 days, the ratio between the unattenuated and the observable luminosities in all the directions approaches unity.

\subsection{SN~1987A at days 615 and 775}
\label{sec_sn1987a}

The spectra of SN~1987A at post-explosion epochs 615 and 775 are presented by \cite{woo93}. Using these data, the masses and the composition of dust formed in the ejecta are constrained \citep{erc07, wesson2015, dwe15, wesson_2021}. 

In this paper, we study the formation and distribution of dust, and their temperatures, in a bottom-up approach. In Fig. \ref{fig_87A}, we plot the IR fluxes from our model alongside the observed fluxes of SN~1987A taken from \cite{wesson2015}. For days 615 and 775, the predicted fluxes for the isotropic model and the three axial directions of the anisotropic model are presented. It is important to note that these figures are not parameter fits, but instead just the co-plotting of the data and the predictions of our model. 
Using our results, we only show the contribution of dust to the fluxes in the IR in Fig. \ref{fig_87A}. The UV-optical fluxes needs to be added to it in order to obtain the complete SED. For instance, at day 615 the observed fluxes at wavelengths smaller than 2 \mic, shown in the figure, can only be accounted when we consider the contribution of the optical luminosities.

The total amount of dust formed in our calculations at days 615 and 775 for the isotropic model are 1.7$\times$10$^{-4}$ \Ms\ and 1.5$\times$10$^{-3}$ \Ms,\ respectively; for the anisotropic case, these are are 2.1$\times$10$^{-4}$ \Ms\ and 9.7$\times$10$^{-4}$ \Ms. A distance of 51.4 kpc was assumed for SN~1987A from the Earth. 

When the ejecta are assumed to be anisotropic, we find that, between days 600 and 800, newly formed dust is mainly confined along the clumps that are aligned close to the $z$ direction. Therefore, the opacities are much enhanced along the $z$ direction at these times, which results in a reduced flux escaping along that direction. On the contrary, when the viewing angle is along the $x$ or $y$ direction, the dusty clumps are aligned in a perpendicular cross-section to the observer, and thus the fluxes from all those clumps reach the observer almost unattenuated. This is visible in the top panel of Fig. \ref{fig_87A}, where along the $x$ or $y$ direction, the predicted fluxes are larger compared to the observations, while the fluxes along the $z$ direction, and also the flux produced by the isotropic ejecta, are smaller than the observed fluxes.


An ellipsoidal ejecta with a tilt of 25$^{\circ}$ to the plane of the sky is best fit to the observations of SN~1987A \citep{kjaer_2010}. Using the polar coordinates given in Fig. \ref{fig_dustcolumn}, we find that for day 615, the fluxes obtained in our calculations are closest to the observations at a viewing angle of $\theta$ = 16$^\circ$, $\phi$ = 65$^\circ$. For day 775, these are at $\theta$ = 21$^\circ$, $\phi$ = 69$^\circ$. Both the viewing angles being close to each other, in a way, justifies their credibility. 

\section{Summary}
\label{sec_summary}

In this study, we have revisited the scenario of dust production in SN ejecta, based on an improved understanding of the ejecta temperature, and their cooling, along with the nature of clumpiness within the ejecta cores. The evolution and distribution of dust in the He core was derived, along with the respective opacities. The emerging IR emission from the dust was modeled, based on the generic, nonuniform distribution of the dust grains that we estimated together with their respective temperature distributions and progressive opacities. For the first time, we show the impact of anisotropy on the chemistry of dust formation by choosing an ellipsoidal geometry for the SN ejecta that imitates SN~1987A. 

In a few points below, we summarize the results of this study: 

\begin{enumerate}[label=(\arabic*), noitemsep, leftmargin=*, align=left, wide = 0pt]
\item The composition of dust in SN ejecta is dominated by O-rich components (silicates and alumina) at all times. For this study, we considered the progenitor star to be 20 \Ms\ at main sequence. \cite{sar13} found that the relative mass of O-rich dust increases with an increase in progenitor mass. 

\item The formation of dust commences at around 580 days post-explosion, if the ejecta are assumed to be isotropic, while it starts at day 450 when an anisotropic, ellipsoidal geometry is adopted for the ejecta. 

\item The rate of dust production is relatively rapid in the period from year 2 to year 5, after the explosion of the SN, typically averaging about 6 $\times$ 10$^{-5}$ \Ms\ day$^{-1}$; following that, the rates decline significantly.  

\item The formation of carbon dust is especially sensitive to the physical conditions of the ejecta. In our model, we find that in the isotropic model, carbon dust forms almost 6.5 years after the explosion. The clumps that form carbon dust are also rich in C$^{18}$O molecules, whose formation precedes the synthesis of the dust grains. 

\item In the spherical ejecta, the mass of dust evolves from less than 10$^{-5}$ \Ms\ at 600 days to about 10$^{-3}$ \Ms\ at 800 days, to $\sim$ 0.06 at 1600 days. In the ellipsoidal scenario, the formation of dust starts earlier. However, post day 800, the total mass of dust present in the ejecta at any given time are comparable to the spherical model, with a variation of no more than 5\%.

\item In this study, we have derived the distribution of dust in the velocity space. For the isotropic scenario, we find that the dust is initially confined in the inner regions only, where the velocities are about 1000 km s$^{-1}$. With time, the distribution spreads toward the outer layers. One exception to this trend is the region where CO molecules form in abundance and induce extra cooling, resulting in early synthesis of dust. In the outermost layer, where the velocities are larger than 2300 km s$^{-1}$, dust forms as late as 2400 days post-explosion. 

\item Even though the total mass of dust in the ellipsoidal ejecta is comparable to the spherical model, the distribution of dust is highly anisotropic in nature. We find that the clumps that are aligned close to the smallest axis of the ellipsoid are more suitable for dust production, both in terms of timescales and efficiency. This results in a larger concentration of dust in the slower clumps at a given time, if the clumps with identical abundances are compared. 

\item To account for the nonuniform distribution of dust and its temperature, we derived the opacities as a function of velocity, thereby defining the progressive escape probability of a photon originating from a clump at velocity $v$ to the outer-edge of the ejecta. The opacities were found to increase due to dust formation until about day 1600, followed by a gradual decline.  

\item Even though the dust is always O-rich, we show that for the isotropic ejecta, the spectral features of silicates or alumina, which are between 10-18 \mic, will not be present in the observed SED after 700 days. This is simply because when there is a local maxima in the emission, the opacities are also large in that wavelength. Moreover, the emission from the inner part of the ejecta is increasingly blocked, as the dust formation progresses outward with time. The effective temperatures of the observable fluxes, estimated from our calculations, are already too low after day 1000 to have strong emission around the 10-\mic\ band. 

\item In the anisotropic ellipsoidal geometry, the polar distribution of dust results in relatively much lower opacities along the faster velocity-axes; in other words, dust forms in clumps, which are in a plane perpendicular to the line of sight, when observed along the large axis of the ellipsoid. Therefore, the predicted observable luminosities are found to be more than an order of magnitude larger than the luminosity along the small axis. 

\item By comparing the fluxes from SN~1987A at days 615 and 775 with our predicted SEDs, we estimate the most likely alignment of the SN ejecta with our line of sight.

\end{enumerate} 
\section{Discussion}
\label{sec_discussion}

In \cite{sar13, sar15}, the gas densities derived from \cite{noz10} were used to construct a homogeneous density model for the ejecta, and a clumpy model was based on a density enhancement to that, using the filling factors given by \cite{jerkstrand2011}. In this paper, we adopted a revised density structure for the ejecta, which is consistent with the dynamics of SN ejecta \citep{tru99}, the nature of clumpiness used by the radiative transfer models of the nebular phase \citep{dessart_2018}, and which are in good agreement with the multidimensional explosion models of SNe \citep{wong15, utrobin_2017, gabler_2021}. We find that the densities derived from \cite{noz10} are actually more consistent with clumpy ejecta. Hence, a density enhancement over that leads to an artificial increase in the densities, thereby suggesting a very rapid formation of dust in the first couple of years in the clumpy model of \cite{sar15}, which saturates at a total budget of 0.13 \Ms. In this paper, our results show that only a small mass of dust is formed in the first couple of years in the clumpy ejecta. Dust formation continues until day 2800, though the first 1600 days mark the most active phase. In addition, compared to \cite{sar13, sar15}, in this paper we have assumed a more realistic evolution of gas and dust temperatures, which are characterized by limited cooling in the first couple of years. The cooling only escalates between year two and year three from the time of explosion, depending on the zone. This results in a delay in the synthesis of dust, thereby lowering densities at the epochs of dust formation and reducing efficiencies compared to \cite{sar15}. However, to be clear, the revised estimations of the gas densities in the clumps are primarily responsible for the lower dust yields in this study (0.06 \Ms), compared to the larger, rapidly growing dust masses predicted by the clumpy model of \cite{sar15}. 

\cite{sluder2018} predict the synthesis of both O-rich and C-rich dust grains in abundance within the first year of explosion; the dust masses are as large as $\sim$ 0.4 \Ms\ within a 1000 days. In this paper, we predict much lower dust masses in the first 1000 days, which does not exceed 0.01 \Ms\  and is almost entirely made up of O-rich dust. We find that dust formation starts in the SN ejecta between days 450 and 550, which is in reasonable agreement with the rapid decrease in the line intensities of Mg or Si about the same time \citep{dan91}, implying the depletion of atoms in the gas phase due to dust formation. The model of \cite{sluder2018} suggests that the rapid increase in dust masses in the first two years from the time of explosion is strongly aided by the accretion of metals \citep{hirashita_2011} on the surface of the small, newly formed grains. In line with \cite{sar13,sar15}, we continue to argue that the surface reactions are likely to be important only at much longer timescales, and when the gas is much cooler. 
Our knowledge of accretion on the grain surfaces is largely based on ISM-like environments \citep{hirashita_2011, hirashita_2012} where chemical equilibrium can be assumed, and thus its application to the much hotter and rapidly evolving SN ejecta is questionable. The rate of sticking on the surface of a grain is expected to considerably decline at high temperatures \citep{blum_2004}. 
Moreover, following the treatment of ISM dust by \cite{guhathakurta_1989}, it can be suggested that the smallest grains are much more prone to lose atoms from its surface at higher temperatures, which are analogous to the time of their genesis from the gas phase. Most importantly, the detection of molecules \citep{lep90, kot09, matsuura2017} and atoms \citep{laming_2020} present in the gas also speaks against rapid growth of dust via accretion of all the residual gas-phase species on the surface of the grains. 

We find that the dust masses in the SN ejecta do not exceed 0.06 \Ms. This is almost an order of magnitude lower than the masses reported to be present in the SNRs when probed in the far-IR and submillimeter wavelengths \citep{gom12, ind14, matsuura_2015}. We do not rule out the possibility of accretion when the gas is cool enough after about five years from explosion. However, the grain growth via cold accretion is expected to have loosely bound surfaces that are easily susceptible to destruction by the reverse shock \citep{dwe15}. The dust destruction efficiency by the reverse shock in Cas A is estimated to be between 70-90\%  \citep{slavin_2020, priestley_2021}, which could be correlated to the amount of dust formed through nucleation compared to the dust that grows through accretion at late times. Another important factor could be a significant cooling of the gas due to dust formation, which might have a snowball effect on the dust production. The likelihood of these scenarios could only be determined through further analysis. 

In the last few years, most papers dealing with dust in SNe have taken it as a general statement that dust formation models always predict a lot of dust to form very early after explosion, which contradicts the observations. 
We would like to clarify here that the actual contradiction between theoretical and observational predictions is that theoretical models do not support a scenario of continuous dust production over decades. In the first 1000 days, the most relevant phase for mid-IR observations of any SN ejecta, the mass of dust produced is not necessarily very large, as shown in this paper. Given the compact nature of the ejecta at these early times, even a small mass of dust is sufficient to make the opacities large, as shown in this paper, thereby suppressing the intensity of the emerging flux as explained by \cite{dwek_2019}. The generic distribution of dust, nonuniform in nature, is also expected to impact the asymmetries in the emission lines \citep{bevan2016} differently, compared to a uniform scenario. In the future, we plan to test that for the isotropic and anisotropic cases. 

In line with \cite{dwe15}, we show in this paper that even if the ejecta dust is largely O-rich (silicates, alumina), the relatively large opacities efficiently suppress the strong silicate features (between 10-20 \mic) of the mid-IR wavelengths. In addition, after about 1200 days, the ejecta dust cools down further so that the emission in that wavelength range is not significant. Therefore, the SED itself cannot be conclusively used to assign the composition of dust in the ejecta. In this study, we stress the importance of the generic, nonuniform distribution of dust, and also the dependence on the viewing angle when estimating the emerging fluxes.

The carbon-rich dust components, silicon carbide and amorphous carbon, are found to form must later than the O-rich ones. amorphous carbon dust appears in the ejecta after about 6-7 years from the time of explosion in the isotropic model. In the anisotropic case, it starts forming after 4 years from explosion, but only in small amounts. The synthesis of SiC grains takes place in two distinctly separate regions of the ejecta: the outermost and the innermost layers of the He core. In the outermost layer, the formation timescale is comparable to that of amorphous carbon, and in the innermost part of the core, a small mass of SiC forms after about day 1000. This result is especially relevant to account for the estimated delay in the formation of the SiC grains of SN origin found in the presolar grains in meteorites \citep{liu_2018, ott_2019}, and also to explain the isotopic ratios of elements found in the presolar grains through mixing between the layers of the SN ejecta \citep{xu_2015}. 

The chemical kinetic models of dust formation in SNe argue that the production of dust grains starting from gas phase reactions is unlikely to take place at very late times, when the gas temperatures in the SN ejecta are low, since several gas phase reactions require sufficient energy to overcome activation barriers \citep{sarangi2018book}. In the network leading to the formation of amorphous carbon dust, the formation of C$_2$ holds the bottleneck; we find that the formation of C$_2$ molecules takes place after about 1000 days post-explosion, once the He$^{+}$ recombines and importantly when the gas is still warm. However, as we explain in Sect. \ref{sec_am_carbon}, due to a very slow rate of CO production, the condensation of the C$_2$ molecules into C chains does not transpire until later than day 2400. However, since C$_2$ was already formed when the gas was still warm, the C chains could condense even at low temperatures to form clusters and eventually carbon dust at such late times. As an additional note, we also want to point out that the conversion of C chains to amorphous carbon dust is still quite swift, once it starts; therefore, this late formation of C dust itself cannot be taken as evidence of the gradual growth of dust over a long span of time. Also, given that the more abundant O-rich dust grains are formed much earlier in time, the late formation of C dust does not provide any significant boost to the overall dust production rate at late times. However, as we mentioned earlier in the text, in the future, we wish to better understand the possibility of accretion of molecules and residual atoms on the surface of these dust grains in cold ejecta, after about a decade from the explosion, to verify if that can amplify the dust formation rates at late times.

In this study, we find the grain sizes to vary between 0.001 to 0.1 \mic. When estimating the fluxes, we did not use a distribution of grain sizes, but instead used a fixed value of 0.01 \mic, since these small grains are within the Rayleigh limit (typically when $a$ is smaller than 0.1 \mic), and we expect only minor differences for the change in grain sizes. In the future, we shall reconsider the possibility of cold accretion and derive the modified grain size distribution taking that into account.

In this paper, we did not test the dependence of net dust production on metallicities, while we acknowledge its importance. The post-explosion elemental and isotopic yields are correlated to the explosion energy and amount of \Ni\ produced at the explosion. Our current understanding of this correlation (\Ni\ mass to explosion energy to element abundances) is largely incomplete. In the future, when we study the effect of metallicity, we want to simultaneously address this issue.

By analyzing a large variety of isotropic and anisotropic clumps in the velocity space, this study is set as a foundation stone to address the chemistry of 3D ejecta. In this approach, we did not account for any mixing between layers. This model in the future can be coupled to the shuffled-shell technique to imitate macroscopic mixing \citep{dessart_2021}. From our understanding of ejecta chemistry, we can predict that even though the mixing may alter the relative timescale for the formation of certain dust species, we do not anticipate a large change in the final dust masses or the timespan for dust production in the ejecta. 
 \\

 \begin{acknowledgements}
We are deeply indebted to Dr. Isabelle Cherchneff, for the development of the SN ejecta chemistry together with Arka Sarangi for previous studies, which is now applied in this paper. \\

We also acknowledge the very useful inputs from Dr. Anders Jerkstrand and Dr. Eli Dwek during the course of this research.  \\

Finally, we like to thank the anonymous referee and the editor of this article for his/her precious time.  \\

This work was supported by a grant from VILLUM FONDEN, PI Jens Hjorth (project number 16599).
\end{acknowledgements}


\bibliographystyle{aa.bst}
\bibliography{Bibliography_sarangi}

\begin{thebibliography}{90}
\expandafter\ifx\csname natexlab\endcsname\relax\def\natexlab#1{#1}\fi

\bibitem[{Arendt {et~al.}(2014)Arendt, Dwek, Kober, Rho, \&
  Hwang}]{arendt_2014}
Arendt, R.~G., Dwek, E., Kober, G., Rho, J., \& Hwang, U. 2014, The
  Astrophysical Journal, 786, 55

\bibitem[{{Barlow} {et~al.}(2010){Barlow}, {Krause}, {Swinyard}, {Sibthorpe},
  {Besel}, {Wesson}, {Ivison}, {Dunne}, {Gear}, {Gomez}, {Hargrave}, {Henning},
  {Leeks}, {Lim}, {Olofsson}, \& {Polehampton}}]{barlow_2010}
{Barlow}, M.~J., {Krause}, O., {Swinyard}, B.~M., {et~al.} 2010, \aap, 518,
  L138

\bibitem[{{Bevan}(2018)}]{bevan_2018}
{Bevan}, A. 2018, \mnras, 480, 4659

\bibitem[{{Bevan} \& {Barlow}(2016)}]{bevan2016}
{Bevan}, A. \& {Barlow}, M.~J. 2016, \mnras, 456, 1269

\bibitem[{{Blum}(2004)}]{blum_2004}
{Blum}, J. 2004, in Astronomical Society of the Pacific Conference Series, Vol.
  309, Astrophysics of Dust, ed. A.~N. {Witt}, G.~C. {Clayton}, \& B.~T.
  {Draine}, 369

\bibitem[{{Bouchet} \& {Danziger}(1993)}]{bou93}
{Bouchet}, P. \& {Danziger}, I.~J. 1993, Astronomy and Astrophysics, 273, 451

\bibitem[{{Burke} \& {Hollenbach}(1983)}]{burke_1983}
{Burke}, J.~R. \& {Hollenbach}, D.~J. 1983, \apj, 265, 223

\bibitem[{{Cherchneff} \& {Dwek}(2009)}]{cherchneff2009}
{Cherchneff}, I. \& {Dwek}, E. 2009, The Astrophysical Journal, 703, 642

\bibitem[{{Cherchneff} \& {Dwek}(2010)}]{cherchneff2010}
{Cherchneff}, I. \& {Dwek}, E. 2010, The Astrophysical Journal, 713, 1

\bibitem[{{Cherchneff} \& {Lilly}(2008)}]{cherchneff2008}
{Cherchneff}, I. \& {Lilly}, S. 2008, The Astrophysical Journal, 683, L123

\bibitem[{{Danziger} {et~al.}(1991){Danziger}, {Lucy}, {Bouchet}, \&
  {Gouiffes}}]{dan91}
{Danziger}, I.~J., {Lucy}, L.~B., {Bouchet}, P., \& {Gouiffes}, C. 1991, in
  Supernovae, ed. S.~E. {Woosley}, 69

\bibitem[{{De Looze} {et~al.}(2019){De Looze}, {Barlow}, {Bandiera}, {Bevan},
  {Bietenholz}, {Chawner}, {Gomez}, {Matsuura}, {Priestley}, \&
  {Wesson}}]{delooze_2019}
{De Looze}, I., {Barlow}, M.~J., {Bandiera}, R., {et~al.} 2019, \mnras, 488,
  164

\bibitem[{{De Looze} {et~al.}(2017){De Looze}, {Barlow}, {Swinyard}, {Rho},
  {Gomez}, {Matsuura}, \& {Wesson}}]{delooze2017}
{De Looze}, I., {Barlow}, M.~J., {Swinyard}, B.~M., {et~al.} 2017, \mnras, 465,
  3309

\bibitem[{{Dessart} \& {Hillier}(2020)}]{dessart_2020}
{Dessart}, L. \& {Hillier}, D.~J. 2020, \aap, 642, A33

\bibitem[{{Dessart} {et~al.}(2021){Dessart}, {Hillier}, {Sukhbold}, {Woosley},
  \& {Janka}}]{dessart_2021}
{Dessart}, L., {Hillier}, D.~J., {Sukhbold}, T., {Woosley}, S.~E., \& {Janka},
  H.~T. 2021, \aap, 652, A64

\bibitem[{{Dessart} {et~al.}(2018){Dessart}, {Hillier}, \&
  {Wilk}}]{dessart_2018}
{Dessart}, L., {Hillier}, D.~J., \& {Wilk}, K.~D. 2018, \aap, 619, A30

\bibitem[{{Donn} \& {Nuth}(1985)}]{don85}
{Donn}, B. \& {Nuth}, J.~A. 1985, Astrophysical Journal, 288, 187

\bibitem[{{Dwarkadas}(2007)}]{dwa07}
{Dwarkadas}, V.~V. 2007, in American Institute of Physics Conference Series,
  Vol. 937, Supernova 1987A: 20 Years After: Supernovae and Gamma-Ray Bursters,
  ed. S.~{Immler}, K.~{Weiler}, \& R.~{McCray}, 120--124

\bibitem[{{Dwek}(1987)}]{dwe87}
{Dwek}, E. 1987, \apj, 322, 812

\bibitem[{Dwek(2006)}]{dwek_2006}
Dwek, E. 2006, Science, 313, 178

\bibitem[{{Dwek} \& {Arendt}(2015)}]{dwe15}
{Dwek}, E. \& {Arendt}, R.~G. 2015, \apj, 810, 75

\bibitem[{{Dwek} {et~al.}(2007){Dwek}, {Galliano}, \& {Jones}}]{dwe07}
{Dwek}, E., {Galliano}, F., \& {Jones}, A.~P. 2007, The Astrophysical Journal,
  662, 927

\bibitem[{{Dwek} {et~al.}(2019){Dwek}, {Sarangi}, \& {Arendt}}]{dwek_2019}
{Dwek}, E., {Sarangi}, A., \& {Arendt}, R.~G. 2019, \apjl, 871, L33

\bibitem[{{Ensman} \& {Burrows}(1992)}]{ens92}
{Ensman}, L. \& {Burrows}, A. 1992, \apj, 393, 742

\bibitem[{{Ercolano} {et~al.}(2007){Ercolano}, {Barlow}, \& {Sugerman}}]{erc07}
{Ercolano}, B., {Barlow}, M.~J., \& {Sugerman}, B.~E.~K. 2007, Monthly Notice
  of the Royal Astronomical Society, 375, 753

\bibitem[{{Gabler} {et~al.}(2021){Gabler}, {Wongwathanarat}, \&
  {Janka}}]{gabler_2021}
{Gabler}, M., {Wongwathanarat}, A., \& {Janka}, H.-T. 2021, \mnras, 502, 3264

\bibitem[{{Gall} \& {Hjorth}(2018)}]{gall_2018}
{Gall}, C. \& {Hjorth}, J. 2018, \apj, 868, 62

\bibitem[{{Gall} {et~al.}(2011){Gall}, {Hjorth}, \& {Andersen}}]{gal11}
{Gall}, C., {Hjorth}, J., \& {Andersen}, A.~C. 2011, The Astronomy and
  Astrophysics Review, 19, 43

\bibitem[{{Gall} {et~al.}(2014){Gall}, {Hjorth}, {Watson}, {Dwek}, {Maund},
  {Fox}, {Leloudas}, {Malesani}, \& {Day-Jones}}]{gal14}
{Gall}, C., {Hjorth}, J., {Watson}, D., {et~al.} 2014, \nat, 511, 326

\bibitem[{{Gomez} {et~al.}(2012){Gomez}, {Krause}, {Barlow}, {Swinyard},
  {Owen}, {Clark}, {Matsuura}, {Gomez}, {Rho}, {Besel}, {Bouwman}, {Gear},
  {Henning}, {Ivison}, {Polehampton}, \& {Sibthorpe}}]{gom12}
{Gomez}, H.~L., {Krause}, O., {Barlow}, M.~J., {et~al.} 2012, The Astrophysical
  Journal, 760, 96

\bibitem[{{Guhathakurta} \& {Draine}(1989)}]{guhathakurta_1989}
{Guhathakurta}, P. \& {Draine}, B.~T. 1989, \apj, 345, 230

\bibitem[{{Hirashita}(2012)}]{hirashita_2012}
{Hirashita}, H. 2012, \mnras, 422, 1263

\bibitem[{{Hirashita} \& {Kuo}(2011)}]{hirashita_2011}
{Hirashita}, H. \& {Kuo}, T.-M. 2011, \mnras, 416, 1340

\bibitem[{{Hollenbach} \& {McKee}(1979)}]{hol79}
{Hollenbach}, D. \& {McKee}, C.~F. 1979, \apjs, 41, 555

\bibitem[{Hou {et~al.}(2022)Hou, Pascazio, Martin, Zhou, Kraft, \&
  You}]{hou_2022}
Hou, D., Pascazio, L., Martin, J., {et~al.} 2022, Journal of Aerosol Science,
  159, 105866

\bibitem[{{Indebetouw} {et~al.}(2014){Indebetouw}, {Matsuura}, {Dwek},
  {Zanardo}, {Barlow}, {Baes}, {Bouchet}, {Burrows}, {Chevalier}, {Clayton},
  {Fransson}, {Gaensler}, {Kirshner}, {Laki{\'c}evi{\'c}}, {Long}, {Lundqvist},
  {Mart{\'{\i}}-Vidal}, {Marcaide}, {McCray}, {Meixner}, {Ng}, {Park},
  {Sonneborn}, {Staveley-Smith}, {Vlahakis}, \& {van Loon}}]{ind14}
{Indebetouw}, R., {Matsuura}, M., {Dwek}, E., {et~al.} 2014, The Astrophysical
  Journal Letters, 782, L2

\bibitem[{{Inoue} {et~al.}(2020){Inoue}, {Hashimoto}, {Chihara}, \&
  {Koike}}]{inoue_2020}
{Inoue}, A.~K., {Hashimoto}, T., {Chihara}, H., \& {Koike}, C. 2020, \mnras,
  495, 1577

\bibitem[{{Jerkstrand} {et~al.}(2011){Jerkstrand}, {Fransson}, \&
  {Kozma}}]{jerkstrand2011}
{Jerkstrand}, A., {Fransson}, C., \& {Kozma}, C. 2011, Astronomy \&
  Astrophysics, 530, A45

\bibitem[{{Kj{\ae}r} {et~al.}(2010){Kj{\ae}r}, {Leibundgut}, {Fransson},
  {Jerkstrand}, \& {Spyromilio}}]{kjaer_2010}
{Kj{\ae}r}, K., {Leibundgut}, B., {Fransson}, C., {Jerkstrand}, A., \&
  {Spyromilio}, J. 2010, \aap, 517, A51

\bibitem[{{Koike} {et~al.}(1995){Koike}, {Kaito}, {Yamamoto}, {Shibai},
  {Kimura}, \& {Suto}}]{koike_1995}
{Koike}, C., {Kaito}, C., {Yamamoto}, T., {et~al.} 1995, \icarus, 114, 203

\bibitem[{{Kotak} {et~al.}(2009){Kotak}, {Meikle}, {Farrah}, {Gerardy},
  {Foley}, {Van Dyk}, {Fransson}, {Lundqvist}, {Sollerman}, {Fesen},
  {Filippenko}, {Mattila}, {Silverman}, {Andersen}, {H{\"o}flich}, {Pozzo}, \&
  {Wheeler}}]{kot09}
{Kotak}, R., {Meikle}, W.~P.~S., {Farrah}, D., {et~al.} 2009, The Astrophysical
  Journal, 704, 306

\bibitem[{{Kozma} \& {Fransson}(1992)}]{kozma_1992}
{Kozma}, C. \& {Fransson}, C. 1992, \apj, 390, 602

\bibitem[{Kozma \& Fransson(1998)}]{kozma_1998}
Kozma, C. \& Fransson, C. 1998, The Astrophysical Journal, 496, 946

\bibitem[{{Laming} \& {Temim}(2020)}]{laming_2020}
{Laming}, J.~M. \& {Temim}, T. 2020, \apj, 904, 115

\bibitem[{{Lepp} {et~al.}(1990){Lepp}, {Dalgarno}, \& {McCray}}]{lep90}
{Lepp}, S., {Dalgarno}, A., \& {McCray}, R. 1990, The Astrophysical Journal,
  358, 262

\bibitem[{{Liljegren} {et~al.}(2020){Liljegren}, {Jerkstrand}, \&
  {Grumer}}]{liljegren_2020}
{Liljegren}, S., {Jerkstrand}, A., \& {Grumer}, J. 2020, \aap, 642, A135

\bibitem[{{Liu} {et~al.}(2018){Liu}, {Nittler}, {Alexander}, \&
  {Wang}}]{liu_2018}
{Liu}, N., {Nittler}, L.~R., {Alexander}, C. M.~O.~D., \& {Wang}, J. 2018,
  Science Advances, 4, eaao1054

\bibitem[{{Liu} \& {Dalgarno}(1995)}]{liu95}
{Liu}, W. \& {Dalgarno}, A. 1995, The Astrophysical Journal, 454, 472

\bibitem[{{Liu} {et~al.}(1992){Liu}, {Dalgarno}, \& {Lepp}}]{liu92}
{Liu}, W., {Dalgarno}, A., \& {Lepp}, S. 1992, The Astrophysical Journal, 396,
  679

\bibitem[{{Lucy} {et~al.}(1989){Lucy}, {Danziger}, {Gouiffes}, \&
  {Bouchet}}]{luc89}
{Lucy}, L.~B., {Danziger}, I.~J., {Gouiffes}, C., \& {Bouchet}, P. 1989, in
  Lecture Notes in Physics, Berlin Springer Verlag, Vol. 350, IAU Colloq. 120:
  Structure and Dynamics of the Interstellar Medium, ed. G.~{Tenorio-Tagle},
  M.~{Moles}, \& J.~{Melnick}, 164

\bibitem[{{Matsuura}(2017)}]{matsuura2017}
{Matsuura}, M. 2017, {Dust and Molecular Formation in Supernovae} (Springer),
  2125

\bibitem[{Matsuura {et~al.}(2015)Matsuura, Dwek, Barlow, Babler, Baes, Meixner,
  Cernicharo, Clayton, Dunne, Fransson, Fritz, Gear, Gomez, Groenewegen,
  Indebetouw, Ivison, Jerkstrand, Lebouteiller, Lim, Lundqvist, Pearson,
  Roman-Duval, Royer, Staveley-Smith, Swinyard, van Hoof, van Loon, Verstappen,
  Wesson, Zanardo, Blommaert, Decin, Reach, Sonneborn, de~Steene, \&
  Yates}]{matsuura_2015}
Matsuura, M., Dwek, E., Barlow, M.~J., {et~al.} 2015, The Astrophysical
  Journal, 800, 50

\bibitem[{{Matsuura} {et~al.}(2011){Matsuura}, {Dwek}, {Meixner}, {Otsuka},
  {Babler}, {Barlow}, {Roman-Duval}, {Engelbracht}, {Sandstrom},
  {Laki{\'c}evi{\'c}}, {van Loon}, {Sonneborn}, {Clayton}, {Long}, {Lundqvist},
  {Nozawa}, {Gordon}, {Hony}, {Panuzzo}, {Okumura}, {Misselt}, {Montiel}, \&
  {Sauvage}}]{mat11}
{Matsuura}, M., {Dwek}, E., {Meixner}, M., {et~al.} 2011, Science, 333, 1258

\bibitem[{{Moriya} {et~al.}(2013){Moriya}, {Maeda}, {Taddia}, {Sollerman},
  {Blinnikov}, \& {Sorokina}}]{moriya_2013}
{Moriya}, T.~J., {Maeda}, K., {Taddia}, F., {et~al.} 2013, \mnras, 435, 1520

\bibitem[{Niculescu-Duvaz {et~al.}(2021)Niculescu-Duvaz, Barlow, Bevan,
  Milisavljevic, \& De{\^A}~Looze}]{maria_2021}
Niculescu-Duvaz, M., Barlow, M.~J., Bevan, A., Milisavljevic, D., \&
  De{\^A}~Looze, I. 2021, Monthly Notices of the Royal Astronomical Society,
  504, 2133

\bibitem[{{Nozawa} \& {Kozasa}(2013)}]{noz13}
{Nozawa}, T. \& {Kozasa}, T. 2013, The Astrophysical Journal, 776, 24

\bibitem[{{Nozawa} {et~al.}(2010){Nozawa}, {Kozasa}, {Tominaga}, {Maeda},
  {Umeda}, {Nomoto}, \& {Krause}}]{noz10}
{Nozawa}, T., {Kozasa}, T., {Tominaga}, N., {et~al.} 2010, The Astrophysical
  Journal, 713, 356

\bibitem[{{Ott} {et~al.}(2019){Ott}, {Stephan}, {Hoppe}, \&
  {Savina}}]{ott_2019}
{Ott}, U., {Stephan}, T., {Hoppe}, P., \& {Savina}, M.~R. 2019, \apj, 885, 128

\bibitem[{{Owen} \& {Barlow}(2015)}]{owen_2015}
{Owen}, P.~J. \& {Barlow}, M.~J. 2015, \apj, 801, 141

\bibitem[{{Pegourie}(1988)}]{pegourie_1988}
{Pegourie}, B. 1988, \aap, 194, 335

\bibitem[{Priestley {et~al.}(2021)Priestley, Arias, Barlow, \&
  Looze}]{priestley_2021}
Priestley, F.~D., Arias, M., Barlow, M.~J., \& Looze, I.~D. 2021, Monthly
  Notices of the Royal Astronomical Society, 509, 3163

\bibitem[{{Priestley} {et~al.}(2020){Priestley}, {Bevan}, {Barlow}, \& {De
  Looze}}]{priestley_2020}
{Priestley}, F.~D., {Bevan}, A., {Barlow}, M.~J., \& {De Looze}, I. 2020,
  \mnras, 497, 2227

\bibitem[{{Rauscher} {et~al.}(2002){Rauscher}, {Heger}, {Hoffman}, \&
  {Woosley}}]{rau02}
{Rauscher}, T., {Heger}, A., {Hoffman}, R.~D., \& {Woosley}, S.~E. 2002, The
  Astrophysical Journal, 576, 323

\bibitem[{{Sarangi} \& {Cherchneff}(2013)}]{sar13}
{Sarangi}, A. \& {Cherchneff}, I. 2013, The Astrophysical Journal, 776, 107

\bibitem[{{Sarangi} \& {Cherchneff}(2015)}]{sar15}
{Sarangi}, A. \& {Cherchneff}, I. 2015, \aap, 575, A95

\bibitem[{{Sarangi} {et~al.}(2018{\natexlab{a}}){Sarangi}, {Dwek}, \&
  {Arendt}}]{sarangi2018}
{Sarangi}, A., {Dwek}, E., \& {Arendt}, R.~G. 2018{\natexlab{a}}, \apj, 859, 66

\bibitem[{{Sarangi} {et~al.}(2018{\natexlab{b}}){Sarangi}, {Matsuura}, \&
  {Micelotta}}]{sarangi2018book}
{Sarangi}, A., {Matsuura}, M., \& {Micelotta}, E.~R. 2018{\natexlab{b}}, \ssr,
  214, 63

\bibitem[{{Sarangi} \& {Slavin}(2022)}]{sarangi_2022_1_arxiv}
{Sarangi}, A. \& {Slavin}, J. 2022, arXiv e-prints, arXiv:2205.08352

\bibitem[{{Seitenzahl} {et~al.}(2014){Seitenzahl}, {Timmes}, \&
  {Magkotsios}}]{seitenzahl_2014}
{Seitenzahl}, I.~R., {Timmes}, F.~X., \& {Magkotsios}, G. 2014, \apj, 792, 10

\bibitem[{{Simpson}(1991)}]{simpson_1991}
{Simpson}, J.~P. 1991, \apj, 368, 570

\bibitem[{{Slavin} {et~al.}(2020){Slavin}, {Dwek}, {Mac Low}, \&
  {Hill}}]{slavin_2020}
{Slavin}, J.~D., {Dwek}, E., {Mac Low}, M.-M., \& {Hill}, A.~S. 2020, \apj,
  902, 135

\bibitem[{{Sluder} {et~al.}(2018){Sluder}, {Milosavljevi{\'c}}, \&
  {Montgomery}}]{sluder2018}
{Sluder}, A., {Milosavljevi{\'c}}, M., \& {Montgomery}, M.~H. 2018, \mnras,
  480, 5580

\bibitem[{{Sugerman} {et~al.}(2006){Sugerman}, {Ercolano}, {Barlow}, {Tielens},
  {Clayton}, {Zijlstra}, {Meixner}, {Speck}, {Gledhill}, {Panagia}, {Cohen},
  {Gordon}, {Meyer}, {Fabbri}, {Bowey}, {Welch}, {Regan}, \&
  {Kennicutt}}]{sug06}
{Sugerman}, B.~E.~K., {Ercolano}, B., {Barlow}, M.~J., {et~al.} 2006, Science,
  313, 196

\bibitem[{{Sutherland} \& {Dopita}(1995)}]{sutherland_1995}
{Sutherland}, R.~S. \& {Dopita}, M.~A. 1995, \apj, 439, 381

\bibitem[{{Szalai} \& {Vink{\'o}}(2013)}]{sza13}
{Szalai}, T. \& {Vink{\'o}}, J. 2013, Astronomy \& Astrophysics, 549, A79

\bibitem[{Szalai {et~al.}(2019)Szalai, Zs{\'{\i}}ros, Fox, Pejcha, \&
  M{\"u}ller}]{szalai_2019}
Szalai, T., Zs{\'{\i}}ros, S., Fox, O.~D., Pejcha, O., \& M{\"u}ller, T. 2019,
  The Astrophysical Journal Supplement Series, 241, 38

\bibitem[{{Temim} \& {Dwek}(2013)}]{temim_2013}
{Temim}, T. \& {Dwek}, E. 2013, \apj, 774, 8

\bibitem[{{Truelove} \& {McKee}(1999)}]{tru99}
{Truelove}, J.~K. \& {McKee}, C.~F. 1999, \apjs, 120, 299

\bibitem[{{Utrobin} {et~al.}(2017){Utrobin}, {Wongwathanarat}, {Janka}, \&
  {M{\"u}ller}}]{utrobin_2017}
{Utrobin}, V.~P., {Wongwathanarat}, A., {Janka}, H.~T., \& {M{\"u}ller}, E.
  2017, \apj, 846, 37

\bibitem[{{van Breemen} {et~al.}(2011){van Breemen}, {Min}, {Chiar}, {Waters},
  {Kemper}, {Boogert}, {Cami}, {Decin}, {Knez}, {Sloan}, \&
  {Tielens}}]{breemen_2011}
{van Breemen}, J.~M., {Min}, M., {Chiar}, J.~E., {et~al.} 2011, \aap, 526, A152

\bibitem[{{V{\'a}rosi} \& {Dwek}(1999)}]{varosi_1999}
{V{\'a}rosi}, F. \& {Dwek}, E. 1999, \apj, 523, 265

\bibitem[{{Weingartner} \& {Draine}(2001)}]{wei01}
{Weingartner}, J.~C. \& {Draine}, B.~T. 2001, \apj, 563, 842

\bibitem[{{Wesson} {et~al.}(2015){Wesson}, {Barlow}, {Matsuura}, \&
  {Ercolano}}]{wesson2015}
{Wesson}, R., {Barlow}, M.~J., {Matsuura}, M., \& {Ercolano}, B. 2015, \mnras,
  446, 2089

\bibitem[{{Wesson} \& {Bevan}(2021)}]{wesson_2021}
{Wesson}, R. \& {Bevan}, A. 2021, \apj, 923, 148

\bibitem[{{Wongwathanarat} {et~al.}(2015){Wongwathanarat}, {M{\"u}ller}, \&
  {Janka}}]{wong15}
{Wongwathanarat}, A., {M{\"u}ller}, E., \& {Janka}, H.-T. 2015, Astronomy \&
  Astrophysics, 577, A48

\bibitem[{{Wooden} {et~al.}(1993){Wooden}, {Rank}, {Bregman}, {Witteborn},
  {Tielens}, {Cohen}, {Pinto}, \& {Axelrod}}]{woo93}
{Wooden}, D.~H., {Rank}, D.~M., {Bregman}, J.~D., {et~al.} 1993, Astrophysical
  Journal Supplement Series, 88, 477

\bibitem[{{Woosley} {et~al.}(1989){Woosley}, {Hartmann}, \&
  {Pinto}}]{woosley1989}
{Woosley}, S.~E., {Hartmann}, D., \& {Pinto}, P.~A. 1989, The Astrophysical
  Journal, 346, 395

\bibitem[{{Xu} {et~al.}(2015){Xu}, {Zinner}, {Gallino}, {Heger}, {Pignatari},
  \& {Lin}}]{xu_2015}
{Xu}, Y., {Zinner}, E., {Gallino}, R., {et~al.} 2015, \apj, 799, 156

\bibitem[{{Zhukovska} {et~al.}(2008){Zhukovska}, {Gail}, \&
  {Trieloff}}]{zhukovska_2008}
{Zhukovska}, S., {Gail}, H.~P., \& {Trieloff}, M. 2008, \aap, 479, 453

\bibitem[{{Zubko} {et~al.}(2004){Zubko}, {Dwek}, \& {Arendt}}]{zub04}
{Zubko}, V., {Dwek}, E., \& {Arendt}, R.~G. 2004, \apjs, 152, 211

\end{thebibliography}

\end{document}